\documentclass[aps,prd,preprint,groupedaddress,preprintnumbers,longbibliography,nofootinbib]{revtex4-1}
\newcommand{\vvev}[1]{\left\langle\kern-0.3em\left\langle #1
    \right\rangle\kern-0.3em\right\rangle}

\usepackage{amsmath}
\usepackage{graphicx}
\usepackage{amsfonts}
\usepackage{caption}
\usepackage{subcaption}

\usepackage{mathrsfs}

\usepackage[cal=esstix,frak=euler,scr=boondox,bb= pazo]{mathalfa}
\newcommand{\xcal}{\mathcal{x}}
\newcommand{\ycal}{\mathcal{y}}
\newcommand{\xs}{X}
\newcommand{\ys}{Y}

\begin{document}

% Use the \preprint command to place your local institutional report
% number in the upper righthand corner of the title page in preprint mode.
% Multiple \preprint commands are allowed.
% Use the 'preprintnumbers' class option to override journal defaults
% to display numbers if necessary
%\preprint{Mainz preprint number}
\preprint{MITP/20-004}

%Title of paper
\title{The Response Field and the Saddle Points of \\
Quantum Mechanical Path Integrals}

% repeat the \author .. \affiliation  etc. as needed
% \email, \thanks, \homepage, \altaffiliation all apply to the current
% author. Explanatory text should go in the []'s, actual e-mail
% address or url should go in the {}'s for \email and \homepage.
% Please use the appropriate macro foreach each type of information

% \affiliation command applies to all authors since the last
% \affiliation command. The \affiliation command should follow the
% other information
% \affiliation can be followed by \email, \homepage, \thanks as well.
%\author{E.~Gozzi}
\author{Ennio Gozzi}
\email[]{gozzi@ts.infn.it}
\affiliation{
INFN, Section of Trieste,
Via Valerio 2, Trieste, 34100, Italy,
and\\
Phys. Dept. Theoretical Section, Univ. of Trieste, Strada Costiera 11, Miramare, Grignano, Trieste, 34152, Italy}
%
%\author{C.~Pagani}
\author{Carlo Pagani}
\email[]{carlo.pagani@lpmmc.cnrs.fr}
%\homepage[]{Your web page}
%\thanks{}
%\altaffiliation{}
\affiliation{
Univ.~Grenoble Alpes, CNRS, LPMMC, 38000 Grenoble, France}
\affiliation{Institut f\"{u}r Physik (THEP),
  Johannes-Gutenberg-Universit\"{a}t\\ Staudingerweg 7, 55099 Mainz,
  Germany}
%
%\author{M.~Reuter}
\author{Martin Reuter}
\email[]{reutma00@uni-mainz.de}
\affiliation{Institut f\"{u}r Physik (THEP),
  Johannes-Gutenberg-Universit\"{a}t\\ Staudingerweg 7, 55099 Mainz,
  Germany}

%Collaboration name if desired (requires use of superscriptaddress
%option in \documentclass). \noaffiliation is required (may also be
%used with the \author command).
%\collaboration can be followed by \email, \homepage, \thanks as well.
%\collaboration{}
%\noaffiliation

%\date{\today}

\begin{abstract}
In quantum statistical mechanics, 
Moyal's equation governs the time evolution of Wigner functions
and of more general Weyl symbols that represent the density matrix
of arbitrary mixed states.
A formal solution to Moyal's equation is given by Marinov's path integral.
In this paper we demonstrate that this path integral can be regarded
as the natural link between several conceptual, geometric, and
dynamical issues in quantum mechanics.
A unifying perspective is achieved by highlighting the pivotal role 
which the response field,
one of the integration variables in Marinov's integral,
plays for pure states even.
The discussion focuses on how the integral's semiclassical approximation
relates to its strictly classical limit;
unlike for Feynman type path integrals,
the latter is well defined in the Marinov case.
The topics covered include 
a random force representation of Marinov's integral 
based upon the concept of ``Airy averaging'',
a related discussion of positivity-violating Wigner functions
describing tunneling processes, 
and the role of the response field in maintaining quantum coherence
and enabling interference phenomena.
The double slit experiment for electrons and the Bohm-Aharonov effect are
analyzed as illustrative examples.
Furthermore, a surprising relationship between the instantons of the 
Marinov path integral over an analytically continued (``Wick rotated'')
response field, and the complex instantons of Feynman-type integrals
is found.
The latter play a prominent role in recent work towards a Picard-Lefschetz theory
applicable to oscillatory path integrals and the resurgence program.
\end{abstract}

% insert suggested PACS numbers in braces on next line
%\pacs{}
% insert suggested keywords - APS authors don't need to do this
%\keywords{}

%\maketitle must follow title, authors, abstract, \pacs, and \keywords
\maketitle

\section{Introduction} \label{sec:introduction}

The present work is dedicated to the phase space formulation of quantum statistical mechanics, and more specifically to the so-called Marinov path integral \cite{MARINOV1991}, which governs the time evolution of the phase space functions that represent density matrices.
While the more familiar Feynman path integral is a formal solution to the Schr{\"o}dinger equation, 
Marinov's integral provides the corresponding solution to the Moyal equation \cite{Moyal:1949sk,Weyl:1927vd,Wigner:1932eb,GROENEWOLD1946405}.
The latter reads $\partial_t \rho = -\left\lbrace \rho, H \right\rbrace_{\rm M}$,
whereby $H$ and $\rho$ are the phase space representatives of the Hamiltonian and the statistical operator (density matrix), respectively,
and $\left\lbrace \cdot , \cdot \right\rbrace_{\rm M}$ denotes the Moyal bracket, a deformation of the Poisson bracket that will be discussed below.

{\noindent \bf (1) Marinov's variables.}
In \cite{MARINOV1991} Marinov derived the path integral by discretization techniques which can be applied to all types of functional integrals, but teach us little about certain rather unusual and intriguing features exhibited by the Marinov integral.
This concerns in particular the physical and geometric interpretation of the variables of integration, 
$\phi \left(t\right)$ and $\xi \left(t\right)$ in the Hamiltonian case,
and $\xs \left(t\right)$ with  $\ys \left(t\right)$ in the Lagrangean case.
In either case, the integration is related to a {\it pair} of trajectories on phase space or configuration space, respectively.
Another remarkable feature of Marinov's path integral is the {\it bi-local} structure of its integrand.

In the present paper we are going to address these issues starting out from a new, physically and geometrically more illuminating derivation of the integral.

{\noindent \bf (2) Complex saddle points of Feynman integrals.}
Recently a lot of work went into the {\it resurgence program} which aims at applying a generalized form of Picard-Lefschetz theory to the notoriously hard to define oscillatory functional integrals that occur in quantum mechanics and quantum field theory, see e.g.~\cite{Behtash:2015loa,Behtash:2015zha,Feldbrugge:2017kzv}.
In this context it has been re-emphasized that, even though a Wick rotation to the imaginary time $\tau \equiv i t $ may have rendered the integral real, also {\it complex} saddle points (instantons) can play an essential rule in the semiclassical limit \cite{Dunne:2015eaa,Dunne:2016nmc,Dunne:2016qix}.

Considering a Feynman-type path integral for one degree of freedom, say,
the rotated integral
$ \int {\cal D} x\left(\cdot \right)\, e^{-\frac{1}{\hbar}S_{\rm E} \left[x\left(\cdot \right) \right]}$,
with $S_{\rm E} \left[x\left(\cdot \right) \right]
\equiv \int d\tau \left\lbrace \frac{1}{2} \left(\partial_\tau x \right)^2 + V\left(x\right) \right\rbrace$
is over real-valued functions $x\left(\tau \right)$.
Nevertheless its semiclassical expansion may receive physically important contributions from saddle points of the complexified path integral.
They are found by solving the {\it holomorphic Newton equation} with inverted potential:
\begin{eqnarray}
\partial^2_\tau z\left(\tau \right) = +\frac{d}{dz}V \left(z\left(\tau \right) \right) \,. \label{eq:1-1}
\end{eqnarray}
Its solutions $z\left(\tau \right) \equiv x\left(\tau \right)+i\,  y\left(\tau \right)$ correspond to a {\it pair} of real functions, $x\left(\tau \right)$ and $y\left(\tau \right)$, obviously.
They satisfy the coupled system of equations
\begin{eqnarray}
\begin{split}
\partial^{2}_\tau x \left(\tau \right) & =
\frac{1}{2} \left[ V^\prime \left(x+i\, y \right)+V^\prime \left(x-i\, y \right) \right] \\
\partial^{2}_\tau y \left(\tau \right) & =
\frac{1}{2i} \left[ V^\prime \left(x+i\, y \right)-V^\prime \left(x-i\, y \right) \right] \,.
\end{split} \label{eq:1-2}
\end{eqnarray} 
To rewrite the equations in a manifestly real form we decompose the holomorphic potential in real and imaginary parts,
$V\left(z\right)=V\left(x+i\, y\right)\equiv V_{\rm R}\left(x,y\right)+i\, V_{\rm I}\left(x,y\right)$,
and disentangle (\ref{eq:1-1}) accordingly:
\begin{eqnarray}
\begin{split}
\partial^{2}_\tau x \left(\tau \right) & =
\partial_x V_{\rm R} \left(x , y \right)\\
\partial^{2}_\tau y \left(\tau \right) & =
-\partial_y V_{\rm R} \left(x , y \right) \,.
\end{split} \label{eq:1-3}
\end{eqnarray} 
Here the Cauchy-Riemann equations have been exploited \cite{Behtash:2015loa,Behtash:2015zha,Tanizaki:2014xba,Cherman:2014sba,Feldbrugge:2017kzv,Witten:2010cx,Witten:2010zr}.

In this framework, the (Euclideanized) path integral under consideration time-evolves wave functions and the doubling of the configuration space variables, $x\rightarrow \left(x,y\right)$, is due to the complexification.
The Marinov path integral, on the other hand, time-evolves Wigner functions or density operators, and it involves a similar doubling of configuration space, $x\rightarrow \left(\xs ,\ys \right)$, from the outset.

One of our present goals will consist in showing that the two settings are very closely related, and that the pairs $\left(x,y \right)$ and $\left(\xs ,\ys \right)$ are almost the same thing.
In particular, we derive and study the $\left(\xs ,\ys \right)$-analogue of the coupled system of equations (\ref{eq:1-3}).

{\noindent \bf (3) The response field.}
Another connection we are going to highlight in this paper is between Marinov's $\ys$-variable and the so-called {\it response field},
which is well known from statistical physics for instance.

Historically, the response field formalism was developed to cast the study of correlation functions related to Langevin equations  
in a path integral form  \cite{Martin:1973zz,Janssen1976,DEDOMINICIS1976}.
In this setting the response field can be looked at as an auxiliary field that allows one to compute the linear response of some field 
under an external perturbation by computing correlation functions with the associated response field.

Technically, the response field is introduced in order to express a functional Dirac delta as an integral over the response field \cite{Tauber:2014}.
This is the very same mechanism which localizes the Marinov path integral on the equation of motion in the strictly classical limit, 
as we shall discuss in Section \ref{subsec:Hamiltonian-path-integral}.
It will also become clear that the Marinov path integral is closely related to the Schwinger-Keldysh formalism after the so called ``Keldysh rotation'' \cite{Kamenev:2011}.

{\noindent \bf (4) Role of the response field in semiclassical quantum mechanics.}
In this paper we also analyze elementary quantum mechanical systems, 
in pure quantum states, 
using Wigner functions that are time-evolved by the Marinov path integral over $\xs \left(t \right)$ and $\ys \left(t \right)$.
In the semiclassical limit we invoke the saddle point approximation and are thus led to study coupled classical equations of motion for two configuration space trajectories, $\xs \left(t\right)$ and $\ys \left(t\right)$. 
By means of various representative examples, including the double slit experiment for electrons and the Bohm-Aharonov effect, we describe the physical role played by the response field $\ys \left(t\right)$.
The latter will turn out essential for the occurrence of interference phenomena and the preservation of quantum coherence during the time evolution.
In this manner it will become clear why the semiclassical limit of quantum (statistical) mechanics must be described by {\it two} configuration- or phase space-trajectories.

{\noindent \bf (5) The strictly classical point $\hbar =0$.}
A familiar textbook argument about the emergence of classical behaviour from quantum mechanics is as follows:
In the semiclassical limit, i.e., when effectively $\hbar \rightarrow 0$, the integrand of the Feynman path integral
\begin{eqnarray}
\int {\cal D} x\left(\cdot \right)\, e^{\frac{i}{\hbar} S\left[x\left(\cdot \right)\right]} \label{eq:1-10}
\end{eqnarray}  
is rapidly oscillating.
Hence contributions from different paths $x\left(t\right)$ mutually cancel by destructive inteference, unless the path $x\left(t\right)$ is a stationary point of the action, in which case it may interfere constructively with its neighbours.
This then leads to the conclusion that the classical trajectories $x^{\rm SP} \left(t\right)$ emerging from the quantum system governed by (\ref{eq:1-10}) are ruled by the saddle point condition
\begin{eqnarray}
\left(\frac{\delta S}{\delta x\left(t\right)}\right)
\left[ x^{\rm SP} \left(\cdot \right) \right]=0
\end{eqnarray}
which is then interpreted as the equation of motion of classical mechanics.

While there is nothing wrong with this argument
(as long as the stationary phase approximation can be justified),
it tends to convey a physical picture that is quite misleading, however.

Namely, since the ``classical path'' $x^{\rm SP} \left(t\right)$ is just one among the infinitely many $x \left(t\right)$ that contribute to (\ref{eq:1-10}), the argument tacitly creates the impression that the trajectories occurring in strictly classical mechanics, henceforth denoted $\xs_{\rm cl}\left(t \right)$, are conceptually of the same nature as the $x \left(t \right)$'s that are integrated over in (\ref{eq:1-10}).
As we are going to show, this impression is false, in a sense that will be made precise below.

Even though the above argument  identifies the correct {\it form}
of the equation of motion in classical mechanics, it suggests the wrong variables for strictly classical mechanics:
The saddle point of the Feynman path integral, $x^{\rm SP} \left(t\right)$, is {\it not} the semiclassical precursor of the variable $\xs_{\rm cl}\left(t \right)$ used in strictly classical mechanics.
The actual quantum ancestor of $\xs_{\rm cl}\left(t \right)$ are the $\xs$-components of saddle points 
$\left(\xs^{\rm SP} \left(t\right),\ys^{\rm SP} \left(t\right) \right)$ pertaining to Marinov's path integral.

A crucial difference between Marinov's and Feynman's path integral is that the former continues to be meaningful when we specialize for {\it strictly} classical mechanics by setting $\hbar =0$ exactly,
while the latter is undefined at $\hbar =0$ due to infinitely rapid oscillations.
In fact, the Marinov path integral evaluated at $\hbar =0$ is closely related to the so-called Classical Path Integral (CPI) that has been studied extensively
\cite{Gozzi:1986ge,Gozzi:1989bf,Gozzi:1989vv,Gozzi:1989xz,Gozzi:1999at,Gozzi:2010iq,Gozzi:2010iu}.

This is yet another connection of Marinov's integral we are going to detail. 
In this manner it will become manifest that the configuration space variable $\xs_{\rm cl}$ employed by classical mechanics must be seen as a symmetric average
\begin{eqnarray}
\xs_{\rm cl} \left(t \right) =
\frac{1}{2} \left[x_+^{\rm SP} \left(t \right) +x_-^{\rm SP} \left(t \right) \right]
\label{eq:1-20}
\end{eqnarray}
where $x_+^{\rm SP} \left(t \right)$ and $x_-^{\rm SP} \left(t \right)$ are saddle points of, respectively, the original Feynman path integral (\ref{eq:1-10}) and the time-reversal thereof.
An analogous statement holds for classical phase space variables.

{\noindent \bf (6) Plan of the paper.}
The rest of this paper is organized as follows.
In Section \ref{sec:preliminaries} we briefly review the main aspects of
the Wigner-Weyl-Moyal framework for the phase space formulation of
quantum (statistical) mechanics.

Then, in Section \ref{sec:Path-integral-representation-Moyal-kernel},
we construct both the Lagrangean and the Hamiltonian versions of the 
Marinov path integral by sewing together two Feynman path integrals,
paying special attention to the interrelations of the various sets of integration variables that are involved.
In Section \ref{sec:Path-integral-representation-Moyal-kernel} we also 
discuss two specific applications, namely, first, we give a detailed account
of how precisely the kinematical variables that appear in the standard classical 
mechanics relate to those that are provided naturally by the various path integrals.
And second, we derive a new random force representation of the Moyal-Marinov
kernel and use it to pin point the condition under which Wigner functions
can assume negative values so that their classical density interpretation breaks down.
Tunneling phenomena will be discussed as a typical example.

Thereafter, in Section \ref{sec:coherence-interference-resp_field},
we demonstrate the relevance of the response field for interference phenomena
and the preservation of quantum coherence,
and in Section \ref{sec:Double-Slit-Experiment-and-BohmAharonov} we 
illustrate this discussion by means of two concrete examples:
the double slit experiment, and
the Bohm-Aharonov effect.
Section \ref{sec:Wick-rotated-instantons} finally is devoted to the 
relationship between complex saddle points of Feynman integrals and
Marinov's path integral,
and section \ref{sec:conclusions} contains the conclusions.

Let us mention that saddle points of the propagation kernel of Wigner functions have been studied also in \cite{Dittrich2006}. 
In the present work we perform a wider analysis and discuss a number of points that have not been studied in detail in the literature so far.

\section{Preliminaries} \label{sec:preliminaries}

We are going to consider an arbitrary quantum system in the following.
We assume that it has $N$ degrees of freedom, and that its time evolution
is governed by the Hamiltonian operator $\widehat{H}$.
\vspace{3mm}\\ 
\noindent{\bf (1) The Feynman kernel.}
Being interested in both pure and mixed states, we describe the dynamics of the system
in terms of a time dependent density operator $\widehat{\rho}\left(t\right)$.
We adopt the Schr{\"o}dinger picture so that the density operator
obeys von Neumann's equation
\begin{eqnarray}
i\hbar \partial_{t}\widehat{\rho} 
& = & 
-\left[\widehat{\rho},\widehat{H}\right]\,.\label{eq:2-1_vonNeumann-evol-density-mat}
\end{eqnarray}
Its formal solution reads
\begin{eqnarray}
\widehat{\rho}\left(t\right) & = & \widehat{{\cal U}}\left(t;t_{0}\right)\widehat{\rho}\left(t_{0}\right)\widehat{{\cal U}}\left(t;t_{0}\right)^{\dagger}\label{eq:2-2-density-mat_via_evol-operator}
\end{eqnarray}
where the time evolution operator $\widehat{{\cal U}}\left(t;t_{0}\right)$
evolves pure states according to $|\psi\left(t\right)\rangle=\widehat{{\cal U}}\left(t;t_{0}\right)|\psi\left(t_{0}\right)\rangle$
and satisfies
\begin{eqnarray}
i\hbar \, \widehat{{\cal U}}\left(t;t_{0}\right) 
& = & 
\widehat{H} \, \widehat{{\cal U}}\left(t;t_{0}\right)\,,\,\,
\widehat{{\cal U}}\left(t_{0};t_{0}\right)=1\,.\label{eq:2-3_evolution-eq_for_U}
\end{eqnarray}
The evolution kernel, i.e., its position matrix elements $\left(x^{\prime},x^{\prime\prime}\in\mathbb{R}^{N}\right)$,
\begin{eqnarray}
K\left(x^{\prime},t;x^{\prime\prime},t_{0}\right) & = & \langle x^{\prime}|\widehat{{\cal U}}\left(t;t_{0}\right)|x^{\prime\prime}\rangle\label{eq:2-4_kernel_as_mat-element_of_U}
\end{eqnarray}
will often be addressed as the {\it Feynman kernel} in the sequel since
the familiar Feynman-type functional integrals provide a path integral representation
of exactly this object.
\vspace{3mm}\\ 
\noindent{\bf (2) The symbol calculus.} Furthermore, we employ
the (Weyl-Wigner-Moyal) phase space formulation of quantum mechanics
and make extensive use of the corresponding symbol calculus \cite{Moyal:1949sk,Weyl:1927vd,Wigner:1932eb,GROENEWOLD1946405,VanHove1951}.
The idea is to set up a linear one-to-one relation, a ``symbol map'', between
operators $\widehat{A},\widehat{B},\cdots$ acting on the quantum mechanical
Hilbert space, and complex valued functions $A,B,\cdots$ defined on
the system's classical phase space manifold, ${\cal M}$. We denote
the symbol which represents the operator $\widehat{A}$ by $A\equiv{\rm symb}\left(\widehat{A}\right)$,
and since the symbol map possesses a well defined inverse we may write
$\widehat{A}={\rm symb}^{-1}\left(A\right)$ for the operator given
by the classical phase space function $A$.

There exists a large variety of different symbol maps. Each one comes
with a specific \emph{star product} which implements the operator
multiplication in the function space of symbols. It is defined by
the requirement that the symbol map be an algebra homomorphism, i.e., that
\begin{eqnarray}
{\rm symb}\left(\widehat{A}\widehat{B}\right) & = & {\rm symb}\left(\widehat{A}\right)*{\rm symb}\left(\widehat{B}\right)\label{eq:2-10_symb_of_product_of_opers}
\end{eqnarray}
for any pair of operators. Hence the star product is non-commutative,
but associative. It can be seen as a ``quantum deformation'' of
the ordinary pointwise product of functions \cite{Moyal:1949sk}.

In this paper we employ the \emph{Weyl symbol} whose symbol map can
be described as follows. If we are given an operator $\widehat{A}$,
represented by means of its position space matrix elements $\langle x^{\prime}|\widehat{A}|x^{\prime\prime}\rangle$,
the associated symbol map is given by
\begin{eqnarray}
A\left(p,q\right) & = & \int d^{N}s\,\langle q+\frac{s}{2}|\widehat{A}|q-\frac{s}{2}\rangle e^{-\frac{i}{\hbar}sp}\,.\label{eq:2-11_Weyl-symb-map}
\end{eqnarray}
Here $p\equiv\left(p^{1},\cdots,p^{N}\right)$ and $q\equiv\left(q^{1},\cdots,q^{N}\right)$
are global Darboux coordinates on phase space which here and in the
following is assumed to be ${\cal M}=\mathbb{R}^{N}\times\mathbb{R}^{N}$.
Throughout the paper, the $\mathbb{R}^{N}$-indices are usually left
implicit, and the corresponding scalar products are understood, i.e., $sp\equiv\sum_{i=1}^{N}s^{i}p^{i}$.
Conversely, given a Weyl symbol $A\left(p,q\right)$ the operator
related to it can be recovered in terms of its position space matrix
elements by means of the integral
\begin{eqnarray}
\langle x^{\prime}|\widehat{A}|x^{\prime\prime}\rangle & = & \int\frac{d^{N}p}{\left(2\pi\hbar\right)^{N}}A\left(p,\frac{x^{\prime}+x^{\prime\prime}}{2}\right)e^{\frac{i}{\hbar}p\left(x^{\prime}-x^{\prime\prime}\right)}\,.\label{eq:2-12_map_from_symb_to_mat-element}
\end{eqnarray}

Picking a specific symbol map means fixing an operator ordering prescription
since it associates a unique operator $\widehat{A}\left(\hat{p},\hat{q}\right)={\rm symb}^{-1}\left(A\left(p,q\right)\right)$
to the classical phase space function $A\left(p,q\right)$. The name
``Weyl symbol'' derives from the fact that in our case the operator
thus obtained is always \emph{Weyl ordered}, e.g., ${\rm symb}^{-1}\left(pq\right)=\frac{1}{2}\left(\hat{p}\hat{q}+\hat{q}\hat{p}\right)$
when $N=1$.

The pertinent star product is most compactly displayed by combining
the position and momentum coordinates in $\phi^{a}\equiv\left(p^{1},\cdots,p^{N},q^{1},\cdots,q^{N}\right)$
and $\partial_{a}\equiv\partial/\partial\phi^{a}$, and adopting the summation
convention for the indices $a,b,\cdots=1,\cdots,2N$:
\begin{eqnarray}
\left(A*B\right)\left(\phi\right) & = & A\left(\phi\right)\exp\left(i\frac{\hbar}{2}\overleftarrow{\partial}_{a}\omega^{ab}\overrightarrow{\partial}_{b}\right)B\left(\phi\right)\,.\label{eq:2-13_def_Weyl_star_product}
\end{eqnarray}
Here $\omega^{ab}$ is the constant Poisson tensor, a $2N\times2N$
block matrix with $\omega^{qp}=-\omega^{pq}=\mathbb{I}$ and $\omega^{pp}=\omega^{qq}=0$.
While this is by no means obvious, it can be verified that the product
(\ref{eq:2-13_def_Weyl_star_product}) is indeed associative. As the
RHS of (\ref{eq:2-13_def_Weyl_star_product}) is analytic in the ``deformation
parameter'' $\hbar$, the star product is a smooth deformation of
the pointwise multiplication: $\left(A*B\right)\left(\phi\right)=A\left(\phi\right)B\left(\phi\right)+O\left(\hbar\right)$.

Furthermore one defines the \emph{Moyal bracket} of two symbols by
\begin{eqnarray}
\left\{ A,B\right\} _{{\rm M}} & = & \frac{1}{i\hbar}\left(A*B-B*A\right)\,=\,{\rm symb}\left(\frac{1}{i\hbar}\left[\widehat{A},\widehat{B}\right]\right)\label{eq:2-14_def_Moyal_bracket}
\end{eqnarray}
or, more explicitly,
\begin{eqnarray}
\left\{ A,B\right\} _{{\rm M}} & = & A\left(\phi\right)\frac{2}{\hbar}\sin\left(\frac{\hbar}{2}
\,\overleftarrow{\partial_{a}}\omega^{ab}\overrightarrow{\partial_{b}}\right)B\left(\phi\right)\,.\label{eq:2-15_Moyal_bracket_via_exp-sin}
\end{eqnarray}
The associativity of the star product implies that the Moyal bracket
satisfies the Jacobi identity. In fact, the Moyal bracket has the
same algebraic properties as the commutator of operators. In the classical
limit $\hbar\rightarrow0$ it approaches the Poisson bracket: $\left\{ A,B\right\} _{{\rm M}}=\left\{ A,B\right\} _{{\rm P}}+O\left(\hbar^{2}\right)$.
In our notation the latter reads $\left\{ A,B\right\} _{{\rm P}}=\partial_{a}A\omega^{ab}\partial_{b}B$.

Applying the symbol map to the density matrix operator $\widehat{\rho}$
we obtain the symbol $\rho\left(p,q\right)$ which has the interpretation
of a \emph{pseudodensity} function on the phase space. For pure states
$\widehat{\rho}=|\psi\rangle\langle\psi|$ this pseudodensity is known
as the \emph{Wigner function} related to the wave function $\psi\left(x\right)=\langle x|\psi\rangle$:
\begin{eqnarray}
W_{\psi}\left(p,q\right) & = & \int d^{N}s\,\psi\left(q+\frac{s}{2}\right)\psi^{*}\left(q-\frac{s}{2}\right)e^{-\frac{i}{\hbar}sp}\,.\label{eq:2-20_def_Wigner_function}
\end{eqnarray}
While arbitrary symbols $\rho\left(p,q\right)$ and Wigner functions
$W_{\psi}\left(p,q\right)$ are not in general positive functions,
the $p$- and $q$-integrals of the latter equal the ordinary densities
on configuration and momentum space, respectively,
\begin{eqnarray}
\int\frac{d^{N}p}{\left(2\pi\hbar\right)^{N}}W_{\psi}\left(p,q\right) & = & \left|\psi\left(q\right)\right|^{2}\,,\label{eq:2-21-funz-d-onda-via-Wigner-func}\\
\int\frac{d^{N}q}{\left(2\pi\hbar\right)^{N}}W_{\psi}\left(p,q\right) & = & \left|\widetilde{\psi}\left(p\right)\right|^{2}\,,\label{eq:2-22-funz-d-onda-P-via-Wigner-func}
\end{eqnarray}
with the Fourier transform $\widetilde{\psi}\left(p\right)\equiv\int d^{N}x\,\psi\left(x\right)\exp\left(-ipx/\hbar\right)$.
\vspace{3mm} \\
{\bf (3) The Moyal-Marinov kernel.}
While the dynamics of $\widehat{\rho}\left(t\right)$ is governed
by von Neumann's equation, the analogous dynamical equation for
its time dependent symbol $\rho\left(p,q,t\right)\equiv\rho\left(\phi,t\right)$
is \emph{Moyal's equation}:
\begin{eqnarray}
\partial_{t}\rho\left(\phi,t\right) & = & -\left\{ \rho,H\right\} _{{\rm M}}\,=\,\frac{2}{\hbar}H\left(\phi\right)\sin\left( \frac{\hbar}{2}
\,\overleftarrow{\partial_{a}}\omega^{ab}\overrightarrow{\partial_{b}}\right)\rho\left(\phi,t\right)\,.\label{eq:2-25_Moyal_evolution-eq_for_rho}
\end{eqnarray}
Here $H\equiv{\rm symb}\left(\widehat{H}\right)$. In the limit $\hbar\rightarrow0$,
Moyal's equation goes over to the standard Liouville equation of classical
statistical physics, $\partial_{t}\rho=-\left\{ \rho,H\right\} _{{\rm P}}$.

We write the formal solution to equation (\ref{eq:2-25_Moyal_evolution-eq_for_rho})
in the form
\begin{eqnarray}
\rho\left(\phi^{\prime},T\right) & = & \int d^{2N}\phi^{\prime\prime}
\, K_{{\rm M}}\left(\phi^{\prime},T;\phi^{\prime\prime},T_{0}\right)\rho\left(\phi^{\prime\prime},T_{0}\right)\label{eq:2-26-formal_def_Moyal_kernel}
\end{eqnarray}
and refer to $K_{{\rm M}}$ as the Moyal-, or Marinov- kernel. In fact,
Marinov \cite{MARINOV1991} has constructed a path integral representation
of $K_{{\rm M}}$. In a slightly symbolic notation\footnote{See ref.~\cite{MARINOV1991} for details concerning the discretization
behind (\ref{eq:2-27_def_PI_for_Moyal_kernel}). Note also that in \cite{MARINOV1991} the variable $\zeta\equiv2\hbar\xi$ is used instead
of our $\xi$.} it reads
\begin{eqnarray}
K_{{\rm M}}\left(\phi^{\prime},T;\phi^{\prime\prime},T_{0}\right) & = & \int{\cal D}\phi^{a}\left(\cdot\right){\cal D}\xi^{a}\left(\cdot\right)\nonumber \\
 &  & \exp\left(-2i\int_{T_{0}}^{T}dt\left\{ \dot{\phi}^{a}\left(t\right)\omega_{ab}\xi^{b}\left(t\right)-\widetilde{H}\left(\phi\left(t\right),\xi\left(t\right)\right)\right\} \right)\label{eq:2-27_def_PI_for_Moyal_kernel}
\end{eqnarray}
Here
\begin{eqnarray}
\widetilde{H}\left(\phi,\xi\right) & \equiv & \frac{1}{2\hbar}\Bigr[H\left(\phi-\hbar\xi\right)-H\left(\phi+\hbar\xi\right)\Bigr]\,,\label{eq:2-28_tilde_H-via-phi_and_xi}
\end{eqnarray}
and $\omega_{ab}$ is the matrix inverse of $\omega^{ab}$, i.e.,
$\omega_{ab}\omega^{bc}=\delta_{a}^{c}$. The functional integration
is over two $2N$-component functions, $\phi^{a}\left(t\right)$ and
$\xi^{a}\left(t\right)$, respectively, whereby the former are constrained
by the boundary conditions $\phi\left(T\right)=\phi^{\prime}$ and
$\phi\left(T_{0}\right)=\phi^{\prime\prime}$. There are no such conditions
on $\xi\left(t\right)$.

In his work, Marinov derived the path integral for $K_{{\rm M}}$
by closely following Feynman's strategy in his derivation of the path
integral for $K$, applying it however to the Moyal's equation (\ref{eq:2-25_Moyal_evolution-eq_for_rho})
rather than the Schr\"odinger equation. Thereby the main problem
consists in the iteration of the evolution kernel for the infinitesimal
time differences. 

In the next section we shall give a different proof of
(\ref{eq:2-27_def_PI_for_Moyal_kernel}) which is simpler and more
illuminating. In particular it will allow us to understand the origin
of the somewhat mysterious finite-difference character of the Hamiltonian
$\widetilde{H}\left(\phi,\xi\right)$
in eq.~(\ref{eq:2-28_tilde_H-via-phi_and_xi}).

\section{Path integral representation of the Moyal kernel \label{sec:Path-integral-representation-Moyal-kernel}}

In this section we derive several variants of a path integral representation
for the Moyal kernel, including Marinov's. We express $K_{{\rm M}}$
as the convolution of two Feynman kernels, then represent each one
of them by the well known functional integral which time-evolves pure
states, and finally perform a crucial change of integration variables.

As it will turn out, this change of variables is of a certain conceptual
significance. In particular it constitutes a well defined point of
contact between the different dynamical variables that we traditionally employ
in quantum and in classical mechanics, respectively. This will shed light
on aspects of their interrelation that get obscured if one restrics
the investigation of the $\hbar\rightarrow0$ limit to pure states
at a too early stage.

We start out from equation (\ref{eq:2-2-density-mat_via_evol-operator})
expressed in terms of position space matrix elements:
\begin{eqnarray}
\langle x^{\prime}|\widehat{\rho}\left(T\right)|x^{\prime\prime}\rangle & = & \int d^{N}y^{\prime}d^{N}y^{\prime\prime}K\left(x^{\prime},T;y^{\prime},T_{0}\right)K\left(x^{\prime\prime},T;y^{\prime\prime},T_{0}\right)^{*}\langle y^{\prime}|\widehat{\rho}\left(T_{0}\right)|y^{\prime\prime}\rangle\,.\label{eq:3-1_rho_x-xp_via_F-kernels}
\end{eqnarray}
We switch to symbols by applying (\ref{eq:2-11_Weyl-symb-map}) and
(\ref{eq:2-12_map_from_symb_to_mat-element}) on the LHS and RHS of
(\ref{eq:3-1_rho_x-xp_via_F-kernels}), respectively, and obtain
\begin{eqnarray}
\rho\left(p,q,T\right) & = & \int\frac{d^{N}p^{\prime}}{\left(2\pi\hbar\right)^{N}}\int d^{N}s\,e^{-\frac{i}{\hbar}sp}\int d^{N}y^{\prime}\,d^{N}y^{\prime\prime}K\left(q+\frac{s}{2},T;y^{\prime},T_{0}\right)\nonumber \\
 &  & \times K\left(q-\frac{s}{2},T;y^{\prime\prime},T_{0}\right)^{*}\rho\left(p^{\prime},\frac{y^{\prime}+y^{\prime\prime}}{2},T_{0}\right)e^{\frac{i}{\hbar}p^{\prime}\left(y^{\prime}-y^{\prime\prime}\right)}\,.\label{eq:3-2_rho-symb_via_F-kernels}
\end{eqnarray}
Now we trade $y^{\prime}$ and $y^{\prime\prime}$ for two new variables
of integration, viz., $q^{\prime}\equiv\left(y^{\prime}+y^{\prime\prime}\right)/2$
and $s^{\prime}\equiv y^{\prime}-y^{\prime\prime}$. This leads to
an equation of the form (\ref{eq:2-26-formal_def_Moyal_kernel}),
i.e.,
\begin{eqnarray}
\rho\left(p,q,T\right) & = & \int d^{N}p^{\prime}d^{N}q^{\prime}\,K_{{\rm M}}\left(p,q,T;p^{\prime},q^{\prime},T_{0}\right)\rho\left(p^{\prime},q^{\prime},T_{0}\right)\label{eq:3-3_rho-T_via_Moyal_kernel}
\end{eqnarray}
wherein the sought-for integral kernel emerges as
\begin{eqnarray}
K_{{\rm M}}\left(p,q,T;p^{\prime},q^{\prime},T_{0}\right) & = & \frac{1}{\left(2\pi\hbar\right)^{N}}\int d^{N}s\int d^{N}s^{\prime}\,e^{\frac{i}{\hbar}\left(s^{\prime}p^{\prime}-sp\right)}K\left(q+\frac{s}{2},T;q^{\prime}+\frac{s^{\prime}}{2},T_{0}\right)\nonumber \\
 &  & \times K\left(q-\frac{s}{2},T;q^{\prime}-\frac{s^{\prime}}{2},T_{0}\right)^{*}\,.\label{eq:3-4_Moyal_kernel_via_Feynman-kernels}
\end{eqnarray}
The convolution-type integral formula (\ref{eq:3-4_Moyal_kernel_via_Feynman-kernels})
allows us to compute the Moyal kernel $K_{{\rm M}}$ from two copies
of the time evolution kernel for pure states, $K$.

For later use we define the function
\begin{eqnarray}
{\cal K}\left(s,q,T;s^{\prime},q^{\prime},T_{0}\right) & \equiv & K\left(q+\frac{s}{2},T;q^{\prime}+\frac{s^{\prime}}{2},T_{0}\right)K\left(q-\frac{s}{2},T;q^{\prime}-\frac{s^{\prime}}{2},T_{0}\right)^{*}\label{eq:3-10_Y-function_via_F-kernels}
\end{eqnarray}
and mention that the integral transformation (\ref{eq:3-4_Moyal_kernel_via_Feynman-kernels})
can be inverted to yield
\begin{eqnarray}
{\cal K}\left(s,q,T;s^{\prime},q^{\prime},T_{0}\right) & = & \frac{1}{\left(2\pi\hbar\right)^{N}}\int d^{N}p\int d^{N}p^{\prime}e^{-\frac{i}{\hbar}\left(s^{\prime}p^{\prime}-sp\right)}K_{{\rm M}}\left(p,q,T;p^{\prime},q^{\prime},T_{0}\right)\,.\label{eq:3-11_Y-function_via_Moyal-kernel}
\end{eqnarray}

The equal-time limit $\lim_{T\rightarrow T_{0}}K\left(q,T;q^{\prime},T_{0}\right)=\delta\left(q-q^{\prime}\right)$
entails correspondingly
\begin{eqnarray}
\lim_{T\rightarrow T_{0}}K_{{\rm M}}\left(p,q,T;p^{\prime},q^{\prime},T_{0}\right) & = & \delta\left(p-p^{\prime}\right)\delta\left(q-q^{\prime}\right)\,,\label{eq:3-12_equal_time_lim_for_K-M}\\
\lim_{T\rightarrow T_{0}}{\cal K}\left(s,q,T;s^{\prime},q^{\prime},T_{0}\right) & = & \delta\left(s-s^{\prime}\right)\delta\left(q-q^{\prime}\right)\,.\label{eq:3-13-equal-time-lim-for-Y-funct}
\end{eqnarray}

We also remark that the new kernel function ${\cal K}$, \emph{at
vanishing $s$-arguments}, is manifestly positive definite, being
the pointwise modulus square of two Feynman kernels:
\begin{eqnarray}
{\cal K}\left(0,q,T;0,q^{\prime},T_{0}\right) & = & \frac{1}{\left(2\pi\hbar\right)^{N}}\int d^{N}p\int d^{N}p^{\prime}\,K_{{\rm M}}\left(p,q,T;p^{\prime},q^{\prime},T_{0}\right)\nonumber \\
 & = & \left|K\left(q,T;q^{\prime},T_{0}\right)\right|^{2}\,.\label{eq:3-14-Y-funct_at_s=00003D0}
\end{eqnarray}

\subsection{Lagrangian Path Integrals \label{subsec:Lagrangian-Path-Integrals}}

{\noindent \bf (1) The functional change of variables.} 
Next we exploit the convolution representation (\ref{eq:3-4_Moyal_kernel_via_Feynman-kernels})
of $K_{{\rm M}}$ in order to derive a path integral for the Moyal
kernel. We begin with systems that are governed by a Hamiltonian quadratic
in the momenta of the form
\begin{eqnarray}
H\left(p,q\right) & = & \frac{1}{2}p^{2}+V\left(q\right)\,.\label{eq:3-19_def_Hamiltonian}
\end{eqnarray}
Their Feynman kernel is given by the well known path integral over
configuration space trajectories $x\left(t\right)$,
see \cite{Schulman:1981vu,Dittrich:1992et,Gozzi-Book}:
\begin{eqnarray}
K\left(q^{\prime},T;q^{\prime\prime},T_{0}\right) & = & \int_{x\left(T_{0}\right)=q^{\prime\prime}}^{x\left(T\right)=q^{\prime}}{\cal D}x\left(\cdot\right)\,\exp\left(\frac{i}{\hbar}\int_{T_{0}}^{T}dt\left\{ \frac{1}{2}\dot{x}^{2}\left(t\right)-V\left(x\left(t\right)\right)\right\} \right)\,.\label{eq:3-20-standard-Feynman-kernel}
\end{eqnarray}

In the main text of this paper all the functional integrals will be
dealt with using the formalism in the continuum notation. A more rigorous
treatment based upon the discretization on a time lattice can be found in
Appendix \ref{sec:Path-integral-on-time-lattice}.

According to eq.~(\ref{eq:3-10_Y-function_via_F-kernels}), the function
${\cal K}$ may be written as the product of two Feynman-type integrals
(\ref{eq:3-20-standard-Feynman-kernel}). Denoting the $N$-component
integration variables by $x_{+}\left(t\right)$ and $x_{-}\left(t\right)$,
respectively, we have
\begin{eqnarray}
{\cal K}\left(s,q,T;s^{\prime},q^{\prime},T_{0}\right) & = & \int_{x_{+}\left(T_{0}\right)=q^{\prime}+\frac{s^{\prime}}{2}}^{x_{+}\left(T\right)=q+\frac{s}{2}}{\cal D}x_{+}\left(\cdot\right)\int_{x_{-}\left(T_{0}\right)=q^{\prime}-\frac{s^{\prime}}{2}}^{x_{-}\left(T\right)=q-\frac{s}{2}}{\cal D}x_{-}\left(\cdot\right)\label{eq:3-21-Y-kernel}\\
 &  & \times \exp\left(\frac{i}{\hbar}\int_{T_{0}}^{T}dt\left\{ \frac{1}{2}\dot{x}_{+}^{2}\left(t\right)-\frac{1}{2}\dot{x}_{-}^{2}\left(t\right)-V\left(x_{+}\left(t\right)\right)+V\left(x_{-}\left(t\right)\right)\right\} \right)\nonumber 
\end{eqnarray}
Now we replace the integration variables $x_{+}\left(t\right)$ and
$x_{-}\left(t\right)$ by their symmetric and antisymmetric linear
combinations
\begin{eqnarray}
\begin{split}
\xs \left(t\right) & =  \frac{1}{2}\left[x_{+}\left(t\right)+x_{-}\left(t\right)\right]\\
\ys \left(t\right) & =  \frac{1}{2}\left[x_{+}\left(t\right)-x_{-}\left(t\right)\right]\,.
\end{split} \label{eq:3-22-symm_and_antisymm-combinations}
\end{eqnarray}
This results in\footnote{See Appendix \ref{sec:Path-integral-on-time-lattice} for an explicit definition of the functional integral
(\ref{eq:3-23-Y-kernel-via-x_and_y}) as the continuum limit of a
discrete multiple integral on a time lattice.}
\begin{eqnarray}
{\cal K}\left(s,q,T;s^{\prime},q^{\prime},T_{0}\right) & = & \int_{\xs \left(T_{0}\right)=q^{\prime}}^{\xs \left(T\right)=q}{\cal D}\xs \left(\cdot\right)\int_{\ys \left(T_{0}\right)=\frac{s^{\prime}}{2}}^{\ys \left(T\right)=\frac{s}{2}}{\cal D}\ys \left(\cdot\right)\label{eq:3-23-Y-kernel-via-x_and_y}\\
 &  & \times \exp\left(2\frac{i}{\hbar}\int_{T_{0}}^{T}dt\,\widetilde{{ L}}\left(\xs \left(t\right),\ys \left(t\right),\dot{\xs }\left(t\right),\dot{\ys }\left(t\right)\right)\right)\nonumber 
\end{eqnarray}
with the Lagrangean
\begin{eqnarray}
\widetilde{{ L}}\left(\xs  ,\ys  ,\dot{\xs } ,\dot{\ys } \right) & \equiv & \dot{\xs }\dot{\ys }-\widetilde{V}\left(\xs ,\ys \right)\label{eq:3-24-widetilde-Lagr_x_and_y}
\end{eqnarray}
and the ``bilocal'' potential
\begin{eqnarray}
\widetilde{V}\left(\xs ,\ys \right) & \equiv & \frac{1}{2}\left[V\left(\xs +\ys \right)-V\left(\xs -\ys \right)\right]\,.\label{eq:3-25-tilde-V}
\end{eqnarray}

By (\ref{eq:3-4_Moyal_kernel_via_Feynman-kernels}) with (\ref{eq:3-10_Y-function_via_F-kernels}),
the actual Moyal kernel equals the Fourier transform of ${\cal K}$ with
respect to its $s$-arguments,
\begin{eqnarray}
K_{{\rm M}}\left(p,q,T;p^{\prime},q^{\prime},T_{0}\right) & = & \frac{1}{\left(2\pi\hbar\right)^{N}}\int d^{N}s\int d^{N}s^{\prime}\,e^{\frac{i}{\hbar}\left(s^{\prime}p^{\prime}-sp\right)}{\cal K}\left(s,q,T;s^{\prime},q^{\prime},T_{0}\right)\,.\label{eq:3-26-K_M-via-Y}
\end{eqnarray}
Therefore inserting (\ref{eq:3-23-Y-kernel-via-x_and_y}) into (\ref{eq:3-26-K_M-via-Y})
provides us with a specific path integral representation of $K_{{\rm M}}$.
As we shall see below, this version of the $K_{{\rm M}}$-integral
is the natural one in order to understand the connection to Marinov's
result. Its other main virtue is that it makes the relationship between
the Moyal integral and a pair of standard Feynman integrals fully
manifest.
\vspace{3mm}\\
{\noindent \bf (2) Euler-Lagrange equations.}
For later use let us also note the Euler-Lagrange equation of the
Lagrangean $\widetilde{L}\left(\xs ,\ys ,\dot{\xs },\dot{\ys }\right)$ that makes
its appearence in the path integral (\ref{eq:3-23-Y-kernel-via-x_and_y}):
\begin{eqnarray}
\begin{split}
\ddot{\xs }\left(t\right)  = & -\frac{1}{2}\left[\nabla V\left(\xs +\ys \right)+\nabla V\left(\xs -\ys \right)\right] \\
\ddot{\ys }\left(t\right)  = & -\frac{1}{2}\left[\nabla V\left(\xs +\ys \right)-\nabla V\left(\xs -\ys \right)\right]\,.
\end{split}
\label{eq:3-27-EOM-x_and_y}
\end{eqnarray}
These equations can be thought of as the result of ``intertwining''
two copies of the classical Newtonian equation of motion,
\begin{eqnarray}
\ddot{x}_{\pm}\left(t\right) & = & -\nabla V\left(x_{\pm}\right)\label{eq:3-28-EOM_via_x_pm}
\end{eqnarray}
by means of the transformation (\ref{eq:3-22-symm_and_antisymm-combinations}).
\vspace{3mm}\\
{\noindent \bf (3) The variant with second time derivatives.}
In order to calculate $K_{{\rm M}}$ in the semiclassical approximation ($\hbar\rightarrow0$),
or to obtain it at the exactly classical point ($\hbar=0$) even,
a different representation of the $K_{{\rm M}}$-integral suggests
itself. Moreover, this second variant will turn out to be the natural
link between the Moyal kernel and $K_{{\rm CPI}}$, i.e., the time
evolution kernel given by the Classical Path Integral \citep{Gozzi:1989bf}.

The integrand of (\ref{eq:3-23-Y-kernel-via-x_and_y}) involves
the action functional $\int_{T_{0}}^{T}dt\,\widetilde{{ L}}$
with the kinetic term $\int_{T_{0}}^{T}dt\,\dot{\xs }\dot{\ys }$. Performing
an integration by parts on this term, carefully keeping track of the
surface terms, we obtain
\begin{eqnarray}
{\cal K}\left(s,q,T;s^{\prime},q^{\prime},T_{0}\right) & = & \int_{\xs \left(T_{0}\right)=q^{\prime}}^{\xs \left(T\right)=q}{\cal D}\xs \left(\cdot\right)\,\exp\left(\frac{i}{\hbar}\left\{ s\dot{\xs }\left(T\right)-s^{\prime}\dot{\xs }\left(T_{0}\right)\right\} \right)\label{eq:3-30-Y-kernel-via_x_and_y-and-int-by-parts}\\
 &  & 
 \times \int_{\ys \left(T_{0}\right)=\frac{s^{\prime}}{2}}^{\ys \left(T\right)=\frac{s}{2}}{\cal D}\ys \left(\cdot\right)\exp\left(-2\frac{i}{\hbar}\int_{T_{0}}^{T}dt\,\left\{ \ys \ddot{\xs }+\widetilde{V}\left(\xs ,\ys \right)\right\} \right)\,.\nonumber 
\end{eqnarray}
In evaluating the boundary terms we made use of the prescribed $Y$-values
at $t=T_{0}$ and $T$, respectively, and this led to the $s$- and
$s^{\prime}$-dependence displayed explicitly in the first exponential
of eq.~(\ref{eq:3-30-Y-kernel-via_x_and_y-and-int-by-parts}). 

Naively comparing (\ref{eq:3-30-Y-kernel-via_x_and_y-and-int-by-parts}) to
(\ref{eq:3-23-Y-kernel-via-x_and_y}) one might suspect that the new
$Y$-integration in (\ref{eq:3-30-Y-kernel-via_x_and_y-and-int-by-parts})
is still subject to the boundary conditions $Y\left(T_{0}\right)=s^{\prime}/2$
and $Y\left(T\right)=s/2$ which could generate further $s,s^{\prime}$-dependencies
of ${\cal K}$. Actually, this is not the case, however: the $Y$-integration
in (\ref{eq:3-30-Y-kernel-via_x_and_y-and-int-by-parts}) is unconstrained,
and the entire $s,s^{\prime}$-dependencies stems from the boundary
terms. For a proof we refer to Appendix \ref{sec:Path-integral-on-time-lattice}, where the integration
by parts is performed at the discretized level.

Lastly we insert (\ref{eq:3-30-Y-kernel-via_x_and_y-and-int-by-parts})
into (\ref{eq:3-26-K_M-via-Y}), interchange the Fourier transformation
with the $X$-integration, and perform the $s$-integrals. This results
in the following rather suggestive functional integral representation
of the Moyal kernel:
\begin{eqnarray}
K_{{\rm M}}\left(p,q,T;p^{\prime},q^{\prime},T_{0}\right) & = & \frac{1}{\left(2\pi\hbar\right)^{N}}\int_{\xs \left(T_{0}\right)=q^{\prime}}^{\xs \left(T\right)=q}{\cal D}\xs \left(\cdot\right)\,\delta\left(\dot{\xs }\left(T\right)-p\right)\delta\left(\dot{\xs }\left(T_{0}\right)-p^{\prime}\right)\nonumber \\
 &  & \int{\cal D}\ys \left(\cdot\right)\exp\left(-2\frac{i}{\hbar}\int_{T_{0}}^{T}dt\,\left\{ \ys \ddot{\xs }+\widetilde{V}\left(\xs ,\ys \right)\right\} \right)\,.\label{eq:3-35-K_M-via-x_and_y-integral-and-deltas}
\end{eqnarray}
While the $Y$-integral is unconstrained, the $\delta$-functions enforce
prescribed \emph{velocities} at the (likewise enforced) terminal positions
of the $X$-trajectories. This imposes a total of $4N$ conditions
on $X\left(t\right)$. Thus, generically, there will exist no classical
trajectory that satisfies all conditions.
\vspace{3mm}\\
{\bf (4) The classical point.} 
A main virtue
of the path integral (\ref{eq:3-35-K_M-via-x_and_y-integral-and-deltas})
comes to light when we look at its classical limit. While the
integrand of Feynman-type integrals such as (\ref{eq:3-20-standard-Feynman-kernel})
oscillates wildly when $\hbar\rightarrow0$, and they become meaningless
at the classical point $\hbar=0$, the Marinov integral behaves in a
much more controlled way. To see this, let us rewrite the $\ys $-integral
from (\ref{eq:3-35-K_M-via-x_and_y-integral-and-deltas}) in terms
of the new integration variable $\widetilde{\ys }\left(t\right)\equiv \ys \left(t\right)/\hbar$:
\begin{eqnarray}
\int{\cal D}\widetilde{\ys }\left(\cdot\right)\,\exp\left(-2i\int_{T_{0}}^{T}dt\left\{ \widetilde{\ys }\ddot{\xs }+\frac{1}{\hbar}\widetilde{V}\left(\xs ,\hbar \, \widetilde{\ys }\right)\right\} \right)\,.\label{eq:3-40-integral-via-tilde-y}
\end{eqnarray}
The kinetic term is perfectly independent of $\hbar$ now, and the
potential term is so to lowest order. In fact, by (\ref{eq:3-25-tilde-V}),
$\widetilde{V}$ is an odd function of $\ys $. 
Analyticity assumed, its
power series has no constant term:
\begin{eqnarray}
\frac{1}{\hbar}\widetilde{V}\left(\xs ,\hbar\widetilde{\ys }\right) & = & 
\frac{1}{\hbar}\sinh\left(\hbar\, \widetilde{\ys }^{k}\partial_{k}\right)V\left(\xs \right) \nonumber \\
& = & \widetilde{\ys }^{k}\partial_{k}V\left(\xs \right)+\frac{1}{6}\hbar^{2}\, \widetilde{\ys }^{k}\widetilde{\ys }^{l}\widetilde{\ys }^{m}
\, \partial_{k}\partial_{l}\partial_{m}V\left(\xs \right)+O\left(\hbar^{4}\widetilde{\ys }^{5}\right)\,. 
\label{eq:3-41-hbar-expansion-of-tilde-V}
\end{eqnarray}
Hence the integrand of (\ref{eq:3-40-integral-via-tilde-y}) continues
to be meaningful exactly at the classical point $\hbar=0$, and the
corresponding integral is nothing but the representation of a delta
functional:
\begin{eqnarray}
\int{\cal D}\widetilde{\ys }\left(\cdot\right)\,\exp\left(-2i\int_{T_{0}}^{T}dt\,\widetilde{\ys }^{k}\left\{ \ddot{\xs }_{k}+\partial_{k}V\left(\xs \right)\right\} \right) & = & \delta\left[\ddot{\xs }_{k}+\partial_{k}V\left(\xs \right)\right]\,.\label{eq:3-42-tilde-y-integral-hbar_0-DiracDelta}
\end{eqnarray}
Thus, all that remains of (\ref{eq:3-35-K_M-via-x_and_y-integral-and-deltas})
is the $\xs $-integral sharply localized on the solutions of the classical
equation of motion, $\ddot{\xs }=-\nabla V$:
\begin{eqnarray}
K_{{\rm M}}\left(p,q,T;p^{\prime},q^{\prime},T_{0}\right) & = & \int{\cal D}\xs \left(\cdot\right)\,\delta\left[\ddot{\xs }+\nabla V\left(\xs \right)\right]\,.\label{eq:3-43-K_M-in-the-hbar_zero-limit}
\end{eqnarray}
The trajectories $\xs \left(t\right)$ contributing to (\ref{eq:3-43-K_M-in-the-hbar_zero-limit})
are subject to the conditions
\begin{eqnarray}
\begin{split}
\xs \left(T_{0}\right)=q^{\prime} & , & \dot{\xs }\left(T_{0}\right)=p^{\prime}\\
\xs \left(T\right)=q & , & \dot{\xs }\left(T\right)=p\,.
\end{split} \label{eq:3-44-initial_and_final-condition} 
\end{eqnarray}
Obviously the functional integral (\ref{eq:3-43-K_M-in-the-hbar_zero-limit})
is closely related to the Classical Path Integral (CPI) in the Lagrange
formalism \cite{Gozzi:1986ge}.
\vspace{3mm}\\
{\bf (5) The semiclassical expansion.}
From this discussion we learn that the ``perturbative'' expansion
of the Lagrangean $\widetilde{L}\left(\xs ,\ys ,\dot{\xs },\dot{\ys }\right)$
in powers of $\ys $ is equivalent to the semiclassical expansion of
the functional integral. 
The lowest, i.e., linear term in $\widetilde{L}$
is of order $O\left(\hbar^{0}\right)$ and gives rise to the singular
delta functional characteristic of strictly classical mechanics. The
(more regular) higher order contributions are systematically generated
by writing (\ref{eq:3-40-integral-via-tilde-y}) as
\begin{eqnarray}
{\cal J} \left[\xs \left(\cdot\right)\right] & \equiv &
\int{\cal D}\widetilde{\ys }\left(\cdot\right)\,\exp\left(-2i\int_{T_{0}}^{T}dt\left\{ \widetilde{\ys }\, \ddot{\xs }+\hbar^{-1}\sinh\left(\hbar\, \widetilde{\ys }^{k}\partial_{k}\right)V\left(\xs \right)\right\} \right)\label{eq:3-50-PI-via-tilde-y_and_x}
\end{eqnarray}
and then expanding out the hyperbolic sine as a power series in $\hbar$.
\vspace{3mm}\\
{\bf (6) The response field in Lagrangean guise.}
The path integral formula (\ref{eq:3-42-tilde-y-integral-hbar_0-DiracDelta})
suggests identifying $\widetilde{\ys}$ with the so called {\it response field}
introduced in the literature to give a path integral representation of
classical (possibly stochastic) systems
\cite{Martin:1973zz,Janssen1976,DEDOMINICIS1976}. 
Indeed, in classical systems one often introduces the response field
by starting out from the classical kernel of propagation on the RHS
of equation (\ref{eq:3-42-tilde-y-integral-hbar_0-DiracDelta}).
The associated path integral formalism is then obtained by introducing
the field $\widetilde{\ys}$, which has the purpose of expressing
the kernel of propagation in a path integral formalism.
In this manner a large arsenal of tools from quantum field theory becomes
applicable.

The identification of the response field will be slightly more straightforward
in the Hamiltonian framework in subsection \ref{subsec:Hamiltonian-path-integral} below.

\subsection{Airy Averaging and Random Force Representation}
%\\{\bf (4)}
In the Marinov path integral
the leading order quantum correction is due to the $O\left(\hbar^2 \right)$ term in the $\widetilde{L}$ of eq.~(\ref{eq:3-41-hbar-expansion-of-tilde-V}); 
it is \emph{cubic} in $\widetilde{\ys }$. For $N=1$, say, (\ref{eq:3-50-PI-via-tilde-y_and_x})
reads at that order:
\begin{eqnarray}
{\cal J}\left[\xs \left(\cdot\right)\right] & = & \int{\cal D}\widetilde{\ys }\left(\cdot\right)\,\exp\left(-2i\int_{T_{0}}^{T}dt\left\{ \widetilde{\ys }\, \ddot{\xs }+\widetilde{\ys }\, V^{\prime}\left(\xs \right)+\frac{\hbar^{2}}{6}V^{\prime\prime\prime}\left(\xs \right)\widetilde{\ys }^{3}\right\} \right)\,.\label{eq:3-51-tilde-y_and_x-PI-order-y3}
\end{eqnarray}
A welcome feature of (\ref{eq:3-51-tilde-y_and_x-PI-order-y3}) is
that the integrand involves no time derivatives of $\widetilde{Y}\left(t\right)$.
As a result, the path integral factorizes. Symbolically,\footnote{The discrete analog of ${\cal J}$ factorizes analogously, see equation (\ref{eq:A-10})
in Appendix \ref{sec:Path-integral-on-time-lattice}.}
\begin{eqnarray}
{\cal J}\left[\xs \left(\cdot\right)\right] & = & \prod_{t\in\left[T_{0},T\right]}I\left(a\left(t\right),b\left(t\right)\right)\,.\label{eq:3-52}
\end{eqnarray}
Here the function $I\left(a,b\right)$, which is defined by an ordinary
integral,
\begin{eqnarray}
I\left(a,b\right) & \equiv & \frac{1}{2\pi}\int_{-\infty}^{\infty}dy\,e^{-i\left(ay+\frac{1}{3}by^{3}\right)}\nonumber \\
 & = & \frac{1}{\pi}\int_{0}^{\infty}dy\,\cos\left(ay+\frac{1}{3}by^{3}\right)\,, \label{eq:3-53}
\end{eqnarray}
is evaluated at the following time dependent parameter values:
\begin{subequations}
\begin{align}
a\left(t\right) & = 2\ddot{\xs }\left(t\right)+2V^{\prime}\left(\xs \left(t\right)\right) \\
b\left(t\right) & = \hbar^2 V^{\prime\prime\prime}\left(\xs \left(t\right)\right)\,.
\end{align} \label{eq:3-54}
\end{subequations}

{\noindent \bf (1) The Airy function.}
Obviously, equation (\ref{eq:3-53}) is essentially the integral representation
of Airy's function \cite{ValleeBook}, so that we are led to the closed-form
result
\begin{eqnarray}
I\left(a,b\right) & = & \frac{1}{\left|b\right|^{1/3}}{\rm Ai}\left(\frac{ab}{\left|b\right|^{4/3}}\right)\nonumber \\
 & = & \frac{1}{\left|b\right|^{1/3}}{\rm Ai}\left({\rm sign}\left(b\right)\frac{a}{\left|b\right|^{1/3}}\right)\,,\label{eq:3-55}
\end{eqnarray}
where ${\rm sign}\left(b\right)=\pm1$ denotes the sign of $b$.

This Airy function describes the universal approach to the classical
limit for arbitrary quantum systems. If $V^{\left(2n+1\geq5\right)}=0$,
eq.~(\ref{eq:3-55}) is exact even, and this includes the important case of the quartic
oscillator.

In the strict classical limit $b\propto\hbar^{2}\rightarrow0$, the
standard property \cite{ValleeBook}
\begin{eqnarray}
\lim_{\alpha\rightarrow0}\frac{1}{\left|\alpha\right|}{\rm Ai}\left(\frac{\xi}{\alpha}\right) & = & \delta\left(\xi\right)\label{eq:3-56}
\end{eqnarray}
implies $I\left(a,b\right)\rightarrow\delta\left(a\right)=\frac{1}{2}\delta\left(\ddot{x}+V^{\prime}\left(x\right)\right)$.
In this way we recover the delta function typical of the CPI,
as it should be.

In the semiclassical regime ($\hbar\neq0$), the function $I\left(a,b\right)$
can be regarded an approximation of this delta function. It is noteworthy
that, depending on the sign of its (real) argument, the Airy function achieves
this approximation in two fundamentally different ways: For large
positive arguments $\xi\rightarrow\infty$, the Airy function ${\rm Ai}\left(\xi\right)$
is exponentially decreasing, while it is rapidly oscillating for large
negative arguments, $\xi\rightarrow-\infty$. 

This behaviour is an example of Stokes phenomenon \cite{Berry:1989proc}.
Moreover, in recent years the Airy function has even become a canonical example of resurgence theory \cite{Dunne:2015eaa,Dunne:2016nmc,Dunne:2016qix}.
At this point we only mention that if one uses the stationary phase approximation to find the asymptotics of the integral ($\xi \in \mathbb{R}$)
\begin{equation}
{\rm Ai} \left(\xi \right) =
\frac{1}{2\pi} \int_{-\infty}^\infty dy \, e^{i \left(y \xi + \frac{1}{3} y^3 \right)} \, ,\label{eq:400}
\end{equation}
the relevant contour deformations and saddle points depend crucially on whether we let $\xi \rightarrow +\infty$ or $\xi \rightarrow -\infty$.
In the former case, the respective contour, $\Gamma_{>}$, passes only through one saddle, leading to
\begin{equation}
{\rm Ai} \left(\xi \right) \sim
\frac{\exp \left( -\frac{2}{3} \xi^{3/2} \right)}{2\sqrt{\pi} \xi^{1/4}}
\qquad \left(\xi \rightarrow +\infty \right)
\label{eq:401}
\end{equation}
while in the latter case
the contour (denoted $\Gamma_{<}$) passes through two saddles, yielding
\begin{equation}
{\rm Ai} \left(\xi \right) \sim
\frac{\sin \left( \frac{2}{3} \left(-\xi\right)^{3/2}+\frac{\pi}{4} \right)}{\sqrt{\pi} \left(-\xi\right)^{1/4}}
\qquad \left(\xi \rightarrow -\infty \right) \, ,
\label{eq:402}
\end{equation}
see \cite{Bleistein:1986,Bender:1999} for the details.

{\noindent \bf (2) Airy averaging.}
As for its general interpretation, it turns out most natural to regard
$I\left(a,b\right)$ as a certain \emph{Airy average}. The notion
of Airy averaging was first discussed in \cite{Englert1984} in a different
context, see also \cite{ValleeBook}. 

For $h\left(\xi\right)$ a function
on the real axis, the Airy average is defined as
\begin{eqnarray}
\langle h\left(\xi\right)\rangle_{{\rm Ai}} & \equiv & \int_{-\infty}^{\infty}d\xi\,h\left(\xi\right){\rm Ai}\left(\xi\right)\,.\label{eq:3-57-def-Airy-average}
\end{eqnarray}
For example, $\langle1\rangle_{{\rm Ai}}=1$, $\langle\xi\rangle_{{\rm Ai}}=0$,
$\langle\xi^{2}\rangle_{{\rm Ai}}=0$, $\langle\xi^{3}\rangle_{{\rm Ai}}=2$,
$\cdots$ , which follows by repeatedly differentiating
the generating function
\begin{eqnarray}
\langle e^{i\eta\xi}\rangle_{{\rm Ai}} & = & \exp\left(-\frac{i}{3}\eta^{3}\right)\,.\label{eq:3-58}
\end{eqnarray}
The differential equation obeyed by the Airy function, $\partial_{\xi}^{2}{\rm Ai}\left(\xi\right)=\xi{\rm Ai}\left(\xi\right)$,
implies the general rule $\langle h^{\prime\prime}\left(\xi\right)\rangle_{{\rm Ai}}=\langle\xi h\left(\xi\right)\rangle_{{\rm Ai}}$.

{\noindent \bf (3) Random force representation.}
Coming back to $I\left(a,b\right)$, let us introduce an auxiliary
(``force'') variable $f$ to rewrite (\ref{eq:3-55}) in the style
of an Airy-averaged delta function:
\begin{eqnarray}
I\left(a,b\right) & = & \int_{-\infty}^{\infty}df\,{\rm Ai}\left(f\right)\delta\left(a-{\rm sign}\left(b\right)\left|b\right|^{1/3}f\right)\,.\label{eq:3-80-I-via-int-force}
\end{eqnarray}
While looking artificially complicated at first sight, this representation
entails a remarkable relationship between the classical and the quantum
time evolution kernels, respectively.

{\noindent \bf(3a)}
The product over all times that appears in equation (\ref{eq:3-52})
converts the ordinary integral (\ref{eq:3-80-I-via-int-force}) to an integral
over functions $f\left(t\right),\,t\in\left[T_{0},T\right]$,
and correspondingly the delta function to a delta functional:
\begin{eqnarray}
{\cal J}\left[\xs \left(\cdot\right)\right] & = & \int{\cal D}f\left(\cdot\right)\,{\cal A}\left[f\left(\cdot\right)\right]
\delta \left[a\left(\cdot\right)-{\rm sign}\left(b\left(\cdot\right)\right)\left|b\left(\cdot\right)\right|^{1/3}\right]\,.\label{eq:3-90}
\end{eqnarray}
The crucial ``weight'' functional
\begin{eqnarray}
{\cal A}\left[f\left(\cdot\right)\right] & \equiv & \prod_{t\in\left[T_{0},T\right]}{\rm Ai}\left(f\left(t\right)\right)\label{eq:3-91}
\end{eqnarray}
implements uncorrelated, independent Airy averages at different times.

{\noindent \bf(3b)}
The intriguing property of the representation (\ref{eq:3-90}) is
that it allows us to express the quantum mechanical Moyal kernel as
a superposition of classical mechanics-type kernels of the form
(\ref{eq:3-43-K_M-in-the-hbar_zero-limit}). To see this, recall that
equation (\ref{eq:3-35-K_M-via-x_and_y-integral-and-deltas}) is tantamount
to
\begin{eqnarray}
K_{{\rm M}}\left(p,q,T;p^{\prime},q^{\prime},T_{0}\right) & = & \int_{\xs\left(T_{0}\right)=q^{\prime},\, \dot{\xs}\left(T_{0}\right)=p^{\prime}}^{\xs\left(T\right)=q,\, \dot{\xs}\left(T\right)=p}{\cal D}\xs\left(\cdot\right)\,{\cal J}\left[\xs\left(\cdot\right)\right]\label{eq:3-92}
\end{eqnarray}
whereby the integration over $\xs\left(t\right)$ is constrained by the
four boundary conditions. If we insert (\ref{eq:3-90})
into (\ref{eq:3-92}) and interchange
the $\xs$- with the $f$-integration, we obtain the following {\it random force representation} of the Moyal-Marinov kernel:
\begin{eqnarray}
K_{{\rm M}}\left(p,q,T;p^{\prime},q^{\prime},T_{0}\right) & = & \int{\cal D}f\left(\cdot\right)\:{\cal A}\left[f\left(\cdot\right)\right]\,
K_{{\rm cl}}\left[f\left(\cdot\right)\right]\left(p,q,T;p^{\prime},q^{\prime},T_{0}\right)\,.\label{eq:3-93-K_M-semicl-as-force-integral-of-K_cl}
\end{eqnarray}
Here $K_{{\rm cl}}$ denotes a modified classical mechanics-kernel depending on $f\left(t\right)$:
\begin{eqnarray}
K_{{\rm cl}}\left[f\left(\cdot\right)\right]\left(p,q,T;p^{\prime},q^{\prime},T_{0}\right) & = & \int_{\xs\left(T_{0}\right)=q^{\prime},\, \dot{\xs}\left(T_{0}\right)=p^{\prime}}^{\xs\left(T\right)=q,\, \dot{\xs}\left(T\right)=p}{\cal D}\xs\left(\cdot\right) \nonumber \\
&\, & \delta\left[\ddot{\xs}+V^{\prime}\left(\xs\right)-\frac{1}{2}\hbar^{2/3}{\rm sign}\left(V^{\prime\prime\prime}\left(\xs\right)\right)\left|V^{\prime\prime\prime}\left(\xs\right)\right|^{1/3}f\right]\,.\label{eq:3-94-K_cl-with-semiclassical-forcing}
\end{eqnarray}

{\noindent \bf(3c)}
Very much like the truly classical Liouville kernel (\ref{eq:3-43-K_M-in-the-hbar_zero-limit}),
its cousin (\ref{eq:3-94-K_cl-with-semiclassical-forcing}) is strictly
localized on the solutions of a certain differential equation. In
the case at hand it is not Newton's equation, but rather a $f$-dependent
modification thereof:
\begin{eqnarray}
\ddot{\xs} & = & -V^{\prime}\left(\xs\right)+\frac{1}{2}\hbar^{2/3}{\rm sign}\left(V^{\prime\prime\prime}\left(\xs\right)\right)\left|V^{\prime\prime\prime}\left(\xs\right)\right|^{1/3}f\,.\label{eq:3-95-semiclassical-Newton-eq}
\end{eqnarray}
We observe that $f\left(t\right)$ has the character of an externally
prescribed, time-dependent, but $\xs$-independent, random force that
is governed by the Airy weight functional ${\cal A}\left[f\left(\cdot\right)\right]$.
The non-classical force term in (\ref{eq:3-95-semiclassical-Newton-eq})
is manifestly non-analytic in $\hbar$.\footnote{
The concept of effective classical trajectories behind the equation (\ref{eq:3-95-semiclassical-Newton-eq}) is different from the one already studied in the literature, see \cite{Razavy:2003}.}

{\noindent \bf (4) Violation of positivity.}
The Weyl symbols of density operators, $\rho \left(p,q;T\right)$, are pseudo-densities, i.e., they can assume negative values in strongly non-classical situations.
If we assume a positive initial distribution $\rho \left(p,q;T_0 \right)>0$,
then a negative value at some later time $T>T_0$ implies that $K_{\rm{M}} \left(p,q,T;p^\prime, q^\prime, T_0 \right)$ must be negative for certain configurations of its arguments.
It is an interesting question under what conditions this can happen, and what precisely is a ``strongly non-classical'' regime in this context.

{\noindent \bf(4a)}
The random force representation of the Moyal kernel, eq.~(\ref{eq:3-94-K_cl-with-semiclassical-forcing}), allows for a fresh look at this problem:

We know that $K_{\rm{cl}} \left[f\left(\cdot \right)\right]$ is simply a special classical mechanics-kernel; as such it amounts to an everywhere non-negative generalized function of $\left(p,q,T;p^\prime,q^\prime,T_0 \right)$ and functional of $f$.
Therefore, by (\ref{eq:3-93-K_M-semicl-as-force-integral-of-K_cl}), a necessary condition for negative values $K_{\rm M} \left(p,q,T;p^\prime,q^\prime,T_0 \right) <0$ is that the ``weight'' functional ${\cal A} \left[f\left(\cdot \right)\right]$ returns negative values for certain $f$'s which contribute to the integral.
Furthermore, because of the factorization (\ref{eq:3-91}), ${\cal A} \left[f\left(\cdot \right)\right] <0$ in turn requires that ${\rm Ai}\left(f\left(t\right)\right)<0$ for certain $t\in \left[T_0,T\right]$.

\begin{figure}
\includegraphics[scale=0.2]{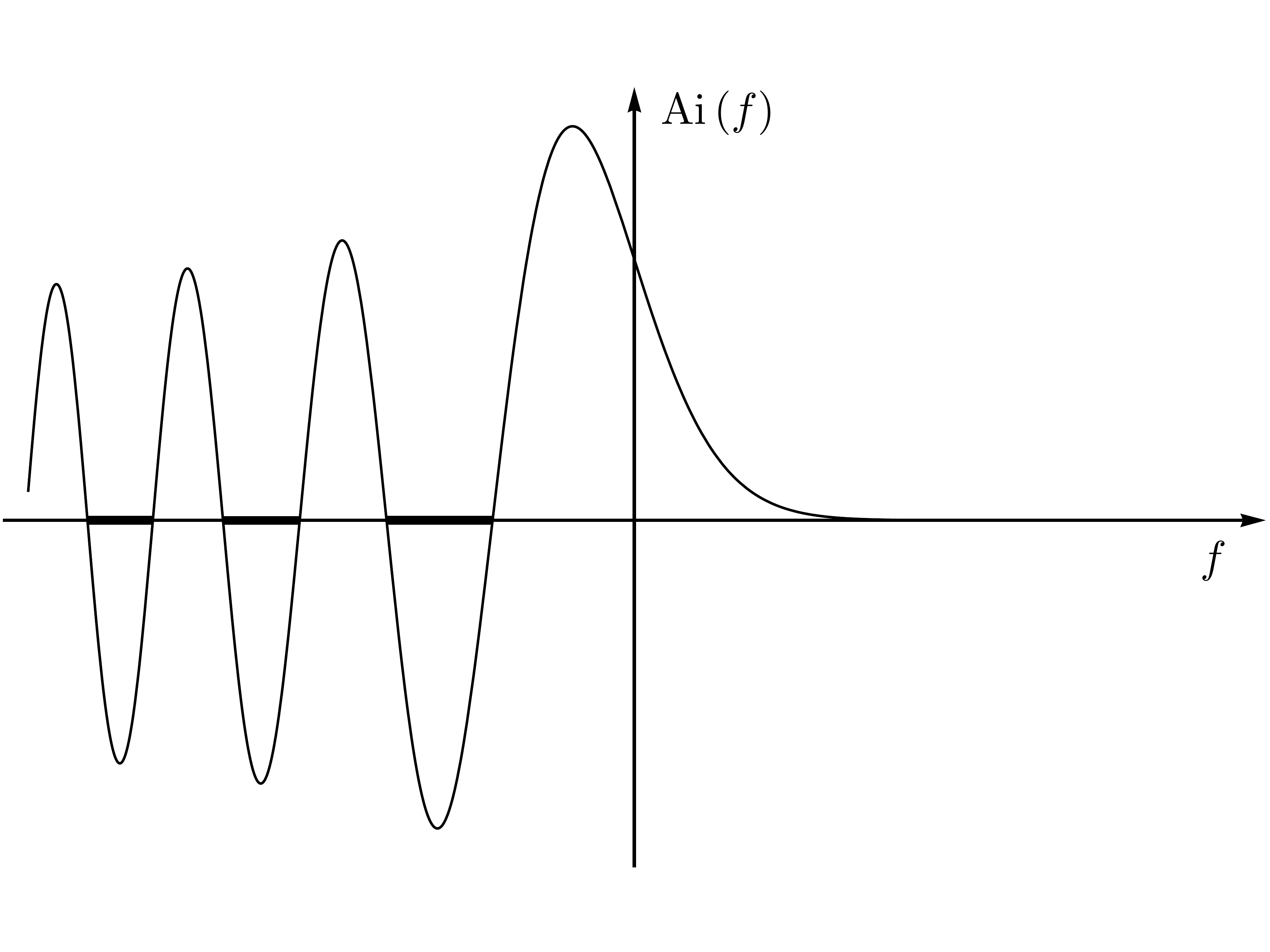}
\caption{
The Airy function ${\rm Ai} \left(f\right)$ for real arguments $f$.
While ${\rm Ai} \left(f\right)$ is strictly positive for positive $f>0$,
there exists an infinity of intervals on the negative $f$-axis where ${\rm Ai} \left(f\right)<0$.
\label{fig:new1}}

\end{figure}

A glance at Figure \ref{fig:new1} reveals that ${\rm Ai}\left(f \right)$ is never negative for positive arguments, $f>0$.
On the negative $f$-axis, however, it possesses infinitely many ``islands''
between adjacent zeros on which ${\rm Ai}\left(f \right)<0$.

In summary, we find the following {\it necessary condition for $K_{\rm M}\left(p,q,T;p^\prime,q^\prime,T_0 \right)<0$}:
To obtain a negative value of the Moyal kernel it is necessary that at least one function $f$, that makes a contribution to the integral 
(\ref{eq:3-93-K_M-semicl-as-force-integral-of-K_cl}) for the boundary condition chosen, assumes at some time $t$ a value $f\left(t\right)$ that lies on the ``islands'' where the Airy function is negative.

Conversely, if the boundary data are such that the differential equation (\ref{eq:3-95-semiclassical-Newton-eq}) possesses a solution only for functions $f$ with $f\left(t \right)>0$  $\forall \, t\in \left[T_0,T \right]$, then it follows that  $K_{\rm M}$ cannot be negative for those boundary data.

{\noindent \bf(4b)}
Thus we see that strong quantum effects which ruin the classical density interpretation of $\rho$ are intimately connected to values of the random force $f\left(t \right)$ lying on the ``negativity islands'' of Airy's function.
Furthermore, we mentioned that the asymptotics of ${\rm Ai} \left(\xi \right)$ for $\xi \rightarrow +\infty$ and $\xi \rightarrow -\infty$, respectively, are in a one-to-one correspondence with the contours, $\Gamma_>$ and $\Gamma_<$, visiting different (sets of) saddle points.

Hence we can say that within the asymptotic expansion,
{\it positivity violating quantum dynamics is possible only if the kinematical data select the $\Gamma_<$ contour with its two saddle points as the relevant one.}
If $\Gamma_>$, having only a single saddle point, is selected instead, positive Wigner functions always evolve into positive ones.
In a way, the system behaves more classically then.

{\noindent \bf(4c)}
These remarks are elementary in the sense that they stem from the classical theory of integration developed for {\it functions}.
However, in the case at hand they readily generalize to {\it functionals}
since the functional integral over the response field factorizes for different times, see eq.~(\ref{eq:3-52}).
Thus it provides us with an explicit example of Picard-Lefschetz and 
resurgence theory {\it applied to path integrals}.
While highly desirable of course,
it remains to be seen if a functional version of Picard-Lefschetz theory can
be established in the general case \citep{Witten:2010cx,Witten:2010zr,Tanizaki:2014xba,Cherman:2014sba,Behtash:2015zha,Behtash:2015loa,Feldbrugge:2017kzv}.

{\noindent \bf (5) Double well potential and tunneling.}
Next we discuss tunneling as an example of a typical quantum process.
It nicely illustrates the relation between strongly non-classical effects and the properties of the Airy's function which governs the random force dynamics.
In fact, it is known that during tunneling events Wigner functions can become negative locally \cite{Razavy:2003}. 

\begin{figure}
\includegraphics[scale=0.2]{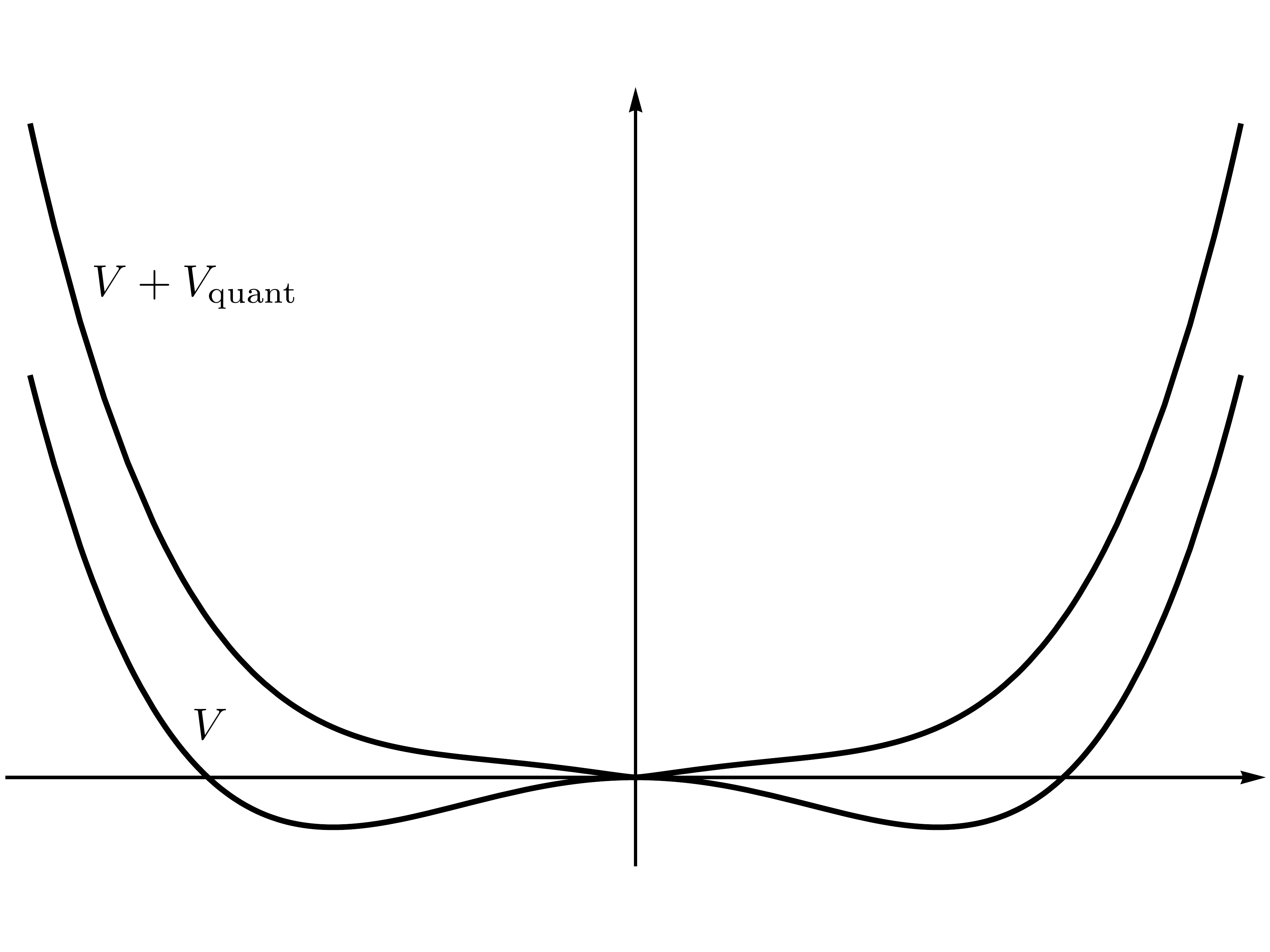}
\caption{
The classical double well potential (lower curve) and the total potential
$V+V_{\rm quant}$ (upper curve) for a particular realization of the random force. 
In this example, $f\left(t \right)$ is taken to be a large negative constant.
\label{fig:new2}}

\end{figure}

{\noindent \bf(5a)}
We consider a particle in the quartic potential
\begin{eqnarray}
V\left(\xs \right) = -\frac{1}{2} \mu^2 \xs^2 +\frac{\lambda}{24} \xs^4 
\label{eq:300}
\end{eqnarray}
with $\lambda,\mu>0$.
It has two degenerate minima at $\pm \xs_0$, with $\xs_0=\sqrt{6\mu^2/\lambda}$,
see Figure \ref{fig:new2}.
The modified Newton equation (\ref{eq:3-95-semiclassical-Newton-eq}) reads for this example:
\begin{eqnarray}
\ddot{\xs} =
\mu^2 \xs - \frac{\lambda}{6} \xs^3+F_{\rm{quant}} \left(\xs;t \right)
\label{eq:301}
\end{eqnarray}
with the random (``quantum'') force
\begin{eqnarray}
F_{\rm{quant}} \left(\xs;t \right) &=& \frac{1}{2} \hbar^{2/3} \lambda^{1/3}
{\rm{sign}}\left(\xs\right) | \xs|^{1/3} f \left(t\right) \nonumber \\
&\equiv & -\frac{d}{d\xs} V_{\rm quant} \left(\xs;t\right)\,. 
\label{eq:302}
\end{eqnarray}
It derives from the potential
\begin{eqnarray}
V_{\rm quant} \left(\xs;t\right) =
-\frac{3}{8} \hbar^{2/3} \lambda^{1/3} |\xs|^{4/3} f\left(t\right)
\label{eq:303}
\end{eqnarray}
which we normalized such that $V_{\rm quant} \left(0;t\right)=0$ for any $f$.

{\bf \noindent (5b)}
The modified Newton equation differs from the classical one in two respects.
First, the corresponding potential $V_{\rm tot}= V+V_{\rm quant}$ includes an explicitly time-dependent correction, $\ddot{\xs}=-V_{\rm tot}^\prime \left(\xs;t\right)$, and second, the differential equation is to be solved with four boundary conditions, namely the initial and final positions and momenta specified by the path integral (\ref{eq:3-92}).
Hence for a fixed generic $f(t)$ there will exist no solution at all typically.
(This is what frequently causes $\delta$-function dependencies of $K_{\rm M}$ on the terminal positions and momenta in the familiar case $f=0$;
see eq.~(\ref{eq:A-21}) for the example of the free particle.)

{\bf \noindent (5c)}
Let us consider a classically forbidden transition now. 
We aim at computing the Moyal-Marinov kernel 
$K_{\rm M} \left(q,p,T;q^\prime ,p^\prime ,T_0 \right)$ for the initial values $q^\prime = -\xs_0,p^\prime =0$ and final values $q = \xs_0,p =0$.
Clearly, the double well potential in Figure \ref{fig:new2} admits no classical solution that describes a particle which moves from the left to the right minimum and has zero velocity at both the terminal points, and this is why we say that the transition is possible by tunneling only.\footnote{
The requirement of vanishing terminal velocities excludes a ``spilling over'' as described in \cite{Balazs1990}.} 

However, the modified Newton equation with the potential $V+V_{\rm quant}$ does admit such a solution, for a wide range of random functions $f\left(t\right)$ even.
As a proof of principle, consider the case where $f\left(t\right)=-k={\rm const}$ is a negative constant, $k>0$.
It is easy to see then that by choosing 
$k>k_{\rm min}$ with $k_{\rm min}$ sufficiently large,
the potential barrier between $-\xs_0$ and $\xs_0$ disappears completely from $V+V_{\rm quant}$.
The desired transition can be realized by a classical trajectory then,
albeit in a non-classical potential $V+V_{\rm quant}$, see Figure \ref{fig:new2}.

As a result, the $f$-integral in (\ref{eq:3-93-K_M-semicl-as-force-integral-of-K_cl}) receives contributions from 
all those trajectories with $k>k_{\rm min}$, as well as from 
an infinity of similar ones with time dependent $f$.
Many of them will give rise to ${\rm Ai} \left(f\right)<0$.
All trajectories together conspire to make up the quantum mechanical tunneling phenomenon.

{\bf \noindent (5d)}
The following points should be noted here:\\
{\bf (i)}
There exists a strict connection between the sign of $f\left(t \right)$
and those random forces that have an effect towards making classically forbidden transitions possible.
The latter forces must be such that the potential near its classical minima gets lifted above its (unchangeable) value at the origin.
But since always, by (\ref{eq:303}), 
${\rm sign} \left(V_{\rm quant} (\xs;t)\right)=-{\rm sign} \left( f(t) \right)$,
we conclude that no positive $f$ is able to achieve this raising,
but a sufficiently large negative $f$ can.\\
{\bf (ii)}
Above we saw that non-positive Wigner functions are possible only thanks to necessarily negative random forces $f\left(t\right)<0$ with ${\rm Ai}\left(f\left(t\right)\right)<0$.
This leads us to the conclusion that, at least in the example,
{\it negative random forces are the indispensable hallmark of both tunneling and phase space densities going negative.}\\
{\bf (iii)}
Since $f<0$ is strictly connected to the $\Gamma_{<}$ contour in the stationary phase analysis it follows furthermore that both tunneling and negative phase space densities can occur only if the terminal conditions pick the saddle points on the $\Gamma_{<}$ contour as the relevant one.

So it emerges the qualitative picture that 
{\it typical quantum behaviour (approximatively classical behaviour) is connected to the steepest descent contours $\Gamma_<$ ($\Gamma_>$) and their concomitant saddle points}.\\
{\bf (iv)}
Contrary to the familiar instanton methods based upon classical trajectories in the inverted potential,
the present random force description of quantum tunneling requires
{\it no Wick rotation to Euclidean time.}

It remains to be seen whether this description can be developed into an efficient quantitative tool.
For different applications of the Marinov integral to tunneling see
also \cite{Marinov:1996da,Marinov:1996ue}.

\subsection{Hamiltonian Path Integral \label{subsec:Hamiltonian-path-integral}}

In this subsection we generalize the discussion towards Hamiltonians
$\widehat{H}$ that are not necessarily quadratic in the momenta.
Admitting arbitrary Weyl symbols $H\left(p,q\right)$ now, we express
the Feynman kernel in terms of the standard phase space path integral
which has form
\begin{eqnarray}
K\left(q^{\prime},T;q^{\prime\prime},T_{0}\right) & = & \int_{x\left(T_{0}\right)=q^{\prime\prime}}^{x\left(T\right)=q^{\prime}}{\cal D}x\left(\cdot\right)\int{\cal D}p\left(\cdot\right) \nonumber \\
 &  & \exp\left(\frac{i}{\hbar}\int_{T_{0}}^{T}dt\left\{ p\left(t\right)\dot{x}\left(t\right)-H\left(p\left(t\right),x\left(t\right)\right)\right\} \right)\,.\label{eq:3-100-QM-PI-in-phase_space}
\end{eqnarray}
Here the integration over the $N$-component momentum variable $p\left(t\right)$
is unconstrained. 

The continuum formula (\ref{eq:3-100-QM-PI-in-phase_space})
should be read as a compact abbreviation of a discretized
integral on a time lattice. 
In order to represent the evolution kernel for the specific
operator $\widehat{H}$ that results from $H\left(p,q\right)$ by
\emph{Weyl ordering} positions and momenta, the discretization
must employ the \emph{mid-point rule} \cite{Berezin:1980xw,Chaichian:2001cz}.

Building
up the Moyal kernel from two integrals of the type (\ref{eq:3-100-QM-PI-in-phase_space}),
involving the variables $p_{+}\left(t\right),x_{+}\left(t\right)$ and $p_{-}\left(t\right),x_{-}\left(t\right)$,
respectively, we have
\begin{eqnarray}
{\cal K}\left(s,q,T;s^{\prime},q^{\prime},T_{0}\right) & = & \int_{x_{+}\left(T_{0}\right)=q^{\prime}+\frac{s^{\prime}}{2}}^{x_{+}\left(T\right)=q+\frac{s}{2}}{\cal D}x_{+}\left(\cdot\right)\int{\cal D}p_{+}\left(\cdot\right)\int_{x_{-}\left(T_{0}\right)=q^{\prime}-\frac{s^{\prime}}{2}}^{x_{-}\left(T\right)=q-\frac{s}{2}}{\cal D}x_{-}\left(\cdot\right)\int{\cal D}p_{-}\left(\cdot\right)\nonumber \\
 &  & \exp\left(\frac{i}{\hbar}\int_{T_{0}}^{T}dt\left\{ p_{+}\dot{x}_{+}+p_{-}\dot{x}_{-}-H\left(p_{+},x_{+}\right)
 +H\left(p_{-},x_{-}\right)\right\} \right)\,.\label{eq:3-101-Y_kernel-in-phase-space}
\end{eqnarray}
The integration is over a total of $4N$ functions now, $p_{\pm}\left(t\right)$
and $x_{\pm}\left(t\right)$. 
\vspace{3mm}\\
{\bf \noindent (1) New phase space variables.}
Again we try to find new integration
variables that would allow for the identification of
a certain ``dynamical'' field alongside with its associated response
field. In addition we require that the new variables should bring
out the symplectic covariance of (\ref{eq:3-101-Y_kernel-in-phase-space})
and of $K_{{\rm M}}$ as far as possible. 

In order to meet these requirements,
we introduce the symmetric linear combinations of the $N$ plus- and
minus-type position and momentum variables, denoted $\xs$ and $\pi_{\xs}$,
respectively, and combine them in a $2N$ component phase space coordinate
$\phi^{a}$, $a=1,2,\cdots,2N$. Likewise we form their antisymmetric
linear combination, $\ys$ and $\pi_{\ys}$, and unite them in the $2N$
component field $\xi^{a}$, times a factor of $\hbar$. Hence, in
our notation that suppresses configuration space indices,
\begin{eqnarray}
\begin{split}
\phi^{a} & \equiv  \left(\phi^{p},\phi^{q}\right)\,=\,\left(\pi_{\xs},\xs\right) \\
\xi^{a} & \equiv  \left(\xi^{p},\xi^{q}\right)\,=\,\left(\pi_{\ys},\ys\right)/\hbar\,,
\end{split} \label{eq:3-102-phi_and_xi-variables-def}
\end{eqnarray}
with the following $N$-component entries:
\begin{eqnarray}
\begin{split}
\xs & =  \frac{1}{2}\left(x_{+}+x_{-}\right)\,=\,\phi^{q} \\
\ys & =  \frac{1}{2}\left(x_{+}-x_{-}\right)\,=\,\hbar\xi^{q} \\
\pi_{\xs} & =  \frac{1}{2}\left(p_{+}+p_{-}\right)\,=\,\phi^{p}\\
\pi_{\ys} & =  \frac{1}{2}\left(p_{+}-p_{-}\right)\,=\,\hbar\xi^{p}\,.
\end{split} \label{eq:3-103-symm_and_asymm-pi_and_x} 
\end{eqnarray}
\vspace{3mm}\\
{\bf \noindent (2) Marinov phase space integral.}
After this transformation of variables, the integral for ${\cal K}$
becomes
\begin{eqnarray}
{\cal K}\left(s,q,T;s^{\prime},q^{\prime},T_{0}\right) & = & \int{\cal D}\phi^{a}\left(\cdot\right){\cal D}\xi^{a}\left(\cdot\right)\,\exp\left(2\frac{i}{\hbar}\int_{T_{0}}^{T}dt\,\widetilde{L}\right)\,.\label{eq:3-104-Y-kernel-via-tilde_L-phi_and_xi}
\end{eqnarray}
It is subject to the boundary conditions
\begin{eqnarray}
\begin{split}
\phi^{q}\left(T_{0}\right)=q^{\prime} & , \,\, \phi^{q}\left(T\right)=q \\
\xi^{q}\left(T_{0}\right)=\frac{s^{\prime}}{2\hbar} & , \,\,  \xi^{q}\left(T\right)=\frac{s}{2\hbar} 
\end{split} \label{eq:3-105-boundary-conditions-for-phi_and_xi}
\end{eqnarray}
and features the Lagrangean
\begin{eqnarray}
\widetilde{L} & = & 
\pi_{\xs}\, \dot{\ys}+\pi_{\ys}\, \dot{\xs}
-\frac{1}{2}\Bigr\lbrace H\left(\pi_{\xs}+\pi_{\ys},\xs+\ys\right)-H\left(\pi_{\xs}-\pi_{\ys},\xs-\ys\right)\Bigr\rbrace \,.\label{eq:3-106-tilde-L_hamiltonian-form}
\end{eqnarray}
In terms of the new variables, and after an integration by parts which
generates a nontrivial boundary term, it yields the action
\begin{eqnarray}
\int_{T_{0}}^{T}dt\,\widetilde{L} & = & \hbar\left[\phi^{p}\xi^{q}\right]_{T_{0}}^{T}-\hbar\int_{T_{0}}^{T}dt\left\{ \dot{\phi}^{a}\omega_{ab}\xi^{b}-\widetilde{H}\left(\phi,\xi\right)\right\} \,.\label{eq:3-107-tilde_L-via_hamiltonian-and_int_by_parts}
\end{eqnarray}
The Hamiltonian $\widetilde{H}$ was defined in eq.~(\ref{eq:2-28_tilde_H-via-phi_and_xi})
already. 

Recall also that $\omega_{ab}$ is the inverse of the Poisson
matrix $\omega^{ab}$, a $2N\times2N$ block matrix with entries $\omega_{qp}=-\omega_{pq}=-\mathbb{I}$
and $\omega_{pp}=\omega_{qq}=0$. It should be interpreted as the
constant coefficient matrix of the symplectic 2-form $\omega\equiv\frac{1}{2}\omega_{ab}d\phi^{a}\wedge d\phi^{b}$
in Darboux coordinates. It gives the phase space ${\cal M}$ the status
of a symplectic manifold. This makes it clear that the ``bulk''
kinetic term that emerged from the integration by parts, $\dot{\phi}^{a}\omega_{ab}\xi^{b}$,
behaves covariantly, namely as a scalar under symplectic diffeomorphisms
when $\dot{\phi}^{a}$ and $\xi^{a}$, respectively, are transformed
as vector components.

The boundary term in (\ref{eq:3-107-tilde_L-via_hamiltonian-and_int_by_parts})
instead does not seem to be covariant. However, it is precisely what
we need in order to achieve full symplectic covariance at the level
of $K_{{\rm M}}$ when we insert (\ref{eq:3-104-Y-kernel-via-tilde_L-phi_and_xi})
into (\ref{eq:3-26-K_M-via-Y}). In fact, after the Fourier transformation
with respect to the $s$-variables we obtain exactly Marinov's result
for $K_{{\rm M}}$ which we anticipated in eq.~(\ref{eq:2-27_def_PI_for_Moyal_kernel}),
with (\ref{eq:2-28_tilde_H-via-phi_and_xi}).
\vspace{3mm}\\
{\bf \noindent (3) Properties.}
Several features of the Moyal-Marinov kernel
(\ref{eq:2-27_def_PI_for_Moyal_kernel}), (\ref{eq:2-28_tilde_H-via-phi_and_xi}) 
are important here:

{\bf \noindent (3a) The doubled phase space ${\cal M}\times {\cal M}$.}
To elucidate the Hamiltonian
structure behind the Marinov path integral, and the operator formalism
it is equivalent to, it is natural to work with the covector $\lambda$,
\begin{eqnarray}
\lambda_{a}\equiv\omega_{ab}\xi^{b} & \Leftrightarrow & \xi^{a}=\omega^{ab}\lambda_{b}\,,\label{eq:3-120-lambda_via_xi-and-viceversa}
\end{eqnarray}
rather than $\xi$. Its components are
\begin{eqnarray}
\lambda_{a} & \equiv & \left(\lambda_{p},\lambda_{q}\right)\,=\,\left(\xi^{q},-\xi^{p}\right)\,.\label{eq:3-121-lambda_components}
\end{eqnarray}
Then, up to boundary terms,
\begin{eqnarray}
-\frac{1}{\hbar}\widetilde{L} & = & \dot{\phi^{a}}\omega_{ab}\xi^{b}-\widetilde{H}\left(\phi,\xi\right)+{\rm b.t.}\nonumber \\
 & = & \lambda_{a}\dot{\phi^{a}}-\widetilde{H}\left(\phi,\omega\lambda\right)+{\rm b.t.}\,.\label{eq:3-122-tildeL-via-lambda}
\end{eqnarray}
The form of this Lagrangean indicates that Marinov's integral,
too, has the structure of a standard phase space path integral. It ``lives'',
however, on the $4N$-dimensional doubled phase space ${\cal M}\times{\cal M}$.
The $\lambda_{a}\dot{\phi}^{a}$-term in (\ref{eq:3-122-tildeL-via-lambda})
implies that the doubled phase space is furnished with the symplectic
structure that treats the $\phi^{a}$'s as $2N$ ``position'' variables,
and the $\lambda_{a}$'s as the $2N$ ``momentum'' variables canonically
conjugate to them.\footnote{See \cite{Gozzi:1993nk,Gozzi:1993nm,Gozzi:1993sm,Pagani:2017scw} for a detailed discussion of Marinov's path
integral from that point of view.}

{\bf \noindent (3b) The response field identified.}
The $p\dot{q}$-type terms in ({\ref{eq:3-122-tildeL-via-lambda}) also
make it fully manifest that $\xi^a$, or $\lambda_a$, respectively,
coincides with the response field introduced in 
\cite{Martin:1973zz,Janssen1976,DEDOMINICIS1976}.
Indeed, $\lambda_a$ is canonically conjugate to $\phi^a$ in much the same way
as the auxiliary response field is introduced in the operatorial Martin-Siggia-Rose formalism \cite{Martin:1973zz}.
Moreover, in the classical limit $\hbar \rightarrow 0$
the Lagrangean ({\ref{eq:3-122-tildeL-via-lambda}) becomes linear in $\lambda_a$.
As we shall see in a moment, this implies that the kernel of propagation is 
localized on the solution of the equation of motion via a functional Dirac delta.
Therefore, $\lambda_a$ plays the same role as the response field, 
which is introduced to ``exponentiate'' the equation of motion in the Janssen-de Dominicis functional formalism, see e.g.~\cite{Tauber:2014}.

{\bf \noindent (3c) The coordinate change on ${\cal M}\times {\cal M}$.} 
The dynamics of $\phi^{a}$ and $\lambda_{a}$ on the
doubled phase space is governed by the Hamiltonian $\widetilde{H}$.
In connection with eq.~(\ref{eq:2-28_tilde_H-via-phi_and_xi}) we
observed already the appearence of the peculiar quantities $\phi^{a}\pm\hbar\xi^{a}$
in the argument of the standard Hamiltonian $H$. They give $\widetilde{H}$
a distinctively non-local appearence in the phase space sense. 
While before
their raison d'\^{e}tre was somewhat mysterious, our derivation explains
them transparently as due to the crucial transformation of variables
in (\ref{eq:3-103-symm_and_asymm-pi_and_x}).

Let us combine the integration variables $x_{\pm}$ and $p_{\pm}$
appearing in the two copies of the phase space path integral (\ref{eq:3-100-QM-PI-in-phase_space})
into the $2N$-component coordinates
\begin{eqnarray}
\phi_{+}^{a}\equiv\left(p_{+},x_{+}\right) & \mbox{ and } & \phi_{-}^{a}\equiv\left(p_{-},x_{-}\right)\,.\label{eq:3-200-phi_+_and_phi_-}
\end{eqnarray}
Then the change of the integration variables, eq.~(\ref{eq:3-103-symm_and_asymm-pi_and_x}),
writes more compactly as
\begin{eqnarray}
\begin{split}
\phi^{a} & = \frac{1}{2}\left(\phi_{+}^{a}+\phi_{-}^{a}\right)  \\
\xi^{a} & = \frac{1}{2\hbar}\left(\phi_{+}^{a}-\phi_{-}^{a}\right)\,.
\end{split} \label{eq:3-201-phi_symm_and_xi_antisymm}
\end{eqnarray}
The inverse transformation becomes
\begin{eqnarray}
\phi_{\pm}^{a} & = & \phi^{a}\pm\hbar\xi^{a}\,,\label{eq:3-202-phi_pm_via_phi_and_xi}
\end{eqnarray}
which makes it clear that the ``peculiar'' quantities appearing
in Marinov's path integral for $K_{{\rm M}}$ are really nothing else
than the two copies of variables pertaining to the pure state evolution
operators.

There are two main motivations for this coordinate change on
${\cal M}\times{\cal M}$,
\begin{eqnarray}
\left(\phi_{+},\phi_{-}\right) & \mapsto & \left(\phi,\xi\right)\,.\label{eq:3-203-phi_pm_mapsto_phi-xi}
\end{eqnarray}
First, as we said already it recasts the combined path integral to a manifestly
canonical form again, displaying the symplectic structure of the doubled
phase space. 

Second, it connects two different sets of dynamical
variables, each of which is the most natural one, and the one which
we routinely use, in its respective field of applications. This is
quantum statistical mechanics with mixed states on one side, and quantum
mechanics of pure states, i.e., wave functions on the other. In the
sequel of this paper it will therefore be interesting to see what
this transformation of variables amounts to in concrete terms.

{\bf \noindent (3d) Hamilton's equations.}
The Euler-Lagrange equations implied by the
Lagrangean $\widetilde{L}\left(\phi,\lambda,\dot{\phi}\right)$ from
eq.~(\ref{eq:3-122-tildeL-via-lambda}), or equivalently Hamilton's
equations of the $\widetilde{H}\left(\phi,\xi\right)$ in (\ref{eq:2-28_tilde_H-via-phi_and_xi}),
assume the form
\begin{eqnarray}
\dot{\phi}^{a} & = & \frac{1}{2}\omega^{ab}\frac{\partial}{\partial\phi^{b}}\Bigr[H\left(\phi-\hbar\xi\right)+H\left(\phi+\hbar\xi\right)\Bigr]\label{eq:3-126-EOM-phi-via-tildeH}\\
\dot{\xi}^{a} & = & -\frac{1}{2\hbar}\omega^{ab}\frac{\partial}{\partial\phi^{b}}\Bigr[H\left(\phi-\hbar\xi\right)-H\left(\phi+\hbar\xi\right)\Bigr]\,.\label{eq:3-127-EOM-xi-via-tildeH}
\end{eqnarray}
They should be compared to the familiar canonical equations of motion
from the classical Hamilton function $H\left(\phi\right)$. In the
present notation they read
\begin{eqnarray}
\dot{\phi}^{a} & = & h^{a}\left(\phi\right)\,=\,\omega^{ab}\partial_{b}H\left(\phi\right)\label{eq:3-190-EOM-phi-via-hamiltonian-vector-field}
\end{eqnarray}
where $h^{a}$ denotes the usual Hamiltonian vector field on phase
space. It is easy to check that, by virtue of the transformations (\ref{eq:3-103-symm_and_asymm-pi_and_x}),
the equations (\ref{eq:3-126-EOM-phi-via-tildeH}) and (\ref{eq:3-127-EOM-xi-via-tildeH})
are equivalent to two identical, decoupled and ``local'' sets of canonical
equations, namely
\begin{eqnarray}
\dot{\phi}_{+}=\omega^{ab}\partial_{b}H\left(\phi_{+}\right) & 
\mbox{ and } & \dot{\phi}_{-}=\omega^{ab}\partial_{b}H\left(\phi_{-}\right)\,,\label{eq:3-128-EOM-hamiltonian-via-phi_pm}
\end{eqnarray}
involving the variables $\phi_{+}=\left(p_{+},x_{+}\right)$ and $\phi_{-}=\left(p_{-},x_{-}\right)$,
respectively. 

Again we observe that the ``phase-space nonlocal'' character of
equations such as (\ref{eq:3-126-EOM-phi-via-tildeH}) and (\ref{eq:3-127-EOM-xi-via-tildeH})
is easy to camouflage: 
in principle it could be cancelled by using the
$\left(\phi_{+},\phi_{-}\right)$-variables rather than $\left(\phi^{a},\lambda_{a}=\omega_{ab}\xi^{b}\right)$.
However, as we are going to discuss in subsection \ref{sub:D-pref_variabl}, this is not done usually,
the reason being that 
{\it the classical limit of quantum mechanics prefers
one set of variables over the other}.

{\bf \noindent (3e) The classical point.} 
Note that the $K_{{\rm M}}$-path
integral (\ref{eq:2-27_def_PI_for_Moyal_kernel}) involves $\hbar$
only via $\widetilde{H}$. Eq.~(\ref{eq:2-28_tilde_H-via-phi_and_xi})
in the limit $\hbar\rightarrow0$ yields the following $O\left(\hbar^{0}\right)$
term:
\begin{eqnarray}
\widetilde{H} & = & -\xi^{a}\partial_{a}H+O\left(\hbar^{2}\xi^{3}\right)\nonumber \\
 & = & \lambda_{a}h^{a}+O\left(\hbar^{2}\xi^{3}\right)\,.\label{eq:3-130-hbar_zero-tilde_H}
\end{eqnarray}
It involves the standard Hamiltonian vector field $h^{a}\equiv\omega^{ab}\partial_{b}H$.
Using (\ref{eq:3-130-hbar_zero-tilde_H}) and (\ref{eq:2-27_def_PI_for_Moyal_kernel})
we therefore obtain a meaningful, and actually correct result for
Marinov's integral at the strictly classical point, $\hbar=0$:
\begin{eqnarray}
K_{{\rm M}}\left(\phi^{\prime},T;\phi^{\prime\prime},T_{0}\right)\Bigr|_{\hbar=0} & = & \int{\cal D}\phi^{a}\left(\cdot\right)\int{\cal D}\lambda_{a}\left(\cdot\right)\exp\left(-2i\int_{T_{0}}^{T}dt\,\lambda_{a}\left(\dot{\phi}^{a}-h\left(\phi\right)\right)\right)\,.\label{eq:3-131-K_M-at-h_0-CPI}
\end{eqnarray}
We see that, like its relative $Y$, the field $\lambda_{a}$ plays
the role of a Lagrange multiplier for the classical equations of
motion, i.e., Newton's equation in the case of $Y$, and the Hamiltonian
equations $\dot{\phi}^{a}=h^{a}\left(\phi\right)$ in the case at
hand:
\begin{eqnarray}
K_{{\rm M}}\left(\phi^{\prime},T;\phi^{\prime\prime},T_{0}\right)\Bigr|_{\hbar=0} & = & \int_{\phi\left(T_{0}\right)=\phi^{\prime\prime}}^{\phi\left(T\right)=\phi^{\prime}}{\cal D}\phi\left(\cdot\right)\,\delta\left[\dot{\phi}-h\left(\phi\right)\right]\,.\label{eq:3-132-K_M_is_delta_at_h_0}
\end{eqnarray}
This path integral governs the time evolution of phase space densities
$\rho\left(\phi\right)$ in strictly classical statistical mechanics.
It is closely related to the CPI discussed in the literature \cite{Gozzi:1989bf,Gozzi:1989xz,Gozzi:1989vv,Gozzi:1993tm,Gozzi:1999at,Deotto:2000ia,Abrikosov:2004cf,Gozzi:2010iq,Gozzi:2010iu,Cattaruzza:2010wc}.
However, it does not share the possibility with the CPI of evolving $p$-form
densities also.

The functional integral (\ref{eq:3-132-K_M_is_delta_at_h_0}) is perfectly
localized on solutions to the classical canonical equations (\ref{eq:3-190-EOM-phi-via-hamiltonian-vector-field}).
Thus it evaluates to
\begin{eqnarray}
K_{{\rm M}}\left(\phi^{\prime},T;\phi^{\prime\prime},T_{0}\right)\Bigr|_{\hbar=0} & = & \delta
\Bigr(\phi^{\prime}-\Phi_{{\rm cl}}\left(T;\phi^{\prime\prime},T_{0}\right)\Bigr)\,,\label{eq:3-133-K_M-at-hbar_0}
\end{eqnarray}
where $\Phi_{{\rm cl}}$ denotes the solution subject to the initial
condition $\Phi_{{\rm cl}}\left(T_0;\phi^{\prime\prime},T_{0}\right)=\phi^{\prime\prime}$.

\subsection{Preferred Kinematical Variables: Quantum vs.~Classical Mechanics \label{sub:D-pref_variabl}}

\noindent {\bf(1) A common language.} Analyzing both the configuration- and phase-space path integrals
for the Moyal kernel we saw explicitly that it approaches a well defined
generalized function (distribution) in the strictly classical limit, and that
\begin{eqnarray}
\lim_{\hbar\rightarrow0}K_{{\rm M}} & \equiv & K_{{\rm M}}^{0}\label{eq:3-400-hbar_to_zero-limit-of-K_M}
\end{eqnarray}
equals the time evolution kernel of the Liouville equation of classical
statistical physics. 

This is markedly different from the behaviour
of Feynman integrals for the evolution of wave functions, $\int{\cal D}x\,\exp\left(iS\left[x\right]/\hbar\right)$,
which display increasingly rapid oscillations when $\hbar\rightarrow0$.
While such path integrals, too, ``know'' about classical mechanics
in the sense that near classical trajectories the condition $\delta S/\delta x=0$
``tames'' the oscillations to a certain degree, the Feynman kernel
per se has no meaningful limit for $\hbar\rightarrow0$, let alone
a limit that would match with a natural object in the formalism of
classical statistical mechanics.

The smooth classical $\leftrightarrow$ quantum transition characteristic 
of the Moyal kernel and of Marinov's path integral reflects the origin of the
phase space formulation of quantum mechanics in the \emph{deformation
quantization} of the corresponding classical structures. For example,
the pointwise product of phase functions, the Poisson bracket, and
Liouville's equation are continuously ``deformed'' into, respectively,
the star product, the Moyal bracket, and Moyal's equation, whereby
$\hbar$ plays the role of the deformation parameter \cite{Bayen:1977ha,Bayen:1977hb}.

Thanks to this continuity at the classical $\leftrightarrow$ quantum
interface it can be said that classical mechanics on the one hand,
and quantum statistical mechanics in the phase space formulation on
the other, employ the same kinematical variables. And more specifically,
as Marinov's path integral is a continuous function of $\hbar$ that
interpolates
between classical ($\hbar=0$) and quantum mechanics ($\hbar>0$),
the kinematical (integration) variables
$\left(\phi,\lambda \right)$ or $\left(\phi,\xi \right)$ 
which it employs belong to a
{\it common language of classical and quantum mechanics}. 

Notably, there exists
no analogous common language if one restricts to the quantum mechanical
side to a theory of pure states only -- as we do more often than
not.
\vspace{3mm}\\
{\bf(2) Identifying the variables of strictly classical mechanics.} 
Keeping the above remarks in mind we observe that
the evolution kernel $K_{{\rm M}}$ at $\hbar=0$ given in eqs.~(\ref{eq:3-132-K_M_is_delta_at_h_0})
or (\ref{eq:3-133-K_M-at-hbar_0}) is expressed in terms of the common
classical and quantum phase space variables
\begin{eqnarray}
\phi^{a} & = & \frac{1}{2}\left(\phi_{+}^{a}+\phi_{-}^{a}\right)\,\equiv\,\phi_{{\rm class.mech.}}^{a} \,.\label{eq:3-400-symm-variables-for-CM}
\end{eqnarray}
They are coordinates on a single copy of ${\cal M}$,
and relate to the first half of the coordinate transformation in (\ref{eq:3-201-phi_symm_and_xi_antisymm}). 

It
is obvious from the localization on classical trajectories featured
by (\ref{eq:3-132-K_M_is_delta_at_h_0}) that the variables (\ref{eq:3-400-symm-variables-for-CM})
are precisely those in terms of which (Hamiltonian) classical mechanics is formulated. Hence the quantities $\left(p,q\right)\equiv\phi^{a}$ employed
in classical mechanics are descendants of neither $\phi_{+}\equiv\left(p_{+},q_{+}\right)$,
nor $\phi_{-}\equiv\left(p_{-},q_{-}\right)$, i.e., of the variables
connected to the time evolution of pure states $|\psi\rangle$ and
their duals $\langle\psi|$, respectively. 

Rather, \emph{the standard
classical phase space variables are to be identified with the symmetric
average (\ref{eq:3-400-symm-variables-for-CM}) of those variables
that are associated with, respectively, the forward- and backward-time
evolution of pure quantum states}.

In this sense, classical dynamics arises from a symmetric superposition
of an evolution forward and backward in time.

From the configuration space path integral (\ref{eq:3-43-K_M-in-the-hbar_zero-limit})
we can reach an equivalent conclusion: the configuration space variable
of classical mechanics must be identified with the linear combination
\begin{eqnarray}
\xs & = & \frac{1}{2}\left(x_{+}+x_{-}\right)\,\equiv\,\xs_{{\rm class.mech.}}\,.\label{eq:3-405-x-symm-of-CM}
\end{eqnarray}
\vspace{1mm}\\ 
{\noindent \bf(3) Status of the response field.} 
Our point of contact between quantum and classical mechanics
are the path integrals (\ref{eq:3-43-K_M-in-the-hbar_zero-limit})
and (\ref{eq:3-132-K_M_is_delta_at_h_0}) which are strictly localized
on classical solutions. They involve only one of the two independent
linear combinations formed in (\ref{eq:3-201-phi_symm_and_xi_antisymm})
from the ``pure state variables'' $\phi_{\pm}$, namely the symmetric
combination $\frac{1}{2}\left(\phi_{+}^{a}+\phi_{-}^{a}\right)=\phi^{a}\equiv\phi_{{\rm class.mech.}}^{a}$.
The other, antisymmetric one is nothing but the response field
\begin{eqnarray}
\frac{1}{2\hbar}\left(\phi_{+}^{a}-\phi_{-}^{a}\right) & = & \xi^{a}\,.\label{eq:3-410-antisymm-xi-variable}
\end{eqnarray}
It is already fully integrated out at the interface of the classical
and quantum formalisms, i.e., in eqs.~(\ref{eq:3-43-K_M-in-the-hbar_zero-limit})
and (\ref{eq:3-132-K_M_is_delta_at_h_0}).
And in the equivalent representations (\ref{eq:3-42-tilde-y-integral-hbar_0-DiracDelta}) and (\ref{eq:3-131-K_M-at-h_0-CPI})
it merely serves as an auxiliary field needed to express the delta functional.

As a result, Marinov's integral at $\hbar\neq0$ does not straightforwardly
suggest a strictly classical counterpart of the quantum response field
with which it would match at $\hbar=0$.

One might worry about the explicit $\hbar$-dependence in (\ref{eq:3-410-antisymm-xi-variable}),
or in the definition of $\xi^{a}$ in the second equation of (\ref{eq:3-201-phi_symm_and_xi_antisymm})
which breaks down when $\hbar\rightarrow0$. However, the classical
limit of $K_{{\rm M}}$ is well defined nevertheless.
This is possible thanks to cancellations
with further, explicit factors of $\hbar$ in the path integral (\ref{eq:3-104-Y-kernel-via-tilde_L-phi_and_xi})
with the action (\ref{eq:3-107-tilde_L-via_hamiltonian-and_int_by_parts}).
And this, in fact, was the very motivation for including the $\hbar$
factors into the definition (\ref{eq:3-103-symm_and_asymm-pi_and_x}).\footnote{In the subsection on Lagrangean path integrals it has been convenient
to adopt notations such that $\ys$ and $\widetilde{\ys}\equiv \ys/\hbar$
correspond to $\hbar\xi^{q}$ and $\xi^{q}$, respectively.}

As eq.~(\ref{eq:3-410-antisymm-xi-variable}) becomes meaningless
at $\hbar=0$, i.e., in \emph{strictly classical} physics, we conclude that
unlike the symmetric combination $\phi^{a}$, which goes over into
the classical variables, the antisymmetric linear combination of $\phi_{\pm}$,
the response field, has no comparable classical descendant.

Nevertheless, as we are going to show, in the {\it semiclassical} regime
of quantum mechanics the response field does play an important role.
There, quantum mechanics can be understood 
in terms of two copies of classical
mechanics, governing the pair $\left(\phi_{+},\phi_{-}\right)$, or
equivalently $\left(\phi,\xi\right)$.

\section{Coherence, Interference, and the Response Field} \label{sec:coherence-interference-resp_field}

In this and the following section we shall observe the response field
``at work'' by applying the Moyal kernel and its path integral to
elementary quantum mechanics. After a number of introductory discussions,
we focus on interference phenomena, the very hallmark of quantum mechanics,
and in section \ref{sec:Double-Slit-Experiment-and-BohmAharonov}
on the double slit experiment and the Bohm-Aharonov effect.

In order to gain some intuitive understanding of the physics related
to the response field, we first determine under what circumstances
large values of $\ys\left(t\right)$ can arise and play an essential
role. More precisely, we ask which kind of information, deduced from
quantum mechanical theory, has a natural description in terms of a
response field whose value is ``nonclassically large''. 
This is to mean
that $K_{{\rm M}}$ should \emph{not} be well approximated by its \emph{strictly}
classical limit (\ref{eq:3-43-K_M-in-the-hbar_zero-limit}).

On the other hand, since it simplifies the discussion and leads to a particularly
clear picture, we usually do assume in this and the next section that
the $K_{{\rm M}}$-integral is dominated by a certain saddle point,
and that it is sufficient to retain the leading non-trivial order
of the saddle point expansion.

Thus our discussion focusses on the differences between the \emph{classical, or
``tree'' approximation of quantum mechanics} on one side, and \emph{strictly
classical mechanics} on the other. Most of the conceptually deep or
puzzling issues of quantum mechanics show up at this level of approximation
already.

Furthermore, we specialize for pure states $\widehat{\rho}\left(t\right)=|\psi\left(t\right)\rangle\langle\psi\left(t\right)|$,
and study their time evolution in terms of the Wigner function:
\begin{eqnarray}
W_{\psi}\left(p,q,T\right) & = & \int d^{N}p^{\prime}d^{N}q^{\prime}\,K_{{\rm M}}\left(p,q,T;p^{\prime},q^{\prime},T_{0}\right)W_{\psi}\left(p^{\prime},q^{\prime},T_{0}\right)\,.\label{eq:4-7-Wigner-function-via-kernel-on-W_init}
\end{eqnarray}
We assume that the time evolution is governed by a Hamiltonian $H=\frac{1}{2}p^{2}+V$,
which is general enough for our present purposes.

\subsection{Nonlocal Correlations and Unbalanced Forward/Backward Sectors}

As a preparation, let us choose the initial state $\widehat{\rho}\left(T_{0}\right)$
to be $\widehat{\rho}\left(T_{0}\right)=|x_{0}\rangle\langle x_{0}|$.
It is sharply localized in configuration space and amounts to the
wave function $\psi\left(x\right)\equiv\langle x|\psi\rangle=\delta\left(x-x_{0}\right)$
for some $x_0 \in\mathbb{R}^{N}$. Its Wigner function is likewise localized
with respect to $q$, but independent of $p$:
\begin{eqnarray}
W\left(p,q\right) & = & \delta\left(q-x_{0}\right)\,.
\end{eqnarray}
For $T>T_{0}$ it evolves under the influence of the potential $V\left(x\right)$
into the final state
\begin{eqnarray}
W\left(p,q,T\right) & = & \int d^{N}p^{\prime}\,K_{{\rm M}}\left(p,q,T;p^{\prime},x_0,T_{0}\right) \nonumber \\
& = & \int d^{N}s\,\exp\left(-\frac{i}{\hbar}sp\right){\cal K}\left(s,q,T;0,x_{0},T_{0}\right)\,.\label{eq:4-10-W_at_T-from-delta_W}
\end{eqnarray}
Hence we obtain, in terms of the functional integral,
\begin{eqnarray}
W\left(p,q,T\right) & = & \int d^{N}s\,\exp\left(-\frac{i}{\hbar}sp\right) \nonumber \\
& \, &
\times \int_{\ys\left(T_{0}\right)=0}^{\ys\left(T\right)=s/2}{\cal D}\ys\left(\cdot\right)\int_{\xs\left(T_{0}\right)=x_{0}}^{\xs\left(T\right)=q}{\cal D}\xs\left(\cdot\right)\exp\left(2\frac{i}{\hbar}\int_{T_{0}}^{T}dt\,\widetilde{L}\right)\,.\label{eq:4-11-W_at_T_from_deltaW-via_PI}
\end{eqnarray}
Several points should be noted here.
\vspace{3mm}\\
{\bf (1) Enforcing large response fields.} 
We observe that the
path integration over $\ys\left(t\right)$ involves the boundary value
$\ys\left(T\right)=s/2$; the latter is large provided the $s$-integral in
(\ref{eq:4-11-W_at_T_from_deltaW-via_PI}) is dominated by large values
of $s$, and this in turn requires the momentum argument of $W$,
i.e., $p$, to be small. Thus the trajectories $\ys\left(t\right)$
that contribute to $W\left(p,q,T\right)$ in the small-$p$ regime
are forced to become large at their terminal point $t=T$, at least.

Now, in view of the irregularity of typical path integral trajectories
(especially in the Hamiltonian case) one might perhaps be hesitant
to conclude rightaway that the same can also be said meaningfully
about $\ys\left(t\right)$ at $t<T$. However, if we now invoke our
assumption that the path integral for the Moyal kernel is dominated
by one or several smooth saddle points, $\left(\xs^{{\rm cl}}\left(t\right),\ys^{{\rm cl}}\left(t\right)\right),\,t\in\left[T_{0},T\right]$,
continuity implies that
{\it in the small-$p$ regime the classical function $\ys^{{\rm cl}}\left(t\right)$
is forced to be large also away from the terminal point}.
\vspace{3mm}\\
{\bf (2) Nonlocal correlations.} 
Equation
(\ref{eq:2-20_def_Wigner_function}) expresses the Wigner function
in terms of the wave function for the same moment of time. Being the
Fourier transform of the bilinear $\psi\left(q+\frac{s}{2}\right)\psi^{*}\left(q-\frac{s}{2}\right)$,
the Wigner function $W_{\psi}\left(p,q\right)$, at small $p$ arguments,
is seen to be determined by this bilinear for a large separation $s\in\mathbb{R}^{N}$
of the two configuration space points $q\pm\frac{s}{2}$.

Hence, loosely speaking, large $q$-separation vectors $s$, and related
to that, large response fields $\ys \left(t\right)$, encode information
about the wave function at distant points in configuration space. This
information is of a special nature, comprising a bi-local correlation
of $\psi$ and $\psi^{*}$, respectively, across large, potentially
macroscopic distances in position space. It is therefore plausible
to suspect that the response field is of special relevance to such
correlations.
\vspace{3mm}\\ 
{\bf (3) Forward/backward asymmetry.} 
If we undo the transformation of the integration
variables (\ref{eq:3-22-symm_and_antisymm-combinations}) for a moment,
we see that a large value of $\ys^{{\rm cl}}\left(t\right)$ amounts
to a large difference $x_{+}^{{\rm cl}}\left(t\right)-x_{-}^{{\rm cl}}\left(t\right)\equiv2\ys^{{\rm cl}}\left(t\right)$,
where $x_{+}^{{\rm cl}}\left(t\right)$ and $x_{-}^{{\rm cl}}\left(t\right)$
are the pertinent saddle points of the two Feynman integrals. 

Since
the ``plus'' and the ``minus'' sectors are related to, respectively,
forward time evolution (or the evolution of ``kets'' $|\cdots\rangle$)
and backward time evolution (the evolution of the ``bras'' $\langle\cdots|$),
we can also say that {\it if for a given set of boundary conditions the
response field is large, this is indicative of a particularly
unbalanced evolution in the forward and backward
sectors, respectively}.\footnote{Of course there is no net physical violation of time reversal invariance
if the potential is real. But this may not be obvious from the contribution
of an individual saddle point, in particular when the boundary conditions
break time reversal invariance.}

\subsection{Interference Terms}

To further illustrate the role of the response field in correlating
different points of configuration space and in interference phenomena
let us consider wave functions that are superpositions of the form
\begin{eqnarray}
\psi\left(x\right) & = & \psi_{1}\left(x\right)+\psi_{2}\left(x\right)\,.\label{eq:4-20-psi_is_psi1+psi2}
\end{eqnarray}
Their density operator reads
\begin{eqnarray}
\widehat{\rho} & = & |\psi_{1}\rangle\langle\psi_{1}|+|\psi_{2}\rangle\langle\psi_{2}|+|\psi_{1}\rangle\langle\psi_{2}|+|\psi_{2}\rangle\langle\psi_{1}|\,.\label{eq:4-21-rho-from-two-state-superposition}
\end{eqnarray}
Applying the symbol map, this equation turns into
\begin{eqnarray}
W_{\psi}\left(q,p\right) & = & W_{1}\left(p,q\right)+W_{2}\left(p,q\right)+C_{12}\left(p,q\right)\,,\label{eq:4-22-W_2states-via-W1-W2-C12}
\end{eqnarray}
where $W_{1}$ and $W_{2}$ are the individual Wigner functions of
$\psi_{1}$ and $\psi_{2}$, respectively, and the cross term $C_{12}\left(p,q\right)$
is given by
\begin{eqnarray}
C_{12}\left(p,q\right) & = & \int d^{N}s\,\,\psi_{1}\left(q+\frac{s}{2}\right)\psi_{2}^{*}\left(q-\frac{s}{2}\right)\exp\left(-\frac{i}{\hbar}sp\right)+{\rm c.c.}\,.\label{eq:4-23-C12-via-psi1_and_psi2}
\end{eqnarray}
The phase space function $C_{12}\left(p,q\right)$ is the symbol
of the non-diagonal terms in the density operator, $|\psi_{1}\rangle\langle\psi_{2}|+|\psi_{2}\rangle\langle\psi_{1}|$,
and so it embodies all interference effects the states $|\psi_{1}\rangle$
and $|\psi_{2}\rangle$ can give rise to. If $C_{12}\left(p,q\right)$
happens to vanish identically no interference occurs, the total Wigner
function equals the sum of $W_{1}$ and $W_{2}$, and the density
operator is a classical mixture, $|\psi_{1}\rangle\langle\psi_{1}|+|\psi_{2}\rangle\langle\psi_{2}|$.

By way of illustration, let us consider two wave functions $\psi_{1}$
and $\psi_{2}$ that are well localized at two distant points in configuration
space\footnote{As always, we assume the configuration space to be $\mathbb{R}^{N}$.},
at $a$ and $b$, say. As a sharp localization will cause no mathematical
difficulties here, we let
\begin{eqnarray}
\psi_{1}\left(x\right)=\delta\left(x-a\right)\,, & \qquad & \psi_{2}\left(x\right)=\delta\left(x-b\right)\,.\label{eq:4-30-psi1-and-psi2-deltas}
\end{eqnarray}
The interference of the situations ``particle sits at point $a$''
and ``particle sits at point $b$'' is described by the following
phase space function then:
\begin{eqnarray}
C_{12}\left(p,q\right) & = & 2\cos\left(p\, \frac{a-b}{\hbar}\right)\delta\left(q-\frac{a+b}{2}\right)\,.\label{eq:4-31-C12-for-2-delta-states}
\end{eqnarray}

Let us now time evolve the Wigner function $W_{\psi}\left(p,q\right)\equiv W_{\psi}\left(p,q,T_{0}\right)$
of (\ref{eq:4-22-W_2states-via-W1-W2-C12}) with (\ref{eq:4-30-psi1-and-psi2-deltas})
from $T_{0}$ to a later time $T$. By linearity we have 
\begin{eqnarray}
W_{\psi}\left(p,q,T\right) & = & W_{1}\left(p,q,T\right)+W_{2}\left(p,q,T\right)+C_{12}\left(p,q,T\right)\,,
\end{eqnarray}
wherein the last term is particularly interesting:
\begin{eqnarray}
\begin{split}
C_{12}\left(p,q,T\right) & =  \int d^{N}s\,\exp\left(-isp/\hbar\right)
\,\Biggr\{{\cal K}\left(s,q,T;a-b,\frac{a+b}{2},T_{0}\right) \\
 &  \qquad +{\cal K}\left(s,q,T;-\left(a-b\right),\frac{a+b}{2},T_{0}\right)\Biggr\}\,.
\end{split} \label{eq:4-40-C12-at-T-via-calY}
\end{eqnarray}
In terms of the $\left(\xs,\ys\right)$-functional integral,
\begin{eqnarray}
C_{12}\left(p,q,T\right) & = & \int d^{N}s\,\exp\left(-isp/\hbar\right)
\,\Biggr\{\int_{\ys\left(T_{0}\right)=\left(a-b\right)/2}^{\ys\left(T\right)=s/2}{\cal D}\ys\left(\cdot\right)+\int_{\ys\left(T_{0}\right)=-\left(a-b\right)/2}^{\ys\left(T\right)=s/2}{\cal D}\ys\left(\cdot\right)\Biggr\}\nonumber \\
 &  & \times \int_{\xs\left(T_{0}\right)=\left(a+b\right)/2}^{\xs\left(T\right)=q}{\cal D}\xs\left(\cdot\right)\,\exp\left(2\frac{i}{\hbar}\int_{T_{0}}^{T}dt\,\widetilde{L}\right)\,.\label{eq:4-41-C12-via-xy-PI}
\end{eqnarray}

Applying the same reasoning as above, equation (\ref{eq:4-41-C12-via-xy-PI})
shows that (besides the $p$-argument) there is another general factor
that affects the size of typical response field values, namely {\it the geometry
of the state} that is evolved. In the example at hand, this geometry
enters via the center of mass and the relative distance of the two
initial localization points, i.e., $\left(a+b\right)/2$ and $a-b$,
respectively. 

If $a-b$ is large, so is $\ys\left(t\right)$, at least
close to the initial point $\left(t=T_{0}\right)$. In view of the
transformation (\ref{eq:3-22-symm_and_antisymm-combinations}), the
boundary conditions on the $\xs$- and $\ys$-integrals in (\ref{eq:4-41-C12-via-xy-PI})
are such that, in the forward/backward-language, both $x_{+}\left(t\right)$
and $x_{-}\left(t\right)$ can ``reach'' the localization points
$a$ and $b$. More generally, any state with relevant structure at
distant points enforces that large, possibly even macroscopically
large response fields $\ys\left(t\right)$ contribute to its time evolution.

It is obvious from our example that such large response field values are needed
for the interference term $C_{12}$ to survive under time evolution
or, stated differently, for maintaining quantum coherence. As such,
they are closely linked to the unitary character of the quantum mechanical
time evolution. 

If one goes beyond our present setting and includes
interactions of the system with environmental degrees of freedom,
or other modifications that can lead to decoherence, a natural decoherence
scenario consists in suppressing or damping the response field in
such a way that $C_{12}\rightarrow0$ ultimately.

\section{Double Slit Experiment and Bohm-Aharonov Effect \label{sec:Double-Slit-Experiment-and-BohmAharonov}}

This section is devoted to two instructive examples illustrating the
role played by the response field in the semiclassical limit: the double
slit experiment, and the Bohm-Aharonov effect.
\begin{figure}
\includegraphics[scale=0.2]{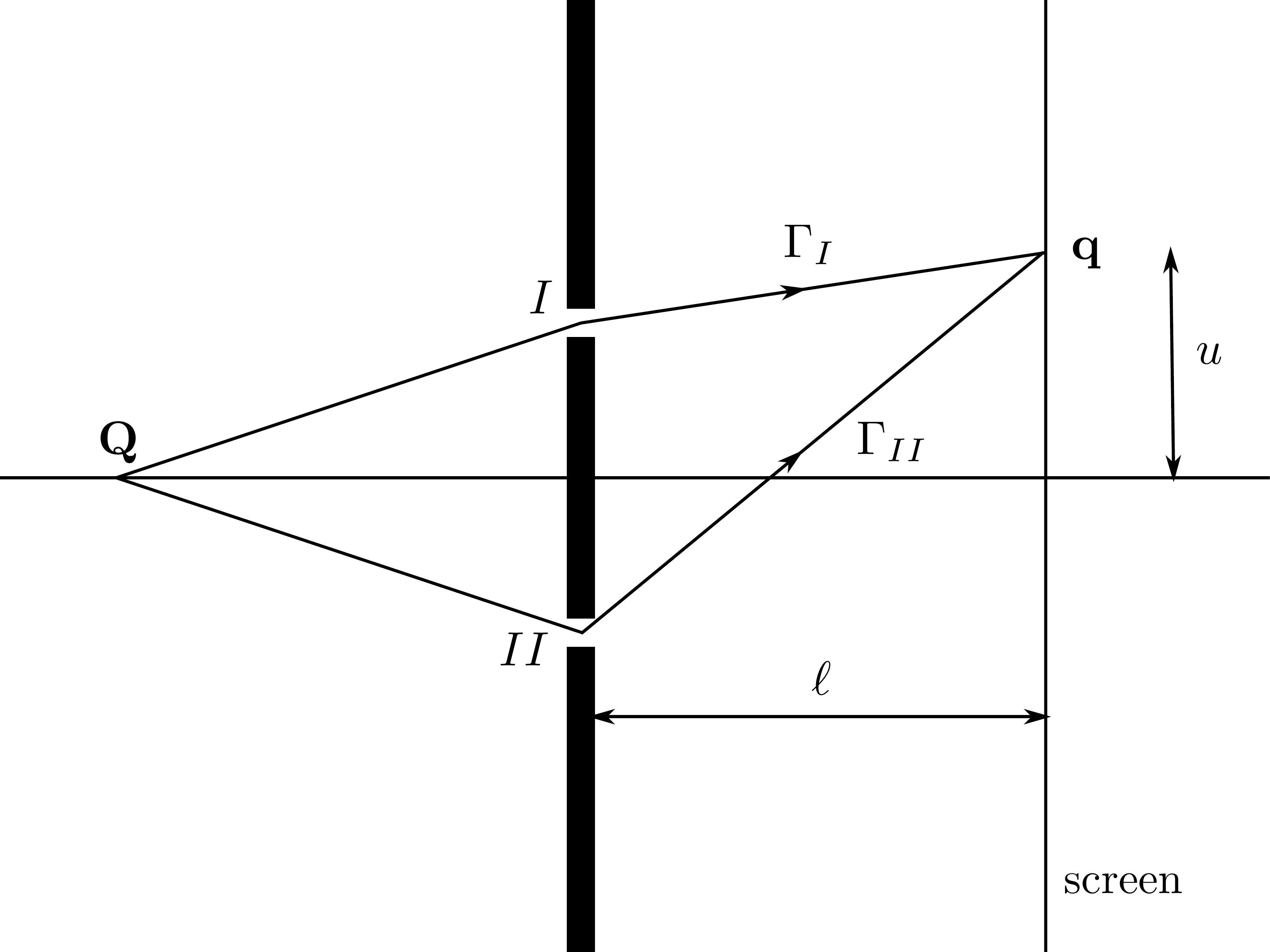}
\caption{The geometry of the double slit experiment for electrons.\label{fig:1}}

\end{figure}

Figure \ref{fig:1} shows a symbolic sketch of a double slit experiment
with electrons. They are emitted from an electron source at the point
${\bf Q}$, then travel along the trajectories $\Gamma_{I}$ or $\Gamma_{II}$,
respectively, passing through slit $I$ or slit $II$, before hitting
the screen at ${\bf q}$. We would like to compute the intensity detected
on the screen in dependence on the point ${\bf q}$, keeping ${\bf Q}$
fixed.\footnote{In this section 3D position and momentum vectors are printed in boldface.}
We model the obstacle that forces the electrons to go through the
slits, assumed infinitesimally narrow, by a potential $V\left(\bf{x}\right)$
which is infinite on the obstacle and vanishes everywhere else. Apart
from the effect of this potential the electrons are considered free
particles.

In terms of the Marinov path integral, the intensity distribution
on the screen is proportional to
\begin{eqnarray}
\left|K\left({\bf q},T;{\bf Q},T_{0}\right)\right|^{2} & = & {\cal K}\left(0,{\bf q},T;0,{\bf Q},T_{0}\right)\nonumber \\
 & = & \int_{{\bf \xs}\left(T_{0}\right)={\bf Q}}^{{\bf \xs}\left(T\right)={\bf q}}{\cal D}{\bf \xs}\left(\cdot\right)\int_{{\bf \ys}\left(T_{0}\right)=0}^{{\bf \ys}\left(T\right)=0}{\cal D}{\bf \ys}\left(\cdot\right)\,\exp\left(2\frac{i}{\hbar}\int_{T_{0}}^{T}dt\,\widetilde{L}\right)\,.\label{eq:5-1-Ksquare-via-PI-for-double-slit}
\end{eqnarray}
We are now going to evaluate this path integral using the lowest non-trivial
order of the saddle point expansion, i.e., the ``classical approximation''
of quantum mechanics.

To shed further light on the significance of the response field we
present the calculation in two different, but equivalent, ways. The
first one employs the variables ${\bf x}_{\pm}$, while the second
highlights the classical- and response fields ${\bf \xs}$ and ${\bf \ys}$,
respectively.

\subsection{The $\left({\bf x}_{+},{\bf x}_{-}\right)$-Perspective}

According to the conventional approach
that does not employ the Marinov integral \cite{Felsager:1981iy}, the kernel
${\cal K}$ in (\ref{eq:5-1-Ksquare-via-PI-for-double-slit}) is regarded
as (literally) the product of two Feynman integrals of the usual form
\begin{eqnarray}
K\left({\bf q},T;{\bf Q},T_{0}\right) & = & \int_{{\bf x}\left(T_{0}\right)={\bf Q}}^{{\bf x}\left(T\right)={\bf q}}{\cal D}{\bf x}\left(\cdot\right)\,e^{\frac{i}{\hbar}S\left[{\bf x}\left(\cdot\right)\right]}\,\approx\,e^{\frac{i}{\hbar}S_{I}}+e^{\frac{i}{\hbar}S_{II}}\,.\label{eq:5-5-kernel-K-via-Feynman-PI}
\end{eqnarray}
In the second step of (\ref{eq:5-5-kernel-K-via-Feynman-PI}) the
saddle point approximation has been invoked, with
\begin{eqnarray}
S_{I,II} & \equiv & S\left[{\bf x}_{{\rm SP}}^{\left(I,II\right)}\left(\cdot\right)\right]\,,\label{eq:5-6-def-S_I-and-S_II}
\end{eqnarray}
where ${\bf x}_{{\rm SP}}^{\left(I\right)}\left(\cdot\right)$ and
${\bf x}_{{\rm SP}}^{\left(II\right)}\left(\cdot\right)$ denote the
classical solutions corresponding to the paths $\Gamma_{I}$ and $\Gamma_{II}$,
respectively. In the double slit geometry they are the only relevant
saddle points. 

When we write down the product of the two $K$'s,
those saddle points
appear both in the ${\bf x}_{+}$- and the ${\bf x}_{-}$-integral:
\begin{eqnarray}
{\cal K}\left(0,{\bf q},T;0,{\bf Q},T_{0}\right) & = & K\left({\bf q},T;{\bf Q},T_{0}\right)^{*}K\left({\bf q},T;{\bf Q},T_{0}\right) \label{eq:5-6-calY-via-x_pm-PI} \\
 & = & \left\{ e^{-\frac{i}{\hbar}S_{I}\left[{\bf x}_{-}={\bf x}_{{\rm SP}}^{\left(I\right)}\right]}+e^{-\frac{i}{\hbar}S_{II}\left[{\bf x}_{-}={\bf x}_{{\rm SP}}^{\left(II\right)}\right]}\right\} \left\{ e^{\frac{i}{\hbar}S_{I}\left[{\bf x}_{+}={\bf x}_{{\rm SP}}^{\left(I\right)}\right]}+e^{\frac{i}{\hbar}S_{II}\left[{\bf x}_{+}={\bf x}_{{\rm SP}}^{\left(II\right)}\right]}\right\} \,.\nonumber
\end{eqnarray}
In the two Feynman integrals representing the $K^{*}K$-product the
respective integration variables ${\bf x}_{+}\left(t\right)$ and
${\bf x}_{-}\left(t\right)$ ``decide'' independently whether they
want to be approximated by one or the other saddle point, making a
total of four cases:
\begin{eqnarray}
{\cal K}\left(0,{\bf q},T;0,{\bf Q},T_{0}\right) & = & \left(e^{-\frac{i}{\hbar}S_{I}}+e^{-\frac{i}{\hbar}S_{II}}\right)\left(e^{\frac{i}{\hbar}S_{I}}+e^{\frac{i}{\hbar}S_{II}}\right)\nonumber \\
 & = & 1+1+e^{\frac{i}{\hbar}\left(S_{I}-S_{II}\right)}+e^{-\frac{i}{\hbar}\left(S_{I}-S_{II}\right)}\,.\label{eq:5-7-calY-via-saddle-point-for-x_PM-PI}
\end{eqnarray}

While in the first line of (\ref{eq:5-7-calY-via-saddle-point-for-x_PM-PI}) the exponentials involving $S_{I}$ and $S_{II}$ are related to the (open)
paths $\Gamma_{I}$ and $\Gamma_{II}$ leading from the source to the screen, each
one of the four terms in the last line of (\ref{eq:5-7-calY-via-saddle-point-for-x_PM-PI})
is linked to a certain \emph{closed} curve, denoted $\Gamma_{I,I}$,
$\Gamma_{II,II}$, $\Gamma_{I,II}$, and $\Gamma_{II,I}$, respectively.
The closed curves arise from combining $\Gamma_{I}$ or $\Gamma_{II}$ in the ``forward'',
i.e., ${\bf x}_{+}$-sector, with $-\Gamma_{I}$ or $-\Gamma_{II}$ in the ``backward'',
i.e., ${\bf x}_{-}$-sector. 
(The minus sign indicates the reversed
orientation.) The closed curves correspond to the following saddle
point combinations:
\begin{eqnarray}
\begin{split}
\Gamma_{I,I}: & \hspace{1cm} {\bf x}_{+}={\bf x}_{{\rm SP}}^{\left(I\right)}\,, \, {\bf x}_{-}={\bf x}_{{\rm SP}}^{\left(I\right)} \\
\Gamma_{II,II}: & \hspace{1cm} {\bf x}_{+}={\bf x}_{{\rm SP}}^{\left(II\right)}\,, \, {\bf x}_{-}={\bf x}_{{\rm SP}}^{\left(II\right)} \\
\Gamma_{I,II}: & \hspace{1cm} {\bf x}_{+}={\bf x}_{{\rm SP}}^{\left(I\right)}\,, \, {\bf x}_{-}={\bf x}_{{\rm SP}}^{\left(II\right)} \\
\Gamma_{II,I}: & \hspace{1cm} {\bf x}_{+}={\bf x}_{{\rm SP}}^{\left(II\right)}\,, \, {\bf x}_{-}={\bf x}_{{\rm SP}}^{\left(I\right)}\,.
\end{split}\label{eq:5-8-Gamma_I-II_and_x_pm}
\end{eqnarray}
These four curves are sketched in Figure \ref{fig:2}.
\begin{figure}
\begin{subfigure}{.5\textwidth}
  \centering
  \includegraphics[width=.8\linewidth]{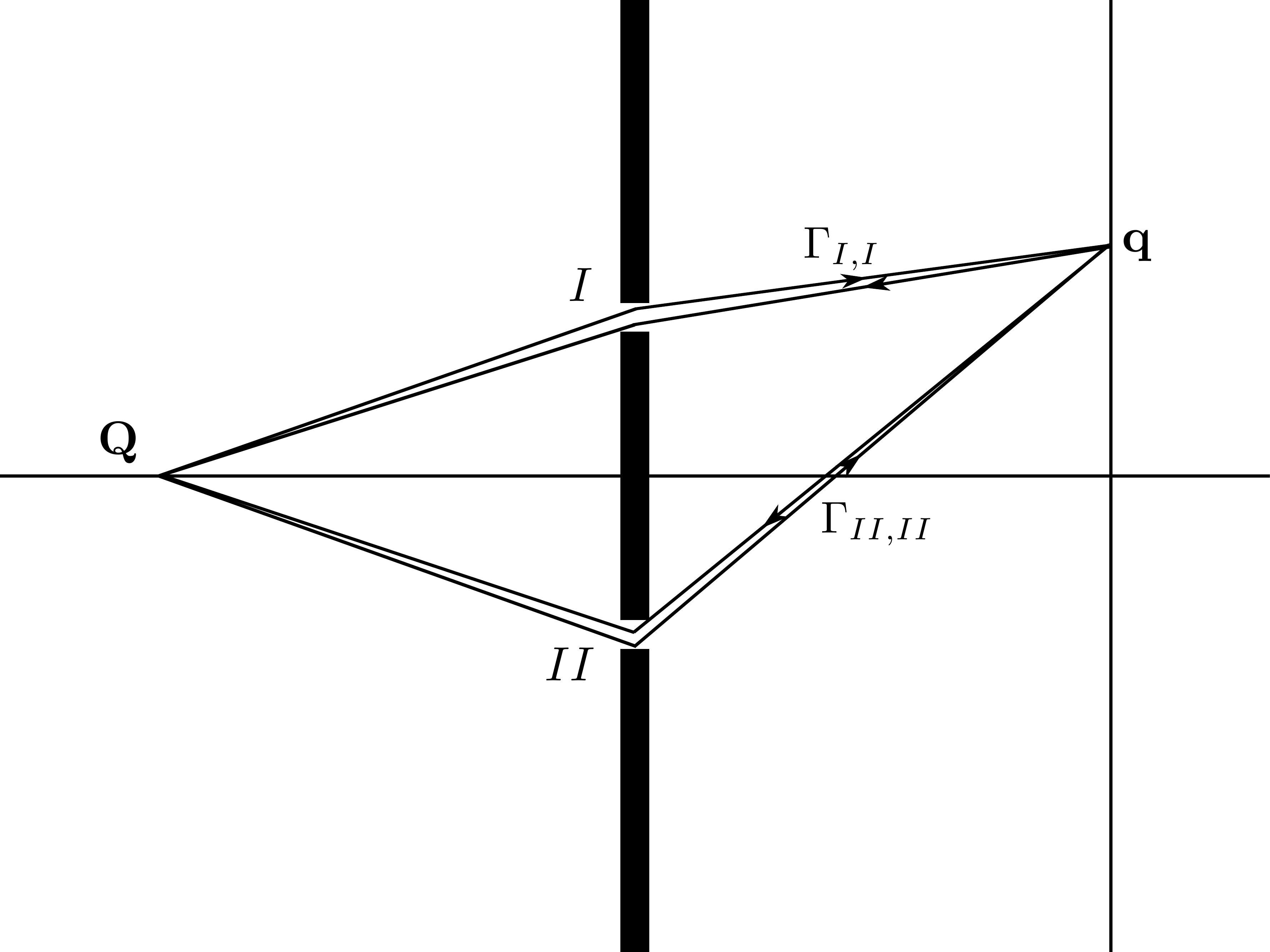}
  \caption{$\,$}
  \label{fig:sfig1}
\end{subfigure}%
\begin{subfigure}{.5\textwidth}
  \centering
  \includegraphics[width=.8\linewidth]{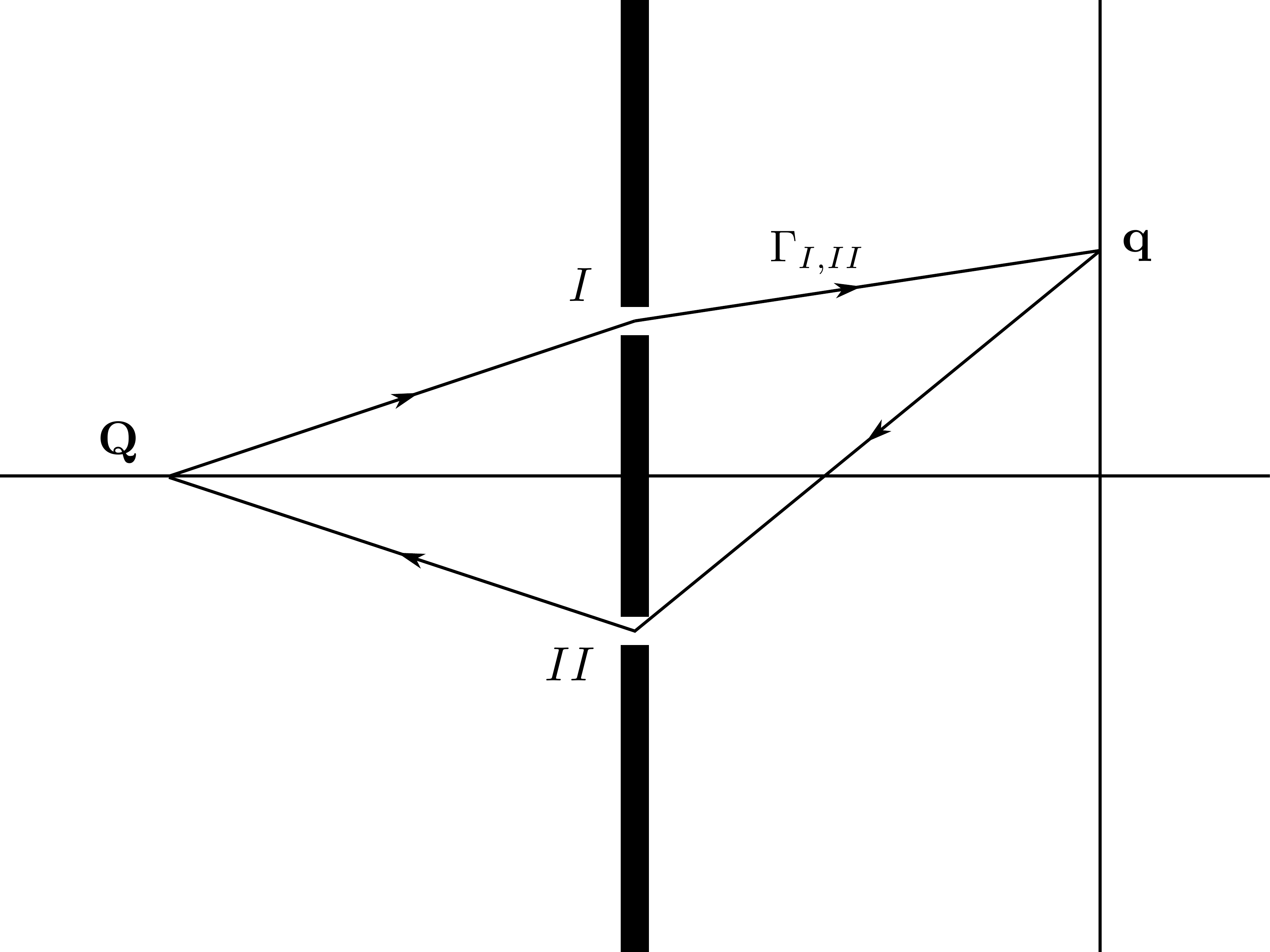}
  \caption{$\,$}
  \label{fig:sfig2}
\end{subfigure}
\caption{The closed curves shown represent the saddle points (\ref{eq:5-8-Gamma_I-II_and_x_pm})
that appear if one adopts the $\left({\bf x}_+,{\bf x}_- \right)$-perspective.
Figure \ref{fig:sfig1}: The backtracking paths $\Gamma_{I,I}$ and $\Gamma_{II,II}$.
Figure \ref{fig:sfig2}: The path $\Gamma_{I,II}$. The saddle point $\Gamma_{II,I}$
is obtained by reversing its orientation.}
\label{fig:2}
\end{figure}

From (\ref{eq:5-7-calY-via-saddle-point-for-x_PM-PI}) we obtain
\begin{eqnarray}
{\cal K}\left(0,{\bf q},T;0,{\bf Q},T_{0}\right) & = & 2+2\cos\left(\Delta S/\hbar\right)\,=\,4\cos^{2}\left(\frac{\Delta S}{2\hbar}\right)\,,\label{eq:5-9-calY-for-double-slit}
\end{eqnarray}
and so it remains to compute the difference of the ``on-shell''
actions, $\Delta S\equiv S_{I}-S_{II}$. Using that 
$S\left[{\bf x}\right]=\frac{1}{2}m\int dt\,\dot{{\bf x}}^{2}$
along the straight sections of $\Gamma_{I,II}$, a simple calculations
yields\footnote{This equation is valid to first order in the deflection angle, i.e.,
in $u$. It treats $v$ and the geometric data of the apparatus as
independent, hence the time difference $T-T_{0}$ is implicitly adjusted
such that $v\left(T-T_{0}\right)=\ell+$(distance obstacle-source).} $\Delta S=mv\left(a/\ell\right)u$ with $v$ and $m$ the velocity
and the mass of the electrons, $a$ the distance of the two slits,
and $\ell$ the distance from the obstacle to the screen. Furthermore,
$u$ is a coordinate on the screen given by the perpendicular distance
of ${\bf q}$ from the central axis.
In evaluating $\Delta S$ we take the limit in which
the distance between the source and the obstacle is much bigger than
any other length in the problem. 

As a result the intensity on the screen is given by
\begin{eqnarray}
I\left(u\right) & \propto & {\cal K}\left(0,{\bf q},T;0,{\bf Q},T_{0}\right)\,=\,4\cos^{2}\left(\frac{mv}{2\hbar}\left(\frac{a}{\ell}\right)u\right)\,.
\label{eq:5-10-I_u-double-slit}
\end{eqnarray}
This is the well known formula which
exhibits the modulation typical of an interference pattern.

\subsection{The $\left({\bf \xs},{\bf \ys}\right)$-Perspective}

Now let us pretend that we are unaware of the hidden product structure
of Marinov's path integral, and let us evaluate the integral (\ref{eq:5-1-Ksquare-via-PI-for-double-slit})
directly, i.e., by integrating over the ``classical variable'' ${\bf \xs}\left(t\right)$
and the response field ${\bf \ys}\left(t\right)$. Equivalently, we
may also start out from its integrated-by-parts version:
\begin{eqnarray}
{\cal K}\left(0,{\bf q},T;0,{\bf Q},T_{0}\right) & = & \int_{{\bf \xs}\left(T_{0}\right)={\bf Q}}^{{\bf \xs}\left(T\right)={\bf q}}{\cal D}{\bf \xs}\left(\cdot\right)\int_{{\bf \ys}\left(T_{0}\right)=0}^{{\bf \ys}  \left(T\right)=0}{\cal D}{\bf \ys}\left(\cdot\right)\nonumber \\
& \, & \times \exp\left(-2\frac{i}{\hbar}\int_{T_{0}}^{T}dt\left\{ {\bf \ys}\cdot\ddot{{\bf \xs}}+\widetilde{V}\left({\bf \xs},{\bf \ys}\right)\right\} \right)\,.\label{eq:5-20-calY-via-xy-PI-and-double-slit}
\end{eqnarray}
Herein the sole purpose of
\begin{eqnarray}
\widetilde{V}\left({\bf \xs},{\bf \ys}\right) & \equiv & \frac{1}{2}\Bigr(V\left({\bf \xs}+{\bf \ys}\right)-V\left({\bf \xs}-{\bf \ys}\right)\Bigr)\label{eq:5-21-tildeV-for-double-slit}
\end{eqnarray}
is to implement the constraint caused by the slit geometry. To avoid
technical issues, let us assume that $V\left({\bf x}\right)$ is actually
an appropriately smoothened variant of the characteristic function
which equals infinity (zero) when the point ${\bf x}$ lies on (off)
the obstacle.
\vspace{3mm}\\
{\bf (1) Response field is not ``small''.} 
A word of caution might be appropriate
here. While we shall invoke the semiclassical expansion again, it
would be quite wrong to apply the approximation of a small response
field at this point. In fact, using
\begin{eqnarray}
\widetilde{V}\left({\bf \xs},{\bf \ys}\right) & = & {\bf \ys}\cdot\nabla V\left({\bf \xs}\right)+O\left({\bf \ys}^{3}\right)\label{eq:5-25}
\end{eqnarray}
would turn (\ref{eq:5-20-calY-via-xy-PI-and-double-slit}) into
\begin{eqnarray}
{\cal K}\left(0,{\bf q},T;0,{\bf Q},T_{0}\right) & = & \int_{{\bf \xs}\left(T_{0}\right)={\bf Q}}^{{\bf \xs}\left(T\right)={\bf q}}{\cal D}{\bf \xs}\left(\cdot\right)\,\delta\left[\ddot{{\bf \xs}}+\nabla V\left({\bf \xs}\right)\right]\,.\label{eq:5-26}
\end{eqnarray}
This integral describes entirely classical physics, however, and ``knows''
nothing about the interference pattern on the screen, which we expect
to find.
\vspace{3mm}\\
{\bf (2) Saddles of Marinov's integral.} Thus, let us be careful and try to find an approximation
in the form
\begin{eqnarray}
{\cal K}\left(0,{\bf q},T;0,{\bf Q},T_{0}\right) & = & \sum_{\left({\bf \xs}_{{\rm SP}},{\bf \ys}_{{\rm SP}}\right)}\exp\left(2\frac{i}{\hbar}\widetilde{S}\left[{\bf \xs}_{{\rm SP}},{\bf \ys}_{{\rm SP}}\right]\right)\label{eq:5-30-calY-via-sum-xy_SP}
\end{eqnarray}
where the sum is over the saddle points $\left({\bf \xs}_{{\rm SP}}\left(t\right),{\bf \ys}_{{\rm SP}}\left(t\right)\right)$
of the action functional
\begin{eqnarray}
\widetilde{S}\left[{\bf \xs},{\bf \ys}\right] & \equiv & \int_{T_{0}}^{T}dt\left\{ \dot{{\bf \xs}}\cdot\dot{{\bf \ys}}-\widetilde{V}\left({\bf \xs},{\bf \ys}\right)\right\} \,,\label{eq:5-31-tildeS-def}
\end{eqnarray}
and we are open-minded as for the magnitude of ${\bf \ys}_{{\rm SP}}\left(t\right)$.

Now, such $\left({\bf \xs},{\bf \ys}\right)$-saddle points should be determined by
solving the coupled system of equations (\ref{eq:3-27-EOM-x_and_y})
which looks rather unwieldy. Even if $V\left({\bf x}\right)$ is the
characteristic function of a fairly simple geometry, the support of
$\widetilde{V}\left({\bf \xs},{\bf \ys}\right)$ on the doubled configuration
space, and the trajectories it allows, are not easily visualized.

However, after a moment of contemplation even the hypothetical physicists
who are unaware of the hidden product structure of Marinov's integral
will discover that by introducing new variables ${\bf x}_{\pm}$ via
\begin{eqnarray}
\begin{split}
{\bf \xs}_{{\rm SP}}\left(t\right) & =  \frac{1}{2}\left[{\bf x}_{+}\left(t\right)+{\bf x}_{-}\left(t\right)\right]\,, \\
{\bf \ys}_{{\rm SP}}\left(t\right) & =  \frac{1}{2}\left[{\bf x}_{+}\left(t\right)-{\bf x}_{-}\left(t\right)\right]\,,
\end{split} \label{eq:5-35}
\end{eqnarray}
the system (\ref{eq:3-27-EOM-x_and_y}) can be decoupled, and that
it boils down to a doubled Newton equation, $\ddot{{\bf x}}_{\pm}=-\nabla V\left({\bf x}_{\pm}\right)$.
Furthermore, these physicists will find out that the standard Newton
equation $\ddot{{\bf x}}=-\nabla V\left({\bf x}\right)$ admits two
solutions consistent with the boundary data, namely ${\bf x}_{{\rm SP}}^{\left(I\right)}$
and ${\bf x}_{{\rm SP}}^{\left(II\right)}$. 
These latter two trajectories
on the (ordinary) configuration space happen to be saddle points of
a Feynman-type path integral, but this fact is of no relevance from
the present perspective.

The complete set of solutions to the \emph{doubled} Newton equation
is found by picking ${\bf x}_{+}\left(t\right)$ and ${\bf x}_{-}\left(t\right)$
independently from the set $\left\{ {\bf x}_{{\rm SP}}^{\left(I\right)}\left(t\right),{\bf x}_{{\rm SP}}^{\left(II\right)}\left(t\right)\right\} $.
Then transforming back to the $\left(\xs,\ys \right)$ language via equation
(\ref{eq:5-35}) gives rise to a total of four open curves on the
doubled configuration space:
\begin{eqnarray}
\left({\bf \xs}_{{\rm SP}},{\bf \ys}_{{\rm SP}}\right):\left[T_{0},T\right]\rightarrow\mathbb{R}^{3}\times\mathbb{R}^{3} & , & t\mapsto\left({\bf \xs}_{{\rm SP}}\left(t\right),{\bf \ys}_{{\rm SP}}\left(t\right)\right)_{\alpha,\beta},\,\,\, \alpha,\beta\in\left\{ I,II\right\} \,.\label{eq:5-36-x_and_y-SP-curves}
\end{eqnarray}
These are the saddle points to be summed over in (\ref{eq:5-30-calY-via-sum-xy_SP}).
We list them in Table \ref{tab:table-1}.
\begin{table}
\begin{tabular}{|c|c|c|c|c|}
\hline 
$\left(\alpha,\beta\right)$ & ${\bf x}_{+}$ & ${\bf x}_{-}$ & ${\bf \xs}_{{\rm SP}}$ & ${\bf \ys}_{{\rm SP}}$\tabularnewline
\hline 
\hline 
$\left(I,I\right)$ & ${\bf x}_{{\rm SP}}^{\left(I\right)}$ & ${\bf x}_{{\rm SP}}^{\left(I\right)}$ & ${\bf x}_{{\rm SP}}^{\left(I\right)}$ & $0$\tabularnewline
\hline 
$\left(II,II\right)$ & ${\bf x}_{{\rm SP}}^{\left(II\right)}$ & ${\bf x}_{{\rm SP}}^{\left(II\right)}$ & ${\bf x}_{{\rm SP}}^{\left(II\right)}$ & $0$\tabularnewline
\hline 
$\left(I,II\right)$ & ${\bf x}_{{\rm SP}}^{\left(I\right)}$ & ${\bf x}_{{\rm SP}}^{\left(II\right)}$ & ${\bf x}_{\Sigma}$ & ${\bf x}_{\Delta}$\tabularnewline
\hline 
$\left(II,I\right)$ & ${\bf x}_{{\rm SP}}^{\left(II\right)}$ & ${\bf x}_{{\rm SP}}^{\left(I\right)}$ & ${\bf x}_{\Sigma}$ & $-{\bf x}_{\Delta}$\tabularnewline
\hline 
\end{tabular}

\caption{The saddle points for $\left({\bf x}_+,{\bf x}_- \right)$ and $\left({\bf \xs},{\bf \ys} \right)$ in terms of 
${\bf x}_{\rm{SP}}^{(I)}$, ${\bf x}_{\rm{SP}}^{(II)}$, ${\bf x}_{\Sigma}$, and ${\bf x}_{\Delta}$. \label{tab:table-1}}

\end{table}

The first two saddle points in the table, $\left(\alpha,\beta\right)=\left(I,I\right)$
and $\left(\alpha,\beta\right)=\left(II,II\right)$, are trivial in
the sense that they have an identically vanishing response field,
$\left({\bf \xs}_{{\rm SP}},{\bf \ys}_{{\rm SP}}\right)_{\alpha,\alpha}=\left({\bf x}_{{\rm SP}}^{\left(\alpha\right)},0\right)$.
The third and the fourth instead, $\left(\alpha,\beta\right)=\left(I,II\right)$
and $\left(\alpha,\beta\right)=\left(II,I\right)$, are of the form
$\left({\bf \xs}_{{\rm SP}},{\bf \ys}_{{\rm SP}}\right)_{\alpha,\beta}=\left({\bf x}_{\Sigma},\pm{\bf x}_{\Delta}\right)$,
with the abbreviations
\begin{eqnarray}
\begin{split}
{\bf x}_{\Sigma} \left(t\right) & \equiv  \frac{1}{2}\left[{\bf x}_{{\rm SP}}^{\left(I\right)}\left(t\right)+{\bf x}_{{\rm SP}}^{\left(II\right)}\left(t\right)\right] \\
{\bf x}_{\Delta} \left(t\right) & \equiv  \frac{1}{2}\left[{\bf x}_{{\rm SP}}^{\left(I\right)}\left(t\right)-{\bf x}_{{\rm SP}}^{\left(II\right)}\left(t\right)\right]\,.
\end{split}\label{eq:5-40-x_Sigma-and-x_Delta}
\end{eqnarray}
The saddle points of Marinov's integral are depicted symbolically
in Figure \ref{fig:3}.
%\begin{eqnarray}
%\begin{split}
%\Gamma_{I,I}: & {\bf x}_{+}={\bf x}_{{\rm SP}}^{\left(I\right)}\,, \, {\bf x}_{-}={\bf x}_{{\rm SP}}^{\left(I\right)} \\
%\Gamma_{II,II}: & {\bf x}_{+}={\bf x}_{{\rm SP}}^{\left(II\right)}\,, \, {\bf x}_{-}={\bf x}_{{\rm SP}}^{\left(II\right)} \\
%\Gamma_{I,II}: & {\bf x}_{+}={\bf x}_{{\rm SP}}^{\left(I\right)}\,, \, {\bf x}_{-}={\bf x}_{{\rm SP}}^{\left(II\right)} \\
%\Gamma_{II,I}: & {\bf x}_{+}={\bf x}_{{\rm SP}}^{\left(II\right)}\,, \, {\bf x}_{-}={\bf x}_{{\rm SP}}^{\left(I\right)}\,.
%\end{split}\label{eq:5-8-Gamma_I-II_and_x_pm}
%\end{eqnarray}
%These four curves are sketched in Figure \ref{fig:2}.
\begin{figure}
\begin{subfigure}{.4\textwidth}
  \centering
  \includegraphics[width=.8\linewidth]{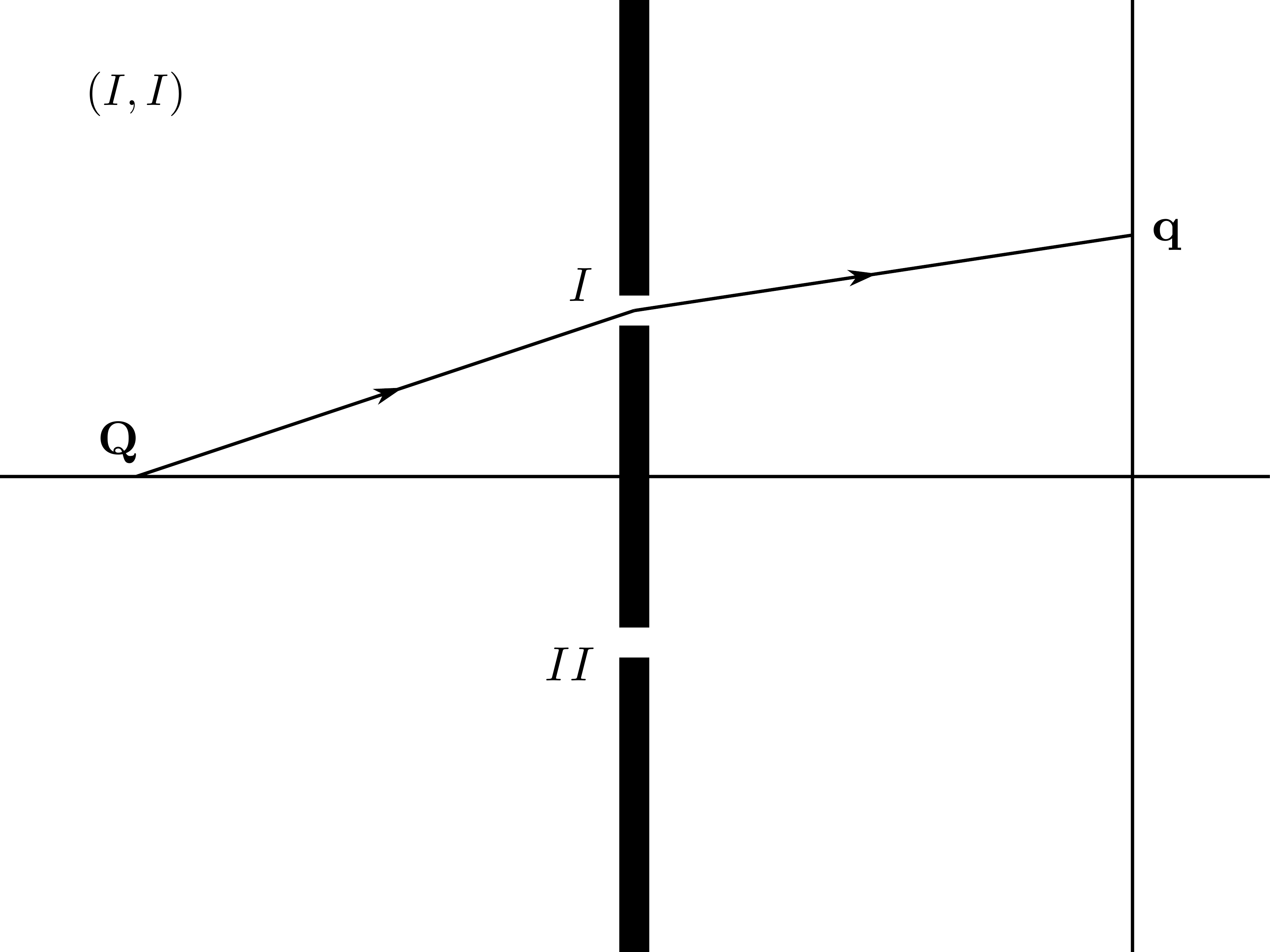}
  \caption{ }
  \label{fig:sfig3a}
\end{subfigure}%
\begin{subfigure}{.4\textwidth}
  \centering
  \includegraphics[width=.8\linewidth]{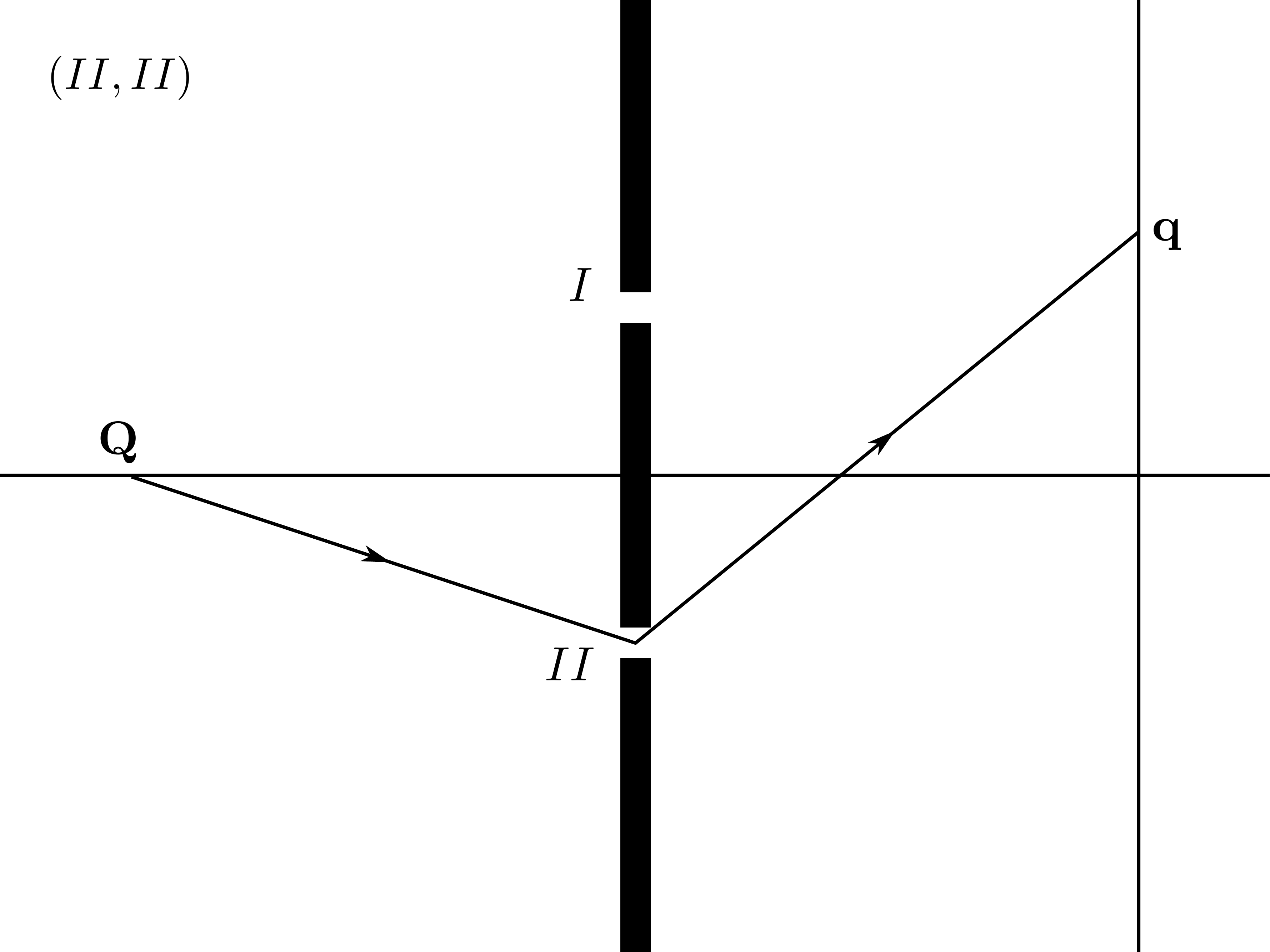}
  \caption{ }
  \label{fig:sfig3b}
\end{subfigure}
\begin{subfigure}{.4\textwidth}
  \centering
  \includegraphics[width=.8\linewidth]{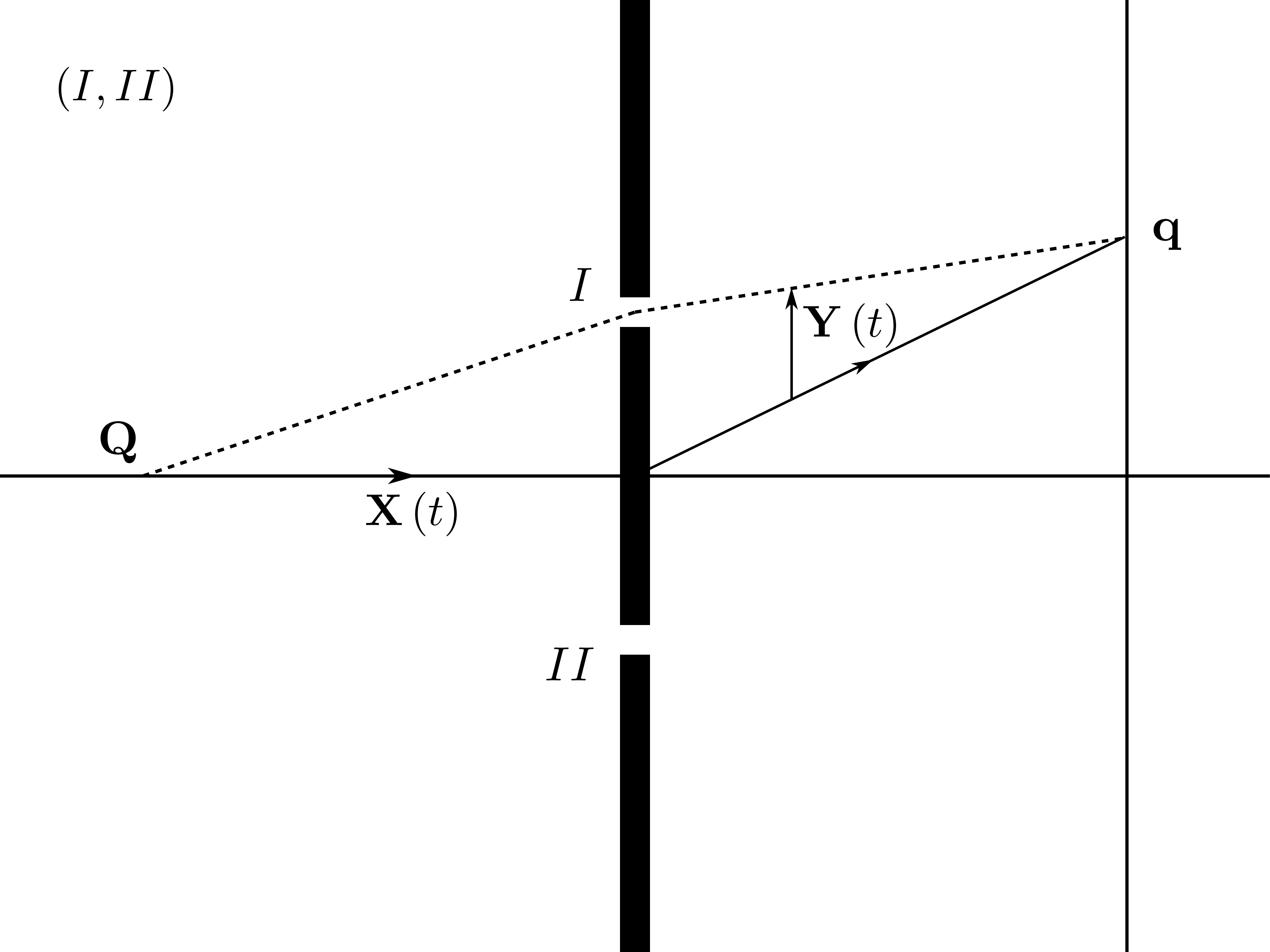}
  \caption{ }
  \label{fig:sfig3c}
\end{subfigure}
\begin{subfigure}{.4\textwidth}
  \centering
  \includegraphics[width=.8\linewidth]{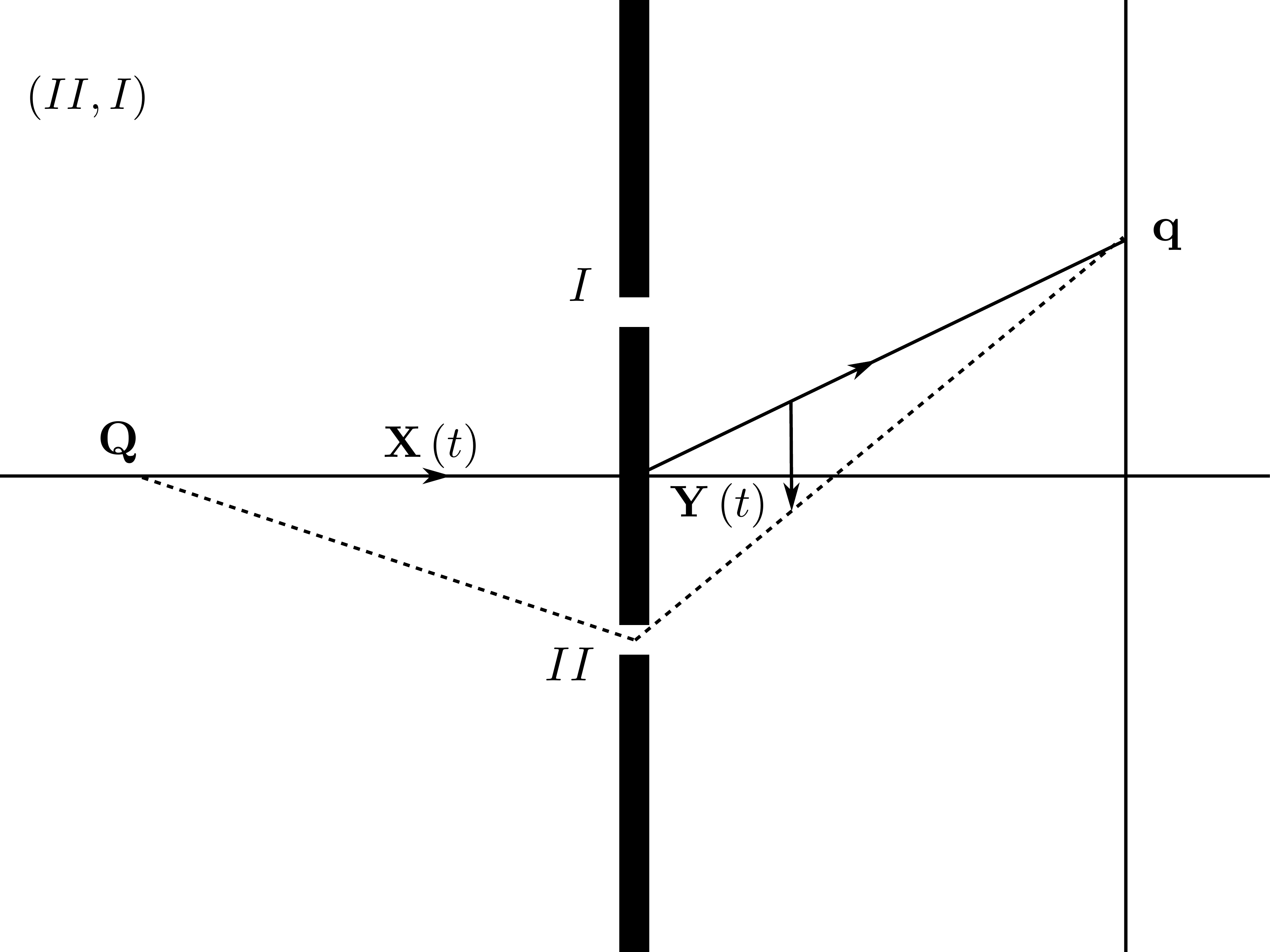}
\caption{ }
  \label{fig:sfig3d}
\end{subfigure}
\caption{The diagrams illustrate the saddle points of Marinov's path integral.
They are open curves on the doubled configuration space.
Figures \ref{fig:sfig3a} and \ref{fig:sfig3b} show projections onto the classical
${\bf \xs}$-space; they are to be combined with ${\bf \ys}\left(t \right)\equiv 0$.
In Figures \ref{fig:sfig3c} and \ref{fig:sfig3d}, the tip of the ${\bf \ys}\left(t \right)$ vector lies on the dashed lines
when added to ${\bf \xs}\left(t \right)$,
which in turn is given by solid lines.}
\label{fig:3}
\end{figure}
\vspace{1mm}\\ 
{\bf (3) The result.} 
Using the entries of Table \ref{tab:table-1}, it is now easy to evaluate
the sum over saddle points in eq.~(\ref{eq:5-30-calY-via-sum-xy_SP}):
\begin{eqnarray}
{\cal K}\left(0,{\bf q},T;0,{\bf Q},T_{0}\right) & = & \sum_{\alpha,\beta\in\left\{ I,II\right\} }\exp\left(2\frac{i}{\hbar}\widetilde{S}\left[\left({\bf \xs}_{{\rm SP}},{\bf \ys}_{{\rm SP}}\right)_{\alpha,\beta}\right]\right)\nonumber \\
 & = & \sum_{\alpha,\beta\in\left\{ I,II\right\} }\exp\left(\frac{i}{\hbar}S\left[\left({\bf x}_{+}\right)_{\alpha,\beta}\right]-\frac{i}{\hbar}S\left[\left({\bf x}_{-}\right)_{\alpha,\beta}\right]\right)\nonumber \\
 & = & 1+1+e^{i\Delta S/\hbar}+e^{-i\Delta S/\hbar}\nonumber \\
 & = & 4\cos^{2}\left(\frac{\Delta S}{2\hbar}\right)\,.\label{eq:5-41-calY-via-sum-SP}
\end{eqnarray}
Thus we recover the result of our first calculation, equation (\ref{eq:5-9-calY-for-double-slit}),
and so (\ref{eq:5-10-I_u-double-slit}) ultimately, as it should be.
\vspace{3mm}\\ 
{\bf (4) Importance of the response field.}
The most interesting aspect of this second calculation are the stationary
points which, besides the ``classical mechanics-field'' ${\bf \xs}\left(t\right)$,
also involve the less familiar response field ${\bf \ys}\left(t\right)$.
It is seen here to co-determine the semiclassical limit and is on a par with ${\bf \xs}\left(t\right)$. 

The time dependence of the stationary points 
is displayed in Figure \ref{fig:3}. 
Several essential points must be noted here.
 
{\bf \noindent (4a)} There exist saddle points of the Marinov path integral
with an identically vanishing response field, ${\bf \ys}\left(t\right)=0$
$\forall t\in\left[T_{0},T\right]$.
In a way, they are ``more classical'' than the other ones, in that
they are fully described by the trajectory of the classical field,
${\bf \xs}\left(t\right)$. In our example, these are the cases $\left(\alpha,\beta\right)=\left(I,I\right)$
and $\left(\alpha,\beta\right)=\left(II,II\right)$, respectively,
describing a perfectly classical electron that travels through slit
$I$ in the first, and slit $II$ in the second case.
 
{\bf \noindent (4b)}
The third and the fourth saddle point have a quite intriguing structure,
$\left({\bf \xs}_{{\rm SP}},{\bf \ys}_{{\rm SP}}\right)=\left({\bf x}_{\Delta},\pm{\bf x}_{\Sigma}\right)$.
Their classical- (response-) field component is given by the vectorial
sum (difference) of the above trajectories passing through slit $I$
and $II$, respectively. 

As a consequence, \emph{the ``classical
mechanics-component'' of the saddle point, i.e., ${\bf X}_{{\rm SP}}\left(t\right)$,
passes through neither of the two slits}: at first it follows the symmetry
axis, then it hits the obstacle, seems to tunnel through it, and gets deflected only then.

Here we can observe quite nicely what is needed in order to promote
classical mechanics to semiclassical quantum mechanics: it is the
additional information provided by the response field. 
The ``improved'' trajectories ${\bf \xs}_{{\rm SP}}\left(t\right)\pm{\bf \ys}_{{\rm SP}}\left(t\right)$
actually do pass through one or the other slit, and are responsible
for the interference pattern ultimately.
 
{\bf \noindent (4c)} The magnitude
of the response field can reach macroscopic values. Along the saddle
point trajectories it can become as large as $a/2$, half of the distance
of the two slits. 

Indeed, perturbatively speaking the final result (\ref{eq:5-10-I_u-double-slit})
sums up arbitrarily high orders of $a$. If one tries to enforce a
small response field by choosing a tiny value of $a$, and Taylor expands (\ref{eq:5-10-I_u-double-slit})
in this quantity, the periodicity of the interference pattern is ruined
at any finite order.

\subsection{The Bohm-Aharonov Effect}

After the following modification the above experimental set up lends
itself for a demonstration of the Bohm-Aharonov effect \cite{Aharonov:1959fk}:
at equal distances from the slits we place a solenoid close to the
obstacle, producing a magnetic field ${\bf B}\left({\bf x}\right)$
that is non-zero in a small domain only and gives rise to a magnetic
flux $\Phi$ through the transverse plane, see Figure \ref{fig:4}. 
\begin{figure}
\includegraphics[scale=0.2]{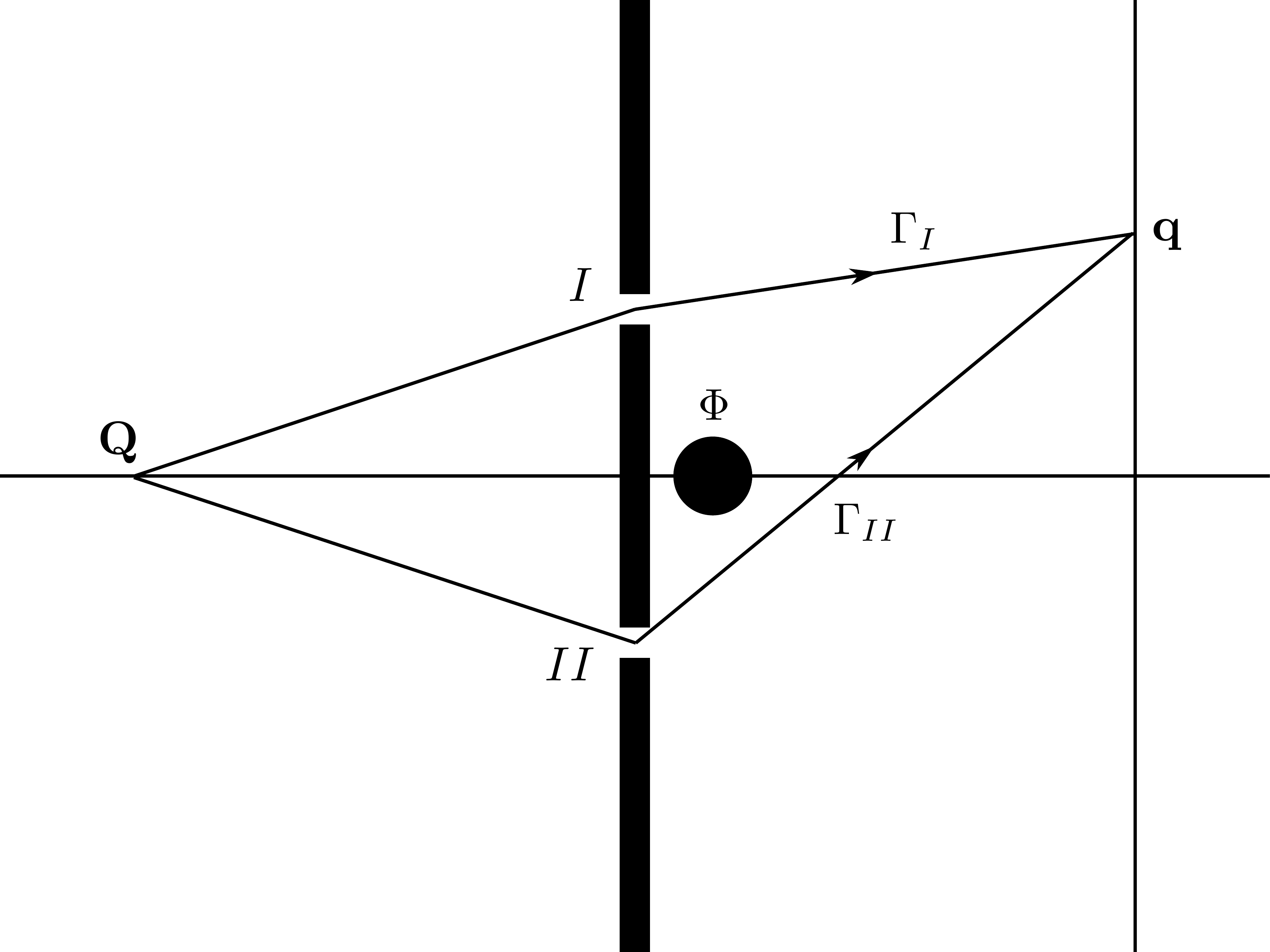}
\caption{Modified double slit experiment to demonstrate the Bohm-Aharonov effect.
The magnetic field is nonzero in the darkened circle only,
giving rise to a magnetic flux $\Phi$.  \label{fig:4}}
\end{figure}
The classical
paths $\Gamma_{I}$ and $\Gamma_{II}$ computed above for ${\bf B}\left({\bf x}\right)\equiv0$
are assumed to pass well outside this domain. 

Let us use the Moyal-Marinov
path integral again to find the modified interference pattern in presence
of the solenoid.
\vspace{3mm}\\
{\bf (1)} 
The ${\cal K}$- and $K_{{\rm M}}$-integrals
for the electrons exposed to an external magnetic field are described
in Appendix \ref{sec:EM-Lagrangean}. They have the by now familiar
structure, except that their Lagrangean $\widetilde{L} \left({\bf \xs},{\bf \ys},\dot{{\bf \xs}},{\bf \dot{\ys}}\right)$
contains additional terms depending on the vector potential, see equations
(\ref{eq:B-2}), (\ref{eq:B-3}). It implies the following Euler-Lagrange
equations for the saddle points:\footnote{For simplicity we omit here the scalar potential $V$ whose sole purpose
is to model the geometric obstructions imposed by the slit geometry.}
\begin{eqnarray}
\begin{split}
\ddot{{\bf \xs}} & =  e\, \dot{{\bf \xs}}\times{\bf B}_{+}\left({\bf \xs},{\bf \ys}\right)+e\, \dot{{\bf \ys}}\times{\bf B}_{-}\left({\bf \xs},{\bf \ys}\right) \\
\ddot{{\bf \ys}} & =  e\, \dot{{\bf \xs}}\times{\bf B}_{-}\left({\bf \xs},{\bf \ys}\right)+e\, \dot{{\bf \ys}}\times{\bf B}_{+}\left({\bf \xs},{\bf \ys}\right)\,.
\end{split}\label{eq:5-60-EOM-with-B}
\end{eqnarray}
They feature the bi-local functions
\begin{eqnarray}
{\bf B}_{\pm}\left({\bf \xs},{\bf \ys}\right) & = & \nabla_{\xs}\times{\bf A}_{\pm}\left({\bf \xs},{\bf \ys}\right)\nonumber \\
 & = & \frac{1}{2}\left[{\bf B}\left({\bf \xs}+{\bf \ys}\right)\pm{\bf B}\left({\bf \xs}-{\bf \ys}\right)\right]\,,\label{eq:5-61-B-via-A}
\end{eqnarray}
with ${\bf B}\left({\bf x}\right)\equiv\nabla\times{\bf A}\left({\bf x}\right)$
denoting the ordinary magnetic field.
\vspace{3mm}\\
{\bf (2)}
In view of the non-local appearance of the equation of motion (\ref{eq:5-60-EOM-with-B})
one must beware of a rather tempting speculation: Might it be that the $\left({\bf \xs},{\bf \ys}\right)$-trajectories
do feel the ${\bf B}$-field,
rather than just the ${\bf A}$-field, 
while the usual ones don't? 

The answer is no, however, since (\ref{eq:5-60-EOM-with-B}) is related to the
decoupled saddle point equations 
$\ddot{{\bf x}}_{\pm}=e\, \dot{{\bf x}}_{\pm}\times{\bf B}\left({\bf x}_{\pm}\right)$
by ${\bf x}_{\pm}={\bf \xs}\pm{\bf \ys}$, and each one of them has the
same form as the classical equation, $\ddot{{\bf x}}=e\, \dot{{\bf x}}\times{\bf B}\left({\bf x}\right)$.
As a result, if the relevant solutions to the latter are unaffected
by the magnetic field, so are ${\bf x}_{+}\left(t\right)$ and ${\bf x}_{-}\left(t\right)$,
and as a consequence, ${\bf \xs}\left(t\right)$ and ${\bf \ys}\left(t\right)$.
\vspace{3mm}\\ 
{\bf (3)}
Thus, the saddle points of the doubled path integral are the same
with and without the magnetic field. What is different, however, is
the value of the action evaluated along those saddle point trajectories. 
A simple
calculation shows that the $\dot{{\bf x}}\cdot{\bf A}$-terms shift
$\Delta S\equiv S_{I}-S_{II}$ by a term proportional to the magnetic
flux, $\Delta S=mv\left(a/\ell\right)u+e\Phi$, so that the intensity
on the screen is now given by
\begin{eqnarray}
I\left(u\right) & \propto & \cos^{2}\left(\frac{mv}{2\hbar}\left(\frac{a}{\ell}\right)u+\frac{e\Phi}{2\hbar}\right)\,.\label{eq:5-70-intensity-for-AB}
\end{eqnarray}
This is indeed the correct result. The ${\bf B}$-field confined to
the interior of the solenoid causes a rigid, i.e., $u$-independent
translation of the interference pattern visible on the screen \cite{Felsager:1981iy}.

\section{Instantons of the ``Wick rotated'' response field} \label{sec:Wick-rotated-instantons}

In the response field formalism, 
it is common to analytically continue, or ``Wick rotate'' the response field $\ys$ by defining a new field $\ys^{\prime}\equiv -i\ys$
and considering $\ys^{\prime}$ real then \cite{Tauber:2014}.

As an example, let us consider an action associated to a stochastic
differential equation, which often takes the form 
$iS=i\int \ys \left[\cdots\right]-\int\frac{1}{2}D \ys^{2}$,
where the dots represent terms specific to the system under consideration.
After introducing $\ys^{\prime}$, the original expression for $\exp\left(iS\right)$ 
becomes 
$\exp \left(-\int \ys^{\prime}\left[\cdots\right]+\int\frac{1}{2}D \ys^{\prime2}\right)$.

Although this change of variables is possibly harmless in perturbation
theory, it is not clear if this is the case in the full path integral.
Indeed, such a ``Wick rotation'' is not always legitimate even in
the saddle point approximation.

In this section we consider only the class of theories for which the ``Wick rotation'' makes sense. For this class we show that the associated instantons
are actually the same as the \emph{complex
instantons found in standard Feynman path integrals}, see e.g. \cite{Witten:2010zr}. 

To see this explicitly
we must first rewrite the action associated to the Moyal kernel 
by introducing the Euclidean time $\tau =i t$
and then perform the transformation $\ys^{\prime}\equiv-i\ys$.
Let us start by considering
\begin{eqnarray}
iS\left[x_{+},x_{-}\right] & = & i\int dt\left\{ \frac{1}{2}\dot{x}_{+}^{2}-V\left(x_{+}\right)\right\} -i\int dt\left\{ \frac{1}{2}\dot{x}_{-}^{2}-V\left(x_{-}\right)\right\} \\
-S_{E}\left[x_{+},x_{-}\right] & = & -\int d\tau\left\{ \frac{1}{2}\dot{x}_{+}^{2}+V\left(x_{+}\right)-\frac{1}{2}\dot{x}_{-}^{2}-V\left(x_{-}\right)\right\} \,,
\end{eqnarray}
where the second line gives the Euclideanized action of the Marinov path
integral. Next we rewrite the Euclidean action $S_{E}$ in terms of
the symmetric and the anti-symmetric combinations, $x_{+}=\xs+\ys$ and
$x_{-}=\xs-\ys$. One finds
\begin{eqnarray}
S_{E}\left[\xs , \ys\right] & = & 
\int d\tau \Bigr\{ 2\dot{\xs}\dot{\ys}+V\left(\xs +\ys \right)-V\left(\xs -\ys \right)\Bigr\} \,.
\end{eqnarray}
Finally we ``Wick rotate'' the response field via $\ys^{\prime}\equiv-i\ys$
and obtain
\begin{eqnarray}
S_{E}\left[\xs ,\ys^{\prime}\right] & = & \int d\tau \Bigr\{2 i\, \dot{\xs}\dot{\ys}^{\prime}+V\left(\xs +i\ys^{\prime}\right)-V\left(\xs -i\ys^{\prime}\right)\Bigr\} \,.
\end{eqnarray}
The equations of motion derived from $S_{E}\left[\xs ,\ys^{\prime}\right]$
read then
\begin{eqnarray}
\begin{split}
\ddot{\xs} & =  \frac{1}{2} \Bigr[ V^{\prime}\left(\xs+i\ys^{\prime}\right)+V^{\prime}\left(\xs-i\ys^{\prime}\right) \Bigr] \\
\ddot{\ys}^\prime & =  \frac{1}{2i} \Bigr[V^{\prime}\left(\xs+i\ys^{\prime}\right)-V^{\prime}\left(\xs-i\ys^{\prime}\right) \Bigr]\,.
\end{split}\label{eq:EOM-Moyal-Wick-rotated-y}
\end{eqnarray}

Surprisingly enough, the equations of motion (\ref{eq:EOM-Moyal-Wick-rotated-y})
appear also in a different context and are particularly relevant in
the resurgence program, see e.g. \cite{Behtash:2015zha}. 

In fact, let us consider instantons
in the standard Feynman path integral. These instantons are
finite action solutions to the equation of motion of the Euclidean action
$S_E=\int d\tau \left(\frac{1}{2}\dot{x}^{2}+V\left(x\right)\right)$, i.e., 
they satisfy $\ddot{x}=V^{\prime}\left(x\right)$.
It turns out important to consider not only real solution,
but also complex solutions to the equation of motion, involving a
holomorphic potential then \cite{Behtash:2015zha}:
\begin{eqnarray}
\frac{d^2}{d\tau^2} z \left(\tau \right)
& = & V^{\prime}\left(z \left(\tau \right)\right)\,,\label{eq:holomorphic-Eucledean-EOM}
\end{eqnarray}
with $z\left(\tau\right)= \xs \left(\tau\right)+i \ys \left(\tau\right)$.
This is the holomorphic Newton equation which we mentioned in the Introduction already.
By comparing the real and the imaginary parts of (\ref{eq:holomorphic-Eucledean-EOM})
and thereby employing the Cauchy-Riemann equations for the holomorphic
potential, it can be checked straightforwardly that (\ref{eq:holomorphic-Eucledean-EOM})
is actually equivalent to (\ref{eq:EOM-Moyal-Wick-rotated-y}).

This establishes an intriguing connection between the instantons of the Moyal
path integral with a ``Wick rotated'' response field on one side, and the complex
instantons in standard Feynman integrals on the other.

It must be emphasized though that
these instantons actually enter in different actions and different
kernels of propagation. It is therefore difficult to state a precise
relation, if any, between the results obtained in the two frameworks.
We hope to be able to come back to this issue in the future.

\section{Summary and conclusions} \label{sec:conclusions}

In this work we studied the Marinov path integral and focused on various aspects of its saddle point approximation.
Particular attention has been paid to the crucial role of the response field, which we discussed in detail.

After outlining the scope of our work in Section \ref{sec:introduction},
in Section \ref{sec:preliminaries} we reviewed the connection between the Moyal approach to quantum mechanics and the path integral introduced by Marinov.
In Section \ref{sec:Path-integral-representation-Moyal-kernel} we provided a 
derivation of the Marinov path integral by combining two Feynman path integrals.
This allowed us to pin down the exact relationship between the kinematical
variables used in quantum and strictly classical mechanics, respectively.
Furthermore, it led to an intriguing random force representation 
of the semiclassical time evolution which is based on concept of Airy
averaging.

(Details regarding the precise definition of the path integrals by discretization on a time lattice have been confined in Appendix \ref{sec:Path-integral-on-time-lattice}.)

In Section \ref{sec:coherence-interference-resp_field} we studied how 
quantum interference and correlation effects emerge in the Marinov path integral and we explored the pivotal role of the response field in this regard.

In Section \ref{sec:Double-Slit-Experiment-and-BohmAharonov} we presented two illustrative applications of the approach advocated in the paper. 
Within the saddle point approximation, 
we displayed how 
the double slit interference and the Bohm-Aharonov effect 
make their appearance in the Marinov path integral
in a novel and somewhat surprising fashion.

(The path integral of a particle interacting with an electro-magnetic field was given in Appendix \ref{sec:EM-Lagrangean}.)

In Section \ref{sec:Wick-rotated-instantons} we highlighted a surprising relationship between the instanton solutions of the Marinov path integral with an analytically continued or  ``Wick rotated'' response field on the one side,
and ordinary complex instantons, 
as studied recently in the resurgence program, for instance, on the other. 
It turned out that after the ``Wick rotation'' of the response field the Euclidean equations of motion are actually the same as those studied in resurgence. 
This implies that the instantons found in one framework can be used also in the other.
This connection might be of interest in the ongoing efforts towards
a Picard-Lefschetz theory applicable to oscillating functional integrals.

The perspective adopted in the present paper offers an alternative approach to the standard saddle point approximation, which is better suited to discuss the semiclassical approximation and the strictly classical limit of the results obtained.
In particular it offers an improved conceptual understanding of the 
quantum-classical interface that cannot be provided by the Feynman integrals
with their escalating oscillations for $\hbar \rightarrow 0$.

It would be very interesting to generalize the present study to the BRS-symmetric extension of the Moyal formalism \cite{Gozzi:1993nk,Gozzi:1993nm,Gozzi:1993sm}. Indeed, this is related to the recent super-extended Schwinger-Keldysh formalism \cite{Crossley:2015evo,Haehl:2016pec}, see \cite{Pagani:2017scw} for a discussion regarding the relationship between the two approaches.

\vspace{1cm}
{\bf Acknowledgements}\\
\noindent We are grateful to Gerald Dunne and Neil Turok for inspiring discussions.

\newpage{}

\appendix

\section{Path integrals on a time lattice \label{sec:Path-integral-on-time-lattice}}

In the main part of the paper the various path integrals are treated
in a somewhat formal continuum notation. However, all manipulations
described there can be justified more rigorously by performing them
on their well defined discrete counterparts defined over a time lattice,
and taking the temporal continuum limit corresponding to an arbitrarily
fine lattice only thereafter. In this appendix we give some details
of these (in general rather cumbersome) calculations, thereby focussing
on the discrete version of the Lagrangean path integrals from Subsection
\ref{subsec:Lagrangian-Path-Integrals}.
\vspace{3mm}\\
{\bf (1) The lattice definition.} 
We construct
the regularized functional integral representation of the Moyal kernel,
i.e., the actual \emph{definition} of its path integral, by combining
two discretized path integrals for the Feynman kernels. For systems
with Lagrangeans $L=\frac{1}{2}\dot{x}^{2}-V\left(x\right)$, the
Feynman kernel is given by the $\nu\rightarrow\infty$ limit of the
following ordinary $\left(\nu-1\right)$-fold integral \cite{Schulman:1981vu}:
\begin{eqnarray}
K\left(x^{\prime},T;x^{\prime\prime},T_{0}\right) & = & \left(2\pi i\hbar\epsilon\right)^{-N\nu/2}\int d^{N}x_{1}\cdots d^{N}x_{\nu-1} \nonumber \\
&\, & \times \exp\left\{ \frac{i\epsilon}{\hbar}\sum_{n=1}^{\nu}\left[\frac{1}{2}\left(\frac{x_{n}-x_{n-1}}{\epsilon}\right)^{2}-V\left(x_{n}\right)\right]\right\} \,.\label{eq:A-1}
\end{eqnarray}
Here we set $x_{0}\equiv x^{\prime\prime}$, $x_{\nu}\equiv x^{\prime}$,
and $\epsilon\equiv\left(T-T_{0}\right)/\nu$ denotes the lattice
constant corresponding to the regularization parameter $\nu$. Inserting
two copies of (\ref{eq:A-1}), for finite and equal values of $\nu$
into (\ref{eq:3-10_Y-function_via_F-kernels}) leads to 
\begin{eqnarray}
&&{\cal K}\left(s,q,T;s^{\prime},q^{\prime},T_{0}\right) \, = \, \left(2\pi\hbar\epsilon\right)^{-N\nu}\int d^{N}x_{1}^{+}\cdots d^{N}x_{\nu-1}^{+}\int d^{N}x_{1}^{-}\cdots d^{N}x_{\nu-1}^{-} \label{eq:A-2} \\
 && \hspace{0.5cm} \times \exp\Biggr\{\frac{i}{2\hbar\epsilon}\sum_{n=1}^{\nu}\left[\left(x_{n}^{+}-x_{n-1}^{+}\right)^{2}-\left(x_{n}^{-}-x_{n-1}^{-}\right)^{2}\right]-\frac{i\epsilon}{\hbar}\sum_{n=1}^{\nu}\left[V\left(x_{n}^{+}\right)-V\left(x_{n}^{-}\right)\right]\Biggr\}\,. \nonumber
\end{eqnarray}
In the sums above, the index values $n=0$ and $n=\nu$ refer to the
terminal positions $x_{n}^{\pm}$. They are not integrated over but
rather are fixed by the boundary conditions:
\begin{eqnarray}
\begin{split}
x_{0}^{+}=q^{\prime}+\frac{1}{2}s^{\prime} & ,\,  x_{0}^{-}=q^{\prime}-\frac{1}{2}s^{\prime} \\
x_{\nu}^{+}=q+\frac{1}{2}s & ,\,  x_{\nu}^{-}=q-\frac{1}{2}s
\end{split}\label{eq:A-3}
\end{eqnarray}
Next we introduce new variables in (\ref{eq:A-2}),
\begin{eqnarray}
\begin{split}
\xs_{n} & = \frac{x_{n}^{+}+x_{n}^{-}}{2} \\
\ys_{n} & = \frac{x_{n}^{+}-x_{n}^{-}}{2}
\end{split} \label{eq:A-4}
\end{eqnarray}
so that $x_{n}^{\pm}=\xs_{n}\pm \ys_{n}$. The transformation
(\ref{eq:A-4}) applies for all $n=0,1,2,\cdots,\nu-1,\nu$. Hereby
the index values $n=1,2,\cdots,\nu-1$ refer to integration variables,
while $n=0$ and $n=\nu$ relate to the boundary variables:
\begin{eqnarray}
\begin{split}
\xs_{0}=q^{\prime} & , \, \xs_{\nu}=q \\
\ys_{0}=\frac{1}{2}s^{\prime} & , \, \ys_{\nu}=\frac{1}{2}s\,.
\end{split}\label{eq:A-5}
\end{eqnarray}
In terms of the new variables, the integral (\ref{eq:A-2}) reads
as follows:
\begin{eqnarray}
&&{\cal K}\left(s,q,T;s^{\prime},q^{\prime},T_{0}\right) \, = \, 2^{-N}\left(\pi\hbar\epsilon\right)^{-N}\int d^{N}\xs_{1}\cdots d^{N}\xs_{\nu-1}\int d^{N}\ys_{1}\cdots d^{N}\ys_{\nu-1} \label{eq:A-6} \\
 && \hspace{0.5cm} \times \exp\left\{ \frac{2i}{\hbar}\epsilon\sum_{n=1}^{\nu}\left[\left(\frac{\xs_{n}-\xs_{n-1}}{\epsilon}\right)\left(\frac{\ys_{n}-\ys_{n-1}}{\epsilon}\right)-\frac{1}{2}\left(V\left(\xs_{n}+\ys_{n}\right)-V\left(\xs_{n}-\ys_{n}\right)\right)\right]\right\} \,.\nonumber
\end{eqnarray}
This multiple integral together with (\ref{eq:A-5}) is our final result for the regularized
path integral. The kernels ${\cal K}$, and after a Fourier transformation,
$K_{{\rm M}}$, are given by the well defined $\nu\rightarrow\infty$
limit of (\ref{eq:A-6}). 

In the \emph{formal} continuum limit (i.e.
when $\nu\rightarrow\infty$ is applied to the integrand rather than
the evaluated integral) the multiple integral (\ref{eq:A-6}) gives
rise to (\ref{eq:3-23-Y-kernel-via-x_and_y}) with (\ref{eq:3-24-widetilde-Lagr_x_and_y})
of the main text.
\vspace{3mm}\\
{\bf (2) Integration by parts.} 
In writing down the alternate variant
of the continuum ${\cal K}$-integral in equation (\ref{eq:3-30-Y-kernel-via_x_and_y-and-int-by-parts})
we performed an integration by parts on the $\dot{\xs}\dot{\ys}$ term
in the action. The justification of this step is slightly more subtle.
First of all one shows that the terms of the discrete kinetic term
in (\ref{eq:A-6}) can be reorganized as follows
\begin{eqnarray}
\sum_{n=1}^{\nu}\left(\xs_{n}-\xs_{n-1}\right)\left(\ys_{n}-\ys_{n-1}\right) & = & -\sum_{n=1}^{\nu-1}\ys_{n}\left(\xs_{n+1}+\xs_{n-1}-2\xs_{n}\right)\nonumber \\
 &  & +\frac{1}{2}s\left(q-\xs_{\nu-1}\right)-\frac{1}{2}s^{\prime}\left(\xs_{1}-q^{\prime}\right)\,.\label{eq:A-7}
\end{eqnarray}
Note the different range of the sums in this identity. 

Multiplying (\ref{eq:A-7}) by $1/\epsilon$ and taking the limit $\nu\rightarrow\infty$ it
would imply the continuum formula
\begin{eqnarray}
\int_{T_{0}}^{T}dt\,\dot{\xs}\left(t\right)\dot{\ys}\left(t\right) & = & -\int_{T_{0}}^{T}dt\,\ys\left(t\right)\ddot{\xs}\left(t\right)+\frac{1}{2}s\, \dot{\xs}\left(T\right)-\frac{1}{2}s^{\prime}\, \dot{\xs}\left(T_{0}\right)\,,\label{eq:A-8}
\end{eqnarray}
which is nothing but the integration by parts performed formally in
Section \ref{sec:Path-integral-representation-Moyal-kernel}.

Instead, the regularized expression which provides the actual definition of the alternative continuum path integral (\ref{eq:3-30-Y-kernel-via_x_and_y-and-int-by-parts})
is given by:\footnote{In writing down (\ref{eq:A-9}) we also omitted a term from the sum
over $\widetilde{V}\left(x_{n},y_{n}\right)$ which gives no contribution
in the $\nu\rightarrow\infty$ limit of the multiple integral.}
\begin{eqnarray}
{\cal K}\left(s,q,T;s,q^{\prime},T_{0}\right) & = & 2^{-N}\left(\pi\hbar\epsilon\right)^{-N\nu}\nonumber \\
 &  & \times \int d^{N}\xs_{1}\cdots d^{N}\xs_{\nu-1}\exp\left\{ \frac{i}{\hbar\epsilon}\left[s\left(q-\xs_{\nu-1}\right)-s^{\prime}\left(\xs_{1}-q^{\prime}\right)\right]\right\} \label{eq:A-9}\\
 &  & \times \int d^{N}\ys_{1}\cdots d^{N}\ys_{\nu-1}\exp
\Biggr\lbrace -\frac{2i\epsilon}{\hbar}\sum_{n=1}^{\nu-1}
\Bigr[\ys_{n}\left(\frac{\xs_{n+1}+\xs_{n-1}-2\xs_{n}}{\epsilon^{2}}\right) \nonumber \\
& & \qquad +\widetilde{V}\left(\xs_{n},\ys_{n}\right)\Bigr] \Biggr\rbrace
 \,.\nonumber 
\end{eqnarray}
While the above sum over $n$ involves the terminal positions $\xs_{0}=q^{\prime}$
and $\xs_{\nu}=q$, respectively, it is independent of $s$ and $s^{\prime}$.
Hence the entire $s,s^{\prime}$-dependence in (\ref{eq:A-9}) is
made explicit by the first exponential function under the integral.
This proves the corresponding claim made in the main text in relation
with equation (\ref{eq:3-30-Y-kernel-via_x_and_y-and-int-by-parts}).

Furthermore, at this point it is straightforward to perform the Fourier
transform that connects ${\cal K}$ to $K_{{\rm M}}$. In this way
we learn that a precise definition of the formal expression (\ref{eq:3-35-K_M-via-x_and_y-integral-and-deltas})
for the Moyal kernel is given by the $\nu\rightarrow\infty$ limit
of
\begin{eqnarray}
K_{{\rm M}}\left(p,q,T;p^{\prime},q^{\prime},T_{0}\right) & = & \left(\pi\hbar\right)^{-N\left(\nu-1\right)}\epsilon^{-N\left(\nu-2\right)} \nonumber \\
 &  & \int d^{N}\xs_{2}\cdots d^{N}\xs_{\nu-2}\int d^{N}\ys_{1}\cdots d^{N}\ys_{\nu-1} \label{eq:A-10} \\
 &\, & \times \exp\left\{ -\frac{2i\epsilon}{\hbar}\sum_{n=1}^{\nu-1}\left[\ys_{n}\left(\frac{\xs_{n+1}+\xs_{n-1}-2\xs_{n}}{\epsilon^{2}}\right)+\widetilde{V}\left(\xs_{n},\ys_{n}\right)\right]\right\} \,,\nonumber 
\end{eqnarray}
whereby the four variables
\begin{eqnarray}
\xs_{0}=q^{\prime} \,, & \hspace{1cm} & \xs_{\nu}=q\label{eq:A-11}\\
\xs_{1}=q^{\prime}+\epsilon p^{\prime} \,, & \hspace{1cm} & \xs_{\nu-1}=q-\epsilon p\label{eq:A-12}
\end{eqnarray}
are determined by the boundary data and also depend on the lattice
constant $\epsilon$. 

The somewhat unusual conditions (\ref{eq:A-12})
are implied by two delta functions that result from the Fourier transformation
and allow performing the $\xs_{1}$- and the $\xs_{\nu-1}$-integrations,
respectively. Together with (\ref{eq:A-11}), those conditions have
the effect of fixing the \emph{velocities} at the terminal points:
\begin{eqnarray}
\begin{split}
p^{\prime}=\frac{\xs_{1}-\xs_{0}}{\epsilon} & \rightarrow  \dot{\xs}\left(T_{0}\right)  \\
p=\frac{\xs_{\nu}-\xs_{\nu-1}}{\epsilon} & \rightarrow  \dot{\xs}\left(T\right)\,.
\end{split} \label{eq:A-13}
\end{eqnarray}
In this way, the $\nu\rightarrow\infty$ limit of (\ref{eq:A-10})
gives a precise meaning to the symbolic notation used in (\ref{eq:3-35-K_M-via-x_and_y-integral-and-deltas}).
\vspace{3mm}\\
{\bf (3) Example: the free particle.}
For simple potentials the nested multiple integrals encountered above
can be calculated explicitly for any given $\nu$ and the limit $\nu\rightarrow\infty$
can be taken. For example, in the case of the free particle, the reader
is invited to evaluate (\ref{eq:A-6}) and (\ref{eq:A-9}) at finite
$\nu$, to verify that both the original and the integrated-by-parts
version of the regularized path integral possess well defined continuum
limits, and to show that those are indeed equal, being
\begin{eqnarray}
{\cal K}\left(s,q,T;s^{\prime},q^{\prime},T_{0}\right) & = & \left[2\pi\hbar\left(T-T_{0}\right)\right]^{-N}\exp\left\{ \frac{i}{\hbar}\frac{\left(q-q^{\prime}\right)\left(s-s^{\prime}\right)}{T-T_{0}}\right\} \,.\label{eq:A-20}
\end{eqnarray}
Plugging (\ref{eq:A-20}) into (\ref{eq:3-26-K_M-via-Y}) then yields
the corresponding Moyal kernel finally:
\begin{eqnarray}
K_{{\rm M}}\left(p,q,T;p^{\prime},q^{\prime},T_{0}\right) & = & \delta\left(q-\left[q^{\prime}+p^{\prime}\left(T-T_{0}\right)\right]\right)\delta\left(p-p^{\prime}\right)\,.\label{eq:A-21}
\end{eqnarray}
As in all systems with a quadratic Hamiltonian, it equals the solution
to the classical Liouville equation.

\section{Particle in a magnetic field \label{sec:EM-Lagrangean}}

Since it is interesting in its own right, and as a background for
our discussion of the Bohm-Aharonov effect, we briefly describe the
Lagrangean Marinov integral and its classical limit for a particle
interacting with an external magnetic field ${\bf B}\left({\bf x}\right)\equiv\nabla \times{\bf A}\left({\bf x}\right)$.\footnote{Letting $N=3$ here, we follow the convention of denoting position
and momentum vectors in $\mathbb{R}^{3}$ by bold face letters.}
\vspace{3mm}\\
{\bf (1) Bilocal potentials.} 
We start out from the standard Lagrangean for a particle
of unit mass and charge $e$, coupled to a vector potential ${\bf A}\left({\bf x}\right)$
and scalar potential $V\left({\bf x}\right)$:
\begin{eqnarray}
L\left({\bf x},\dot{{\bf x}}\right) & = & \frac{1}{2}\dot{{\bf x}}^{2}+e\, \dot{{\bf x}}\cdot{\bf A}\left({\bf x}\right)-V\left({\bf x}\right)\,.\label{eq:B-1}
\end{eqnarray}
Following the steps outlined in Subsection \ref{subsec:Hamiltonian-path-integral}
we are again led to the formula (\ref{eq:3-23-Y-kernel-via-x_and_y})
for ${\cal K}\left({\bf s},{\bf q},T;{\bf s}^{\prime},{\bf q}^{\prime},T_{0}\right)$,
but with (\ref{eq:3-24-widetilde-Lagr_x_and_y}) replaced by
\begin{eqnarray}
\widetilde{L}\left({\bf \xs},{\bf \ys},\dot{{\bf \xs}},\dot{{\bf \ys}}\right) & = & \dot{{\bf \xs}}\cdot\dot{{\bf \ys}}+e\, \dot{{\bf \xs}}\cdot{\bf A}_{-}\left({\bf \xs},{\bf \ys}\right)+e\, \dot{{\bf \ys}}\cdot{\bf A}_{+}\left({\bf \xs},{\bf \ys}\right)-V\left({\bf \xs},{\bf \ys}\right)\,.\label{eq:B-2}
\end{eqnarray}
Herein the three bi-local potentials are given by
\begin{eqnarray}
{\bf A}_{+}\left({\bf \xs},{\bf \ys}\right) & = & \frac{1}{2}\left[{\bf A}\left({\bf \xs}+{\bf \ys}\right)+{\bf A}\left({\bf \xs}-{\bf \ys}\right)\right]\nonumber \\
{\bf A}_{-}\left({\bf \xs},{\bf \ys}\right) & = & \frac{1}{2}\left[{\bf A}\left({\bf \xs}+{\bf \ys}\right)-{\bf A}\left({\bf \xs}-{\bf \ys}\right)\right]\label{eq:B-3}\\
\widetilde{V}\left({\bf \xs},{\bf \ys}\right) & = & \frac{1}{2}\left[V\left({\bf \xs}+{\bf \ys}\right)-V\left({\bf \xs}-{\bf \ys}\right)\right]\,.\nonumber 
\end{eqnarray}
Given the (still exact) path integral for ${\cal K}\left({\bf s},{\bf q},T;{\bf s}^{\prime},{\bf q}^{\prime},T_{0}\right)$,
the pertinent exact Moyal kernel is obtained by the usual Fourier
transformation, equation (\ref{eq:3-26-K_M-via-Y}).
\vspace{3mm}\\
{\bf (2) Occurrence of the gauge field.} 
The semiclassical approximation of $K_{{\rm M}}$ is governed by the small-$\ys$
expansion of the bi-local potentials. At leading order,
\begin{eqnarray}
{\bf A}_{+}\left({\bf \xs},{\bf \ys}\right) & = & {\bf A}\left({\bf \xs}\right)+O\left({\bf \ys}^{2}\right)\nonumber \\
{\bf A}_{-}\left({\bf \xs},{\bf \ys}\right) & = & \left({\bf \ys}\cdot\nabla_{{\bf \xs}}\right){\bf A}\left({\bf \xs}\right)+O\left({\bf \ys}^{3}\right)\label{eq:B-4}\\
\widetilde{V}\left({\bf \xs},{\bf \ys}\right) & = & \left({\bf \ys}\cdot\nabla_{{\bf \xs}}\right)V\left({\bf \xs}\right)+O\left({\bf \ys}^{3}\right)\,.\nonumber 
\end{eqnarray}
The Lagrangean $\widetilde{L}$ reads at this order:
\begin{eqnarray}
\widetilde{L} & = & \frac{d}{dt}\left\{ {\bf \ys}\cdot\left[\dot{{\bf \xs}}+e{\bf A}\left({\bf \xs}\right)\right]\right\} \nonumber \\
 &  & +{\bf \ys}\cdot\left[-\ddot{{\bf \xs}}+e\dot{{\bf \xs}}\times{\bf B}\left({\bf \xs}\right)-\nabla V\left({\bf \xs}\right)\right] \label{eq:B-5} \\
 &  & +O\left({\bf \ys}^{3}\right)\,.\nonumber 
\end{eqnarray}
Proceeding as in Section \ref{sec:Path-integral-representation-Moyal-kernel}
we find in this approximation after performing the ${\bf \ys}$-integration:
\begin{eqnarray}
{\cal K}\left({\bf s},{\bf q},T;{\bf s}^{\prime},{\bf q}^{\prime},T_{0}\right) & = & \int_{{\bf \xs}\left(T_{0}\right)={\bf q}^{\prime}}^{{\bf \xs}\left(T\right)={\bf q}}{\cal D}{\bf \xs}\left(\cdot\right)\exp\left(\frac{i}{\hbar}\left\{ {\bf s}\cdot\left[\dot{{\bf \xs}}\left(T\right)+e{\bf A}\left({\bf q}\right)\right]-{\bf s}^{\prime}\cdot\left[\dot{{\bf \xs}}\left(T_{0}\right)+e{\bf A}\left({\bf q}^{\prime}\right)\right]\right\} \right)\nonumber \\
 &  & \times \delta\Bigr[\ddot{{\bf \xs}}-e\, \dot{{\bf \xs}}\times{\bf B}\left({\bf \xs}\right)+\nabla V\left({\bf \xs}\right)\Bigr]\,.\label{eq:B-6}
\end{eqnarray}
Note that the first term on the RHS of (\ref{eq:B-5}), a total derivative,
has augmented the terminal velocities in the phase factors under the
integral to $\dot{{\bf \xs}}+e{\bf A}\left({\bf \xs}\right)$, evaluated
at the initial and final point, respectively. This combination is
necessary to guarantee the gauge covariance of the formalism at the
level of the function ${\cal K}$.
\vspace{3mm}\\
{\bf (3) Classical kernel.} 
The strictly classical special case
of the kernel $K_{{\rm M}}$ is obtained by inserting
(\ref{eq:B-6}) into (\ref{eq:3-26-K_M-via-Y}) and performing the
Fourier transformation with respect to ${\bf s}$ and ${\bf s}^{\prime}$.
This yields the classical Moyal kernel in the form
\begin{eqnarray}
\lim_{\hbar \rightarrow 0}
K_{{\rm M}}\left({\bf p},{\bf q},T;{\bf p}^{\prime},{\bf q}^{\prime},T_{0}\right) & = & \int{\cal D}{\bf \xs}\left(\cdot\right)\delta \Bigr[\ddot{{\bf \xs}}-e\dot{{\bf \xs}}\times{\bf B}\left({\bf \xs}\right)+\nabla V\left({\bf \xs}\right)\Bigr]\,.\label{eq:B-7}
\end{eqnarray}
Hereby the functional integration over ${\bf \xs}\left(t\right)$ is
subject to the following four boundary conditions: 
\begin{eqnarray}
\begin{split}
{\bf \xs}\left(T\right)={\bf q} & , \qquad \dot{{\bf \xs}}\left(T\right)={\bf p}-e{\bf A}\left({\bf q}\right) \\
{\bf \xs}\left(T_{0}\right)={\bf q}^{\prime} & , \qquad \dot{{\bf \xs}}\left(T_{0}\right)={\bf p}^{\prime}-e{\bf A}\left({\bf q}^{\prime}\right)\,.
\end{split}\label{eq:B-8}
\end{eqnarray}

We see that the integrand of (\ref{eq:B-7}) is strictly localized
on solutions of the classical equation of motion, $\ddot{{\bf \xs}}=e\,  \dot{{\bf \xs}}\times{\bf B}-\nabla V$,
which feels the magnetic field only via ${\bf B}$, the gauge invariant
curl of ${\bf A}$. Instead, the boundary conditions (\ref{eq:B-8})
depend on ${\bf A}$ directly. There$,$ the presence of the vector
potential is a consequence of the crucial total derivative terms in
the $\widetilde{L}$ of equation (\ref{eq:B-5}). Thus the classical
limit of $K_{{\rm M}}$ as given by (\ref{eq:B-7}) with (\ref{eq:B-8})
depends on the terminal momentum only via the combination ${\bf p}-e{\bf A}\left({\bf q}\right)$,
the counterpart of a covariant derivative in the phase space formulation.
\vspace{3mm}\\
{\bf (4) Gauge invariant probabilities.}
As a final consistency check we mention that equation (\ref{eq:3-14-Y-funct_at_s=00003D0})
continues to be true in presence of a vector potential. Together with
(\ref{eq:B-6}), evaluated at vanishing ${\bf s}$-arguments, it implies
that in the classical limit
\begin{eqnarray}
\left|K\left({\bf q},T;{\bf q}^{\prime},T_{0}\right)\right|^{2} & = & \int_{{\bf \xs}\left(T_{0}\right)={\bf q}^{\prime}}^{{\bf \xs}\left(T\right)={\bf q}}{\cal D}{\bf \xs}\left(\cdot\right)\delta\left[\ddot{{\bf \xs}}-e\dot{{\bf \xs}}\times{\bf B}\left({\bf \xs}\right)+\nabla V\left({\bf \xs}\right)\right]\,.\label{eq:B-9}
\end{eqnarray}
As it must be, the path integral in (\ref{eq:B-9}) is indeed seen
to be a positive and manifestly gauge independent function of the
position and time arguments, at least formally \cite{Schulman:1981vu,Dittrich:1992et,Gozzi-Book}.

% Create the reference section using BibTeX:
\bibliography{paper}

\end{document}